\documentclass{aa} 
\usepackage{graphicx}
\usepackage{txfonts}
\usepackage{subfigure}
\usepackage{natbib}
\usepackage{ stmaryrd }
\usepackage{multirow}

\begin{document}
   \title{CO outflows from high-mass Class 0 protostars in Cygnus-X}

   \subtitle{}

\author{A. Duarte-Cabral\inst{1,2}
  \and S. Bontemps\inst{1,2}
  \and F. Motte\inst{3}
  \and M. Hennemann\inst{3}
  \and N. Schneider\inst{1,2}
  \and Ph. Andr\'e\inst{3}
  }

\offprints{Ana Duarte Cabral, \email{Ana.Cabral@obs.u-bordeaux1.fr}}

 \institute{Univ. Bordeaux, LAB, UMR 5804, F-33270, Floirac, France
 \and CNRS, LAB, UMR 5804, F-33270, Floirac, France
 \and Laboratoire AIM, CEA/DSM-CNRS-Universit\'e Paris Diderot, IRFU/Service d'Astrophysique, C.E. Saclay, Orme de merisiers, 91191 Gif-sur-Yvette, France}
 
\date{Received 2 March 2013; accepted 17 August 2013}


  \abstract
   {
   It is unclear whether high-mass cores in monolithic collapse exist or not, and what the accretion process and origin of the material feeding the precursors of high-mass stars are. As outflows are natural consequences of the accretion process, they represent one of the few (indirect) tracers of accretion.}
   {We aim to search for individual outflows from high-mass cores in Cygnus X and to study the characteristics of the detected ejections.} 
   {We used CO (2-1) PdBI observations towards six massive dense clumps, containing 9 high-mass cores. We estimated their bolometric luminosities and masses, and measured the energetics of outflows. We compared our sample to low-mass objects studied in the literature and developed simple evolutionary models to reproduce the observables.}
   {We find that 8 out of 9 high-mass cores are driving clear individual outflows, representing true equivalents of Class 0 protostars in the high-mass regime. The remaining core has only a tentative outflow detection. It could be amongst the first examples of a true individual high-mass prestellar core. We also find that the momentum flux of high-mass objects has a linear relation to the reservoir of mass in the envelope, as a scale-up of the relations found for low-mass protostars. This suggests a fundamental proportionality between accretion rates and envelope masses.}
   {We conclude that high-mass Class 0 protostars exist, and their collapsing envelopes have similar sizes and a similar fragmentation scale to the low-mass equivalents, with enough mass to directly form high-mass stars from a monolithic collapse. If the pre-collapse evolution is quasi-static, the fragmentation scale is expected to limit the size of the initial mass reservoirs for all masses leading to higher densities at birth and therefore shorter free-fall times for higher mass stars. However, we find the collapse timescales to be similar for both low- and high-mass objects. This implies that in a quasi-static view, we would require significant turbulent/magnetic support to slow down the collapse of the more massive envelopes. But with this support still to be discovered, and with the indications of large dynamics in pre-collapse gas for high-mass star formation, we propose that such an identical collapse timescale implies that the initial densities, which should set the duration of the collapse, should be similar for all masses. This suggests that the mass that ultimately incorporates massive stars has to have been accreted from larger scales than those of low-mass stars, and in a dynamical way. }


   \maketitle
%


\section{Introduction}

Molecular outflows are important to better understand star formation. Collapsing protostars are so embedded in their cores and envelopes that few characteristics can be measured directly. Only the total luminosity, the envelope mass, and the power of the outflows do in practice characterise the young protostars. Molecular outflows are believed to be mostly made up of gas entrained by a fast jet and wind that cannot be observed directly \citep[e.g.,][]{1994ApJ...422..616S}. These jets and winds would be formed as the result of a complex magnetohydrodynamic (MHD) process coupling rotation, magnetic field, and gravity, to extract angular momentum from the accreting material \citep[e.g.,][]{1982MNRAS.199..883B,1992ApJ...394..117P,1995A&A...295..807F,2008A&A...477....9H,2011ApJ...742...56V}. Outflows and ejections are natural consequences and tracers of accretion. It is by observing high-velocity wings in the CO lines that protostellar ejections are best investigated. These CO outflows are used to trace the existence of accretion, and measuring their momentum fluxes ($F_{\rm{co}}$) provides (indirect) measurements of the accretion rates ($\dot{M}_{\rm acc}$), assuming that $F_{\rm{CO}} \propto \dot{M}_{\rm acc}$.

It is widely accepted that low-mass stars form by the collapse of cores stemming from the fragmentation of molecular clouds \citep[e.g.,][]{1997MNRAS.288..145P,1998A&A...336..150M,2000prpl.conf...59A,2005ApJ...620..786K,2008ApJ...684..395H}. The precise origin of the fragmentation is not fully understood, but the most recent observations indicate that molecular clouds are dominated by turbulent motions that create a population of gravitationally unstable cores that may then collapse. These cores have typically the local Jeans mass. Magnetic fields may regulate this process by providing additional support and by guiding turbulent flows to primarily form sheets and filaments that can themselves be gravitationally unstable \citep[e.g.,][]{1996ApJ...467..321N,1999ApJ...510..274N}. This scenario explains the formation of low-mass stars well.  

The origin of high-mass stars, defined as ionising OB stars with more than $\sim 8\,$M$_\odot$, is much more enigmatic since their final masses significantly exceed the Jeans masses in dense regions of molecular clouds. There are two lines of theoretical ideas to overcome this enigma. Either the local Jeans mass is increased on rare occasions owing to higher internal support, for instance from turbulence \citep[e.g.,][]{2008ApJ...684..395H} or a mix of turbulence and magnetic fields \citep[e.g.,][]{2003ApJ...585..850M}, or the cores at the centre of a protocluster clump continue to gather mass from their surroundings while collapsing \citep[e.g.,][]{2006MNRAS.370..488B}. Studies of outflows can help distinguish between these scenarios. In fact, well collimated outflows can provide hints of the number of individual protostars present inside a high-mass star-forming clump, by pin-pointing the individual collapsing cores, i.e. the Class 0 protostars. Outflows as tracers of accretion can also indicate whether a particular high-mass core is protostellar or prestellar, the latter not being expected to exist in a competitive accretion scenario \citep[e.g.,][]{2001MNRAS.324..573B}. 
Finally, from a statistical approach, outflows trace accretion rates \citep[e.g.,][]{1996A&A...311..858B} and together with other observables such as the mass in the core and total luminosity, they help to constrain the evolutionary models of the formation of high-mass stars. For high-mass protostars, outflows are in fact the only accretion tracer since their luminosities are rapidly dominated by their stellar luminosities and not by their accretion luminosities \citep[e.g.,][]{1993ApJ...418..414P}.

\begin{table*}[!ht]
\caption{Observing parameters and obtained beam sizes and rms.}
\label{obs_summary}
\renewcommand{\footnoterule}{}  
\begin{tabular}{c | c c c c c  | c l c}
\hline \hline
 Source          &    \multicolumn{2}{c}{Phase centre  (J2000)}  & Synthesised beam & P.A. &  r.m.s.$^{(a)}$ & \multicolumn{2}{c}{Number of fragments$^{(b)}$} & \multirow{2}{*}{Morphology} \\
name     & R.A.     &  Dec.                    & [arcsec$\,\times\,$arcsec]      & [degrees]    & [{\rm  mJy/beam}]    & Total & \, \, High-mass & \\
\hline
 CygX-N3   & 20 35 34.1   &  42 20 05.0      &  1.25$\,\times\,$1.07   & 57   &  36  &    4 	& \, \,  2 ($>$\,10\,M$_\odot$) 	& Elongated \\
 CygX-N12  & 20 36 57.4   &  42 11 27.5      &  1.33$\,\times\,$0.95  & 24   &  32  &    4	& \, \, 2 ($>$\,15\,M$_\odot$)   & Elongated \\
 CygX-N40  & 20 38 59.8   &  42 23 42.0      &  1.24$\,\times\,$1.07  & 57   &  35  &    1 	& \, \, 0     				& Diffuse\\
 CygX-N48  & 20 39 01.5   &  42 22 04.0      &  1.27$\,\times\,$1.03  & 39   &  51  &    5	& \, \, 2 ($>$\,10\,M$_\odot$)	& Clustered \\
 CygX-N53  & 20 39 03.1   &  42 25 50.0      &  1.27$\,\times\,$1.03  & 38   &  51  &    7	& \, \,  2 ($>$\,20\,M$_\odot$)	& Centrally condensed \\
 CygX-N63  & 20 40 05.2   &  41 32 12.0      &  1.34$\,\times\,$0.94  & 23   &  32  &    3	& \, \,  1 ($>$\,40\,M$_\odot$) 	& Centrally condensed \\
\hline
\end{tabular}
\\
$^{(a)}$ 1$\sigma$ r.m.s. estimated in 0.4 km\,s$^{-1}$ channels.\\
$^{(b)}$ From \citet[][]{2010A&A...524A..18B}.
\end{table*}

Outflows in high-mass star-forming regions have often been observed at low spatial resolution (i.e. physically large scales due to the large distances) resulting in mapping low-collimated lobes associated with cluster-forming clumps \citep[e.g.,][]{2002A&A...383..892B,2004ApJ...604..258S,2005ApJ...625..864Z}. Detections and studies of individual massive protostellar outflows are rare \citep[e.g.,][]{2002A&A...387..931B,2009ApJ...696...66Q,2011ApJ...728....6Q,2011ApJ...735...64W}. From these it appears that when the scale of individual cores can be reached, the CO outflows are collimated and have similar properties to the better studied outflows from low-mass protostars. They are also systematically more powerful, as expected from the need for high accretion rates for high-mass protostars ($> 10^{-5}$ M$_{\odot}$\,yr$^{-1}$), which are essential to surpass the radiation pressure and continue accreting beyond 10$\,$M$_\odot$  \citep[][]{1987ApJ...319..850W}.

We present here the results of a high angular resolution study of the CO outflows towards six IR-quiet massive dense clumps (hereafter MDCs, see \citealp{2007A&A...476.1243M})\footnote{Owing to their small sizes (of the order of 0.1\,pc), these clumps have so far been considered as cores (massive dense cores; e.g. see  \citealp{2007A&A...476.1243M}). However, following the nomenclature by \citet{2000prpl.conf...97W}, we adopt the designation of clumps for these 0.1\,pc regions, because these are sub-fragmented into several individual cores \citep{2010A&A...524A..18B}, and will therefore form a small cluster of stars.} in Cygnus X at a distance of 1.4~kpc \citep[][]{2012A&A...539A..79R} containing nine high-mass cores \citep{2010A&A...524A..18B}. An overview of the Cygnus-X region and a more detailed description of the sample of MDCs we study here can be found in Sect.~\ref{sec:cygnusX}, and in Sect.~\ref{sec:obs} we detail the observations. Section~\ref{sec:results} presents the results on the spectral energy distribution (SED) fittings and outflow properties for our sample of nine high-mass cores, which are then analysed in Sect.~\ref{sec:analysis}, with a comparison with the low-mass protostellar properties. Section~\ref{sec:model} presents a set of three evolutionary models that we attempt to constrain with observables. We discuss the implications of our results in Sect.~\ref{discussion}, and our conclusions are outlined in Sect.~\ref{concl}.


\section{Massive dense clumps in Cygnus X}
\label{sec:cygnusX}

Cygnus X is the richest high-mass star-forming molecular cloud complex located at less than 3 kpc from the Sun. It contains $4 \times 10^{6}$\,M$_{\odot}$ of molecular gas \citep[][]{2006A&A...458..855S} extending over $\sim$~100~pc in diameter, and it hosts a number of H{\small II} regions, the product of recent high-mass star formation \citep[e.g.,][]{1991A&A...241..551W}. It is associated with several young OB associations \citep[e.g.,][]{2001A&A...371..675U}, including one of the largest in our Galaxy, Cyg-X OB2 \citep[e.g.,][]{2000A&A...360..539K}.

A survey of the 1.2~mm emission in Cygnus X was made by \citet[][]{2007A&A...476.1243M}, where several MDCs, with more than 40\,M$_{\odot}$ in 0.1\,pc, were found in the region. Among these, \citet[][]{2007A&A...476.1243M} identified 17 IR-quiet MDCs, which correspond to sources with a flux of less than 10\,Jy at 21$\mu$m (i.e. with a bolometric luminosity of less than 10$^{3}$ L$_{\odot}$). With high masses and low luminosities, these IR-quiet MDCs are the best candidates for embedding the young massive protostars. As such, a flux-limited selection of six IR-quiet MDCs (Cyg-X N3, N12, N40, N48, N53, and N63) were followed up by \citet[][]{2010A&A...524A..18B} with the IRAM\footnote{IRAM is supported by INSU/CNRS (France), MPG (Germany), and IGN (Spain).} Plateau de Bure Interferometer (hereafter PdBI) in the 1.3 and 3.5\,mm continuum emission. 
This sample of MDCs is that studied here, and can be considered to be representative of high-mass star formation in a single complex. 

Despite the similar properties of these six MDCs on a spatial scale of $\sim$0.1\,pc \citep[][]{2007A&A...476.1243M}, they span a range of different environments. While CygX-N40, N48 and N53 are located along the massive DR21 filament/ridge, a highly dynamical star formation site \citep[][]{2010A&A...520A..49S,2012A&A...543L...3H}, CygX-N63 is a relatively isolated clump to the south of DR21. On the other hand, CygX-N3 and N12 are situated to the west of DR21, in the DR17 region, and they both show morphologies consistent with being influenced by the winds from the nearby OB clusters \citep[][]{2006A&A...458..855S}. CygX-N3 in particular sits at the tip of a DR17 pillar.
 
The interferometric continuum observations of these six MDCs by \citet[][]{2010A&A...524A..18B} have shown that they are fragmented into a number of cores (see the last three columns of Table~\ref{obs_summary} for details). Among the cores detected with the PdBI, nine have masses between 10 and 50~M$_{\odot}$ within $\sim$4000\,AU. These are CygX-N3 MM1 and MM2, CygX-N12 MM1 and MM2, CygX-N48 MM1 and MM2, CygX-N53 MM1 and MM2, and CygX-N63 MM1, and they represent excellent candidates to be individual high-mass Class 0 protostars. Four out of nine are detected at 24$\,\mu$m but with weak fluxes below 0.5~Jy. Only in CygX-N40 are there no high-mass fragments (only one core of $<$ 2~M$_{\odot}$ is detected), and most of the single-dish continuum emission is extended and filtered out in PdBI. 


\section{Observations}
\label{sec:obs}




The six MDCs have been observed in 2004 with the PdBI in the 1.3\,mm and 3.5\,mm continuum emission and in four spectral units covering the $^{12}$CO\,($2-1$), SiO\,($2-1$), H$^{13}$CO$^+$\,($1-0$), and H$^{13}$CN\,($1-0$) lines at 230.54, 86.85, 86.75, and 86.34 GHz, respectively. The observations in the continuum and in the H$^{13}$CO$^+$\,($1-0$) and H$^{13}$CN\,($1-0$) lines are reported in \citet[][]{2010A&A...524A..18B} and \citet[][]{2011A&A...527A.135C}, respectively. We present here the $^{12}$CO\,($2-1$) observations dedicated to CO outflows driven by young protostars.

The observations 
were performed in track-sharing mode with two targets per track for the following pairs:
CygX-N48/CygX-N53, CygX-N3/CygX-N40, and CygX-N12/CygX-N63. The D configuration track observations were performed between June and October 2004 with five antennas with baselines ranging from 24 to 82\,m. The C configuration tracks were obtained in November and December 2004 with six antennas in 6Cp with baselines ranging from 48 to 229\,m. As a phase calibrator, we mostly used the bright nearby quasar 2013+370 and as a flux calibrator the evolved star MWC349 located in Cygnus\,X. 

To recover all spatial scales, the zero-spacing was obtained with the IRAM 30m telescope as part of an observing run dedicated to the DR21 filament (see  \citealp[][]{2010A&A...520A..49S} for more details). OTF maps of $^{12}$CO\,($2-1$) of CygX-N3, N12 and N63 were observed on 5 June 2007 using the A230 receiver with the VESPA correlator. The average system temperature was $\sim$275 K. Sources CygX-N48, N53, and N40 were observed between 2 and 5 June 2007 as part of the large-scale OTF mapping of the DR21 filament, and these maps have average system temperatures around 600 K. The data were corrected for the main beam efficiency of 0.52. The half power beamwidth (HPBW) at this frequency is $\sim$11".

After combining the zero-spacing information from the IRAM 30m with the PdBI data, the maps were cleaned using the natural weighting to favour the highest sensitivity. The resulting synthesised beam and rms in the continuum are summarized in Table~\ref{obs_summary}, together with the field names and centres of phase. The cleaning components were searched across the whole area of the primary beams. No support for cleaning was used to avoid introducing any bias into the resulting maps of emission.



\section{Results}
\label{sec:results}

\subsection{Bolometric luminosities and envelope masses}
\label{sec:sample}

To estimate the properties of our sample of nine high-mass cores, we have estimated their masses and bolometric luminosities by constructing SEDs using the 24$\mu$m from Spitzer MIPS, the 70, 160, 250, 350, and 500$\mu$m from \emph{Herschel} PACS and SPIRE observed as part of the HOBYS programme\footnote{The \emph{Herschel} imaging survey of OB Young Stellar objects (HOBYS) is a \emph{Herschel} key programme. See http://hobys- herschel.cea.fr} (\citealp{2010A&A...518L..77M}; \citealp{2012A&A...543L...3H}; in prep), the 1.2mm from MAMBO \citep[][]{2007A&A...476.1243M}, and the 1.3mm and 3.5mm emission from PdBI \citep[][]{2010A&A...524A..18B}.  Since our sample of sources consists of IR-quiet cores, the flux below 8$\mu$m is always negligible, and integrating the SEDs down to 24$\mu$m is enough to get a good estimate of the bolometric luminosity. Full details on the flux determinations and SED fittings can be found in Appendix~\ref{ap:SEDs}. The source CygX-N40 MM1 is not included in the nine high-mass cores under study, not only because it is a low-mass core, but also because we were unable to recover a good SED fitting for estimating its bolometric luminosity. 

\begin{figure}[!t]
	\centering
	{\renewcommand{\baselinestretch}{1.1}
	\includegraphics[width=0.48\textwidth]{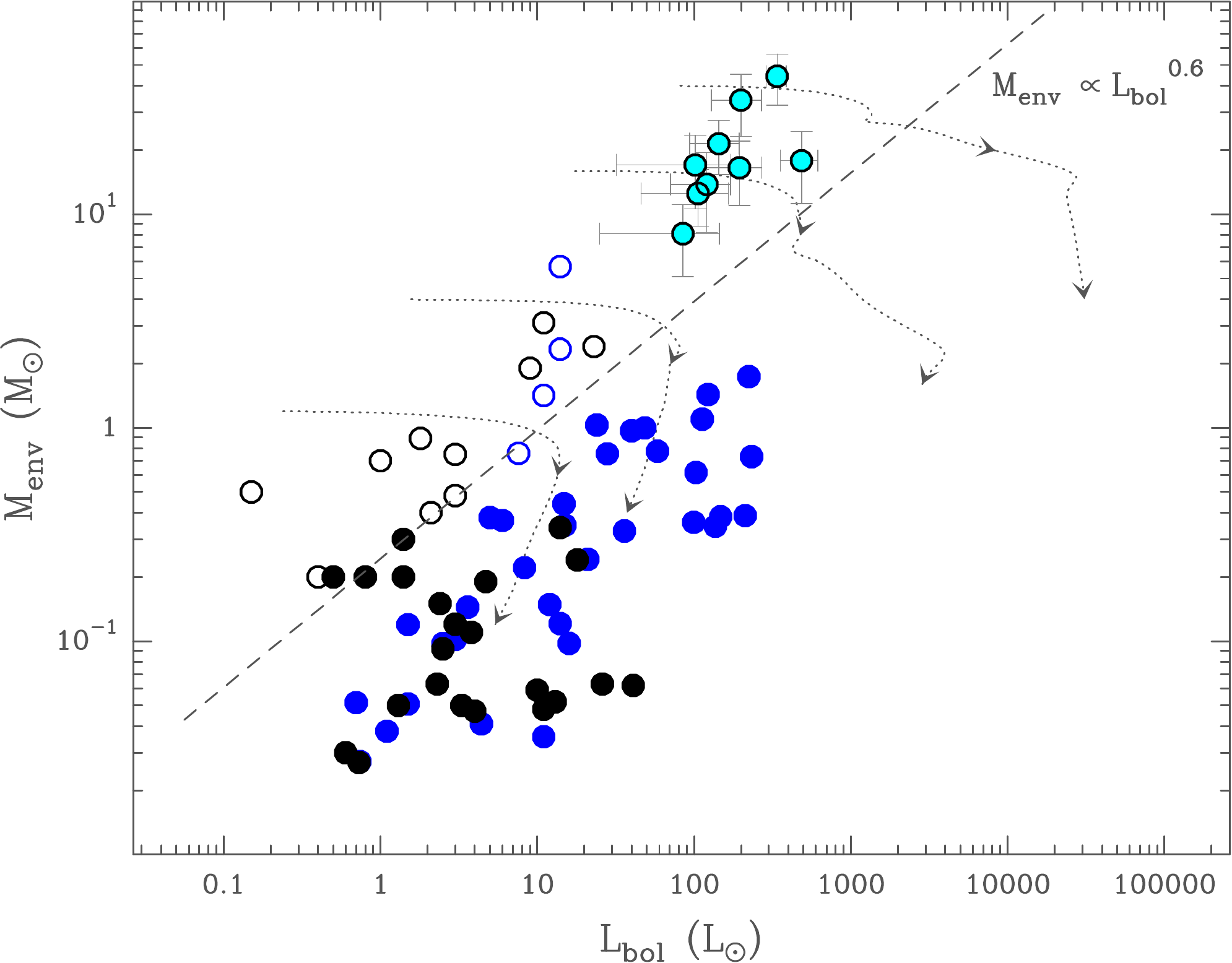}
	\caption[]{\small{Envelope/core mass with respect to the bolometric luminosity. For the massive protostars studied here, in light blue circles, the values are as estimated from the SED fittings (see Sect.~\ref{sec:sample} and Appendix~\ref{ap:SEDs} for more details). The remaining filled and empty symbols are the low-mass Class I and Class 0 objects, respectively (from \citealt{1996A&A...311..858B} in black, and from \citealt{2001A&A...365..440M} and \citealt{2000prpl.conf...59A}  in dark blue). The dotted curves show four examples of evolutionary tracks with decreasing accretion rates, for envelopes with initial masses of 0.6, 2, 8, and 20~M$_{\odot}$ (see Sect.~\ref{sec:model}). The arrows show the positions in each track where 50\% and 90\% of the envelope mass has been accreted onto the protostar. The dashed line represents the $M_{\rm  env} \propto L_{\rm  bol}^{0.6}$ relation. The placement of the curve at $M_{\rm  env} / L_{\rm  bol}^{0.6} = 0.25 M_{\odot}/L_{\odot}^{0.6}$ comes as an empirical border line between Class 0 and Class I evolutionary stages.}}
	\label{fig:mass_lum_nomodel}}
\end{figure} 

The mean separation between the high-mass cores detected by \citet[][]{2010A&A...524A..18B} is of the order of 3000-5000\,AU. We use the average separation of 4000\,AU to define the size (full width at half maximum, FWHM) of the cores for which we estimate the mass and bolometric luminosities, so as to be able to derive the properties of regions where the envelopes of different sources do not overlap.  Interestingly, these sizes are similar to the core sizes found in low-mass star-forming regions \citep[e.g. $\rho$-Ophiuchi,][]{1998A&A...336..150M, 2000prpl.conf...59A}. Because the PdBI 1.3mm emission filters out the emission that arises from outside the central $\sim$1000\,-\,2000\,AU, the 1.3mm fluxes used to construct the SEDs are rescaled to envelope sizes of 4000\,AU FWHM, by assuming a density profile, outside the central $\sim$1000\,AU, as $\rho \propto r^{-2}$ \citep[see][and Appendix~\ref{ap:SEDs} for more details]{2010A&A...524A..18B}. The only exception is CygX-N63 MM1, an extremely compact core, where all the single-dish 1.2mm emission \citep[MAMBO,][]{2007A&A...476.1243M} is in fact recovered by rescaling the flux to simply $\sim 2500$\,AU. A summary of the inferred masses and bolometric luminosities from Appendix~\ref{ap:SEDs} are shown in Table~\ref{tab:luminosities}. For reference, the last row of Table~\ref{tab:luminosities} shows an example of a low-mass Class 0 protostar, VLA1623 \citep[e.g.,][]{1993ApJ...406..122A,1999ApJ...513L..57A,1996A&A...311..858B}.

\begin{table*}[!t]
\small
\caption{Source properties, velocities, and outflow momentum fluxes.}
\begin{tabular}{l | c c | c c c | c}
\hline 
\hline 
\multirow{2}{2cm}{Source} 	&	\multirow{1}{2cm}{\centering $M_{\rm env}$}	 &	\multirow{1}{2cm}{\centering $L_{\rm bol}$} & \multirow{1}{1cm}{\centering$v_{o}$ }	& \multirow{1}{1cm}{\centering $v_{peak}$}		& \multirow{1}{1cm}{\centering $\sigma_v$} 		& \multirow{1}{3cm}{\centering  $F_{\rm{co}}$ $/ 10^{-5}$ } \\
		& 	\multirow{1}{*}{(M$_{\odot}$) }	 &	\multirow{1}{*}{(L$_{\odot}$)}  					&  \multicolumn{3}{c |}{(km\,s$^{-1}$)} 	&  (M$_{\odot}$\,km\,s$^{-1}$\,yr$^{-1}$)\\  
\hline
CygX-N3 MM1	& 	12.5		$\pm$ 3.7		&	106	$\pm$	60 	& \multirow{2}{*}{15.1}	& \multirow{2}{*}{15.5} 	& \multirow{2}{*}{2.1} 	& \, 131 \\
CygX-N3 MM2	&	13.8		$\pm$ 5.6		&	121	$\pm$	50 	&					&  					& 				 	& \,\,\, 72 \\
\hline
CygX-N12 MM1	&	17.7		$\pm$ 6.9		&	481	$\pm$	130	&\multirow{2}{*}{15.2}	& \multirow{2}{*}{16.1} 	& \multirow{2}{*}{3.0} 	& $>$ 36 \\
CygX-N12 MM2	&	16.4		$\pm$ 5.8		&	185	$\pm$	75	&					& 					& 				 	& $>$ 12 \\
\hline
CygX-N40 MM1	&	$<$ 2				&	-				&  -3.5				& 	-2.5				& 	3.0				& \,\,\,\,\,\,\,  7 \\
\hline
CygX-N48 MM1	&	17.0		$\pm$ 6.4		&	102	$\pm$	70	&\multirow{2}{*}{-3.5}		& \multirow{2}{*}{-3.6} 	& \multirow{2}{*}{2.3} 	& \,\, 135 \\
CygX-N48 MM2	&	8.1		$\pm$ 3.0		&	85	$\pm$	60	&					& 					& 					& \,\,\,\, 45 \\
\hline
CygX-N53 MM1	&	34.2		$\pm$ 11.1	&	199	$\pm$	70	&\multirow{2}{*}{-4.1}		& \multirow{2}{*}{-5.5}	& \multirow{2}{*}{3.2}		& \,\, 412 \\
CygX-N53 MM2	&	21.4		$\pm$ 6.1		&  	144 $\pm$	50	&					& 					& 					& $<$ 121 \\
\hline
CygX-N63 MM1	&	44.3		$\pm$ 11.9	&	339	$\pm$	50	&-4.3				&	-3.3				& 	2.3				& \,\, 291\\
		   	&				  	&					&				  	& 	 			     	&            				&  \\
VLA1623		&		0.7$^{(a)}$	&	1.0$^{(a)}$		&			-		&			-		&		-			& \,\,\,\,  14$^{(b)}$ \\	
\end{tabular}
\\

 {$^{(a)}$ From \citet[][]{1993ApJ...406..122A,1999ApJ...513L..57A}; $^{(b)}$ From \citet[][]{1996A&A...311..858B} } 
\label{tab:summary_velo_energetics}
\label{tab:luminosities}
\end{table*}

Figure~\ref{fig:mass_lum_nomodel} shows the correlations of the core mass with respect to the bolometric luminosity for the sample of nine high-mass protostars studied here. The low-mass samples of Class 0 and Class I objects from \citet[][]{1996A&A...311..858B}, \citet[][]{2001A&A...365..440M} and \citet[][]{2000prpl.conf...59A} are also shown. The values of mass and luminosity of some of the sources from \citet[][]{1996A&A...311..858B} have been updated according to the estimate from \citet[][]{2000prpl.conf...59A}, and we have not included sources whose estimate of the outflow momentum flux consisted of an upper limit. This figure also shows evolutionary tracks for several envelope masses \citep[adapted from][see Sect.~\ref{sec:model}]{1996A&A...311..858B,2000prpl.conf...59A,2008A&A...490L..27A}, and the curve of $M_{\rm  env} \propto L_{\rm  bol}^{0.6}$, marking a conceptual border between Class 0 and Class I sources.

The nine Cygnus X individual high-mass cores of our sample had been interpreted by \citet[][]{2010A&A...524A..18B} to correspond to true single high-mass protostellar objects. The absence or very little emission in the near-IR would place these objects as prestellar cores, where no stellar embryo has formed yet, or Class 0 protostars, where the envelope masses are still higher than the correspondent stellar masses.  
And in fact, they are not found to be very luminous (for high-mass star precursors) with bolometric luminosities ranging from $\sim 100$ to 500$\,$L$_\odot$. More evolved high-mass protostars have typical luminosities in the range $10^4$ to $10^5\,$L$_\odot$ (e.g. Kurtz et al. 2000). They are, however, much more luminous than low-mass Class 0 protostars \citep[$\sim 0.3$ to 30$\,$L$_\odot$, see e.g.,][]{2000prpl.conf...59A}, and their location in the  $M_{\rm  env} - L_{\rm  bol}$ diagram, together with the evolutionary tracks, actually places them as early phase protostars (Class~0), which will form stars with masses ranging from $\sim 10$ to possibly more than 25$\,$M$_\odot$. 

\begin{figure*}[!t]
	\centering
	{\renewcommand{\baselinestretch}{1.1}
	\includegraphics[width=0.4\textwidth]{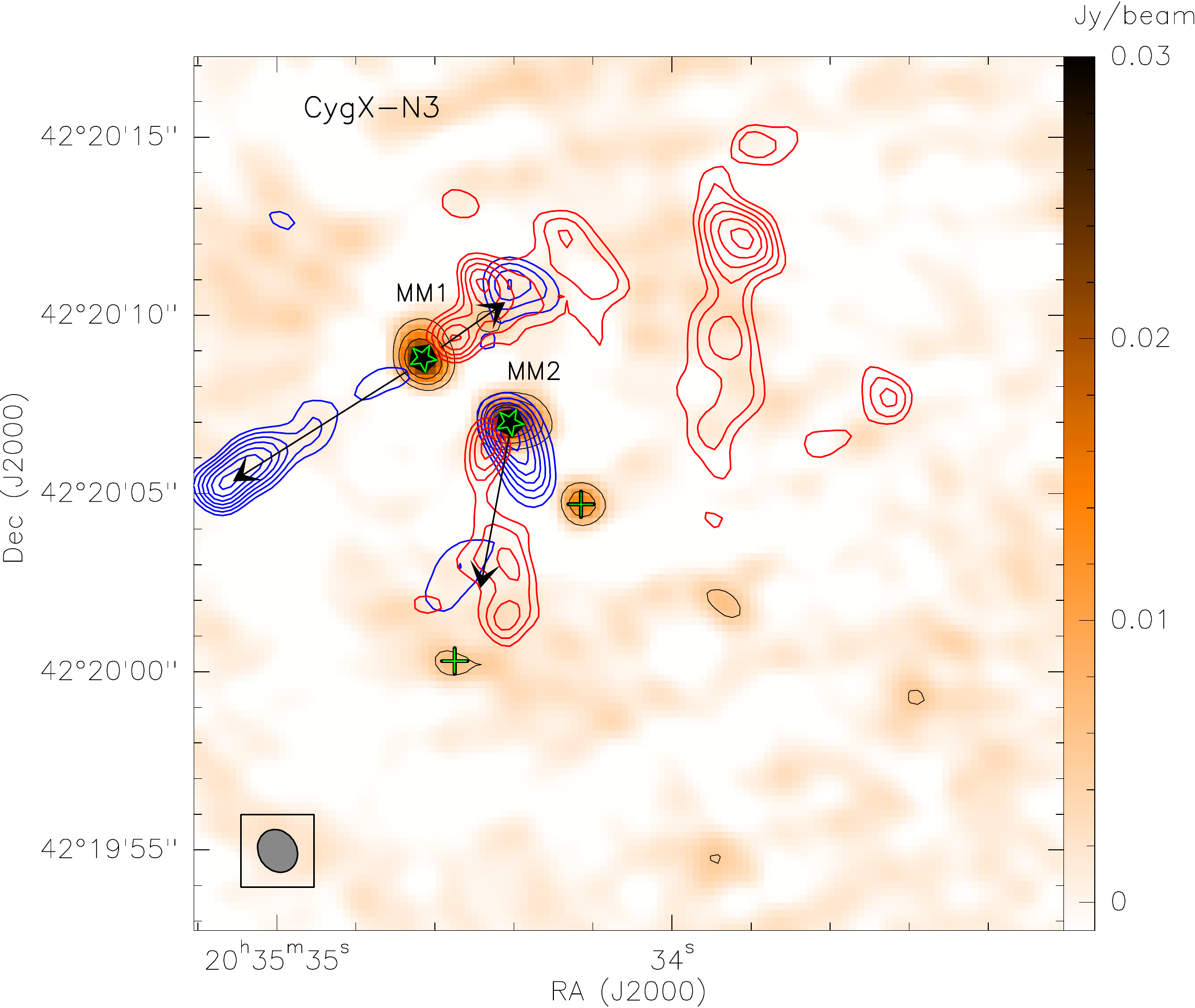}
	\hspace{1cm}
	\includegraphics[width=0.4\textwidth]{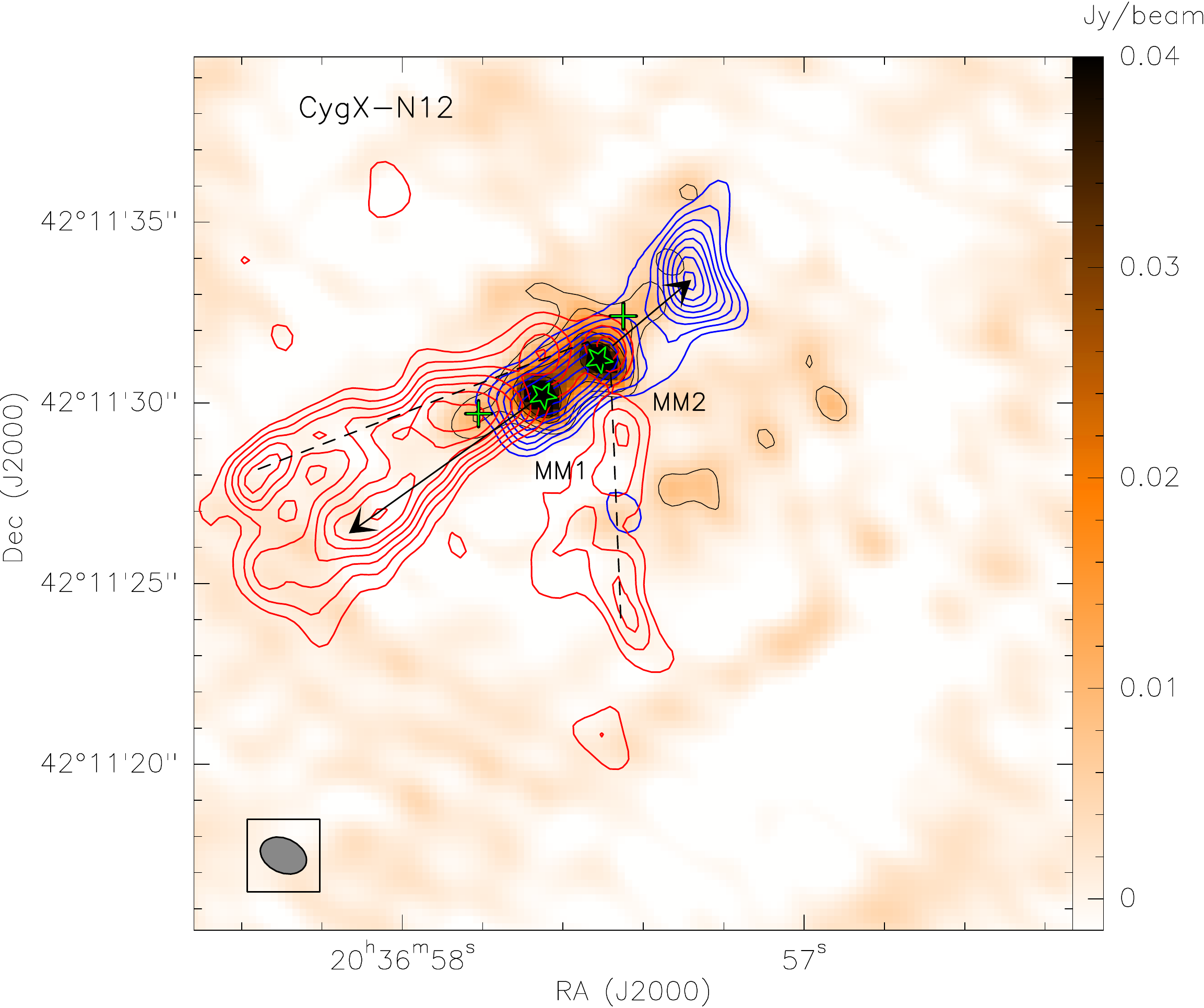}\\
	\vspace{0.3cm}
	\includegraphics[width=0.4\textwidth]{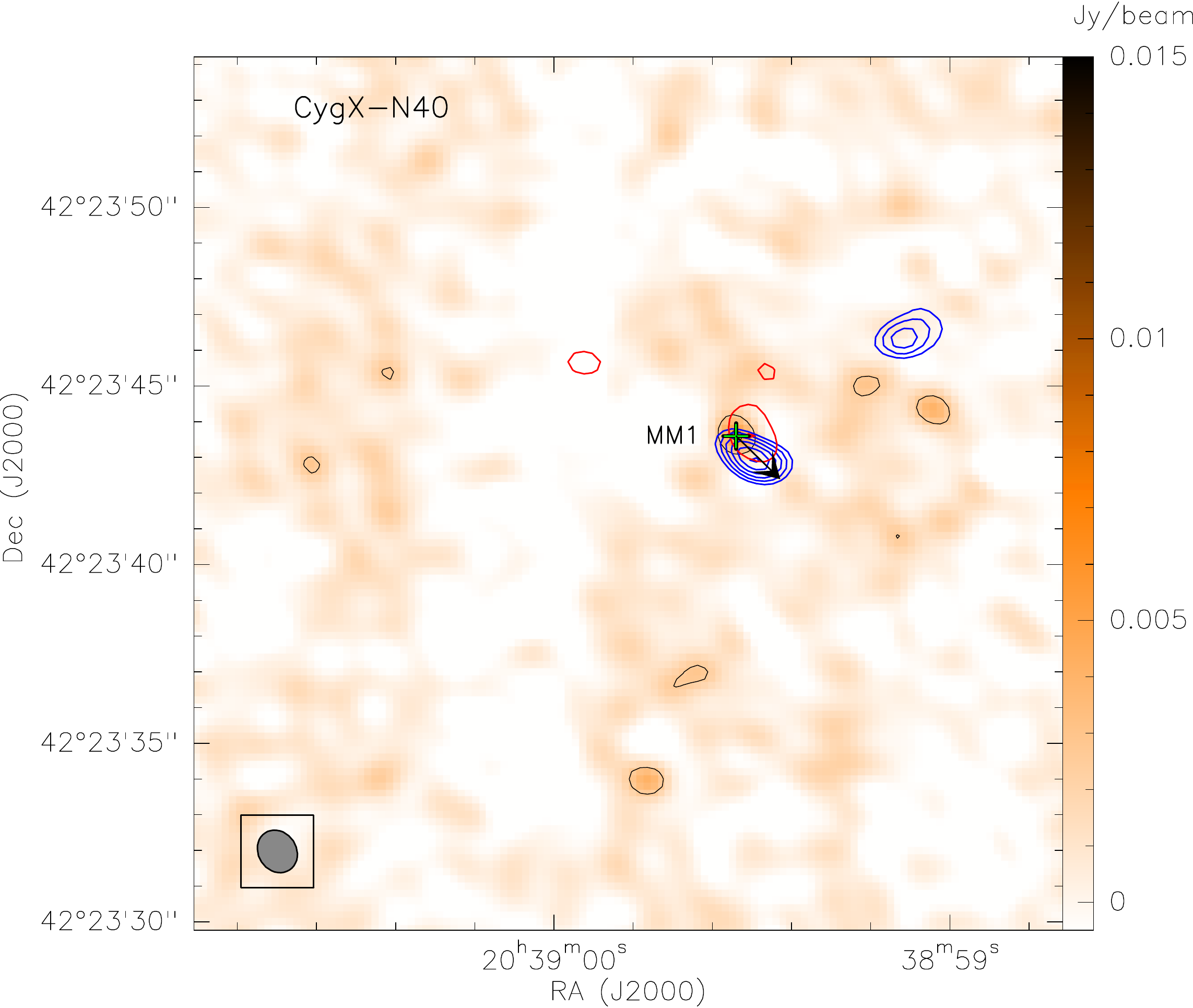}
	\hspace{1cm}
	\includegraphics[width=0.4\textwidth]{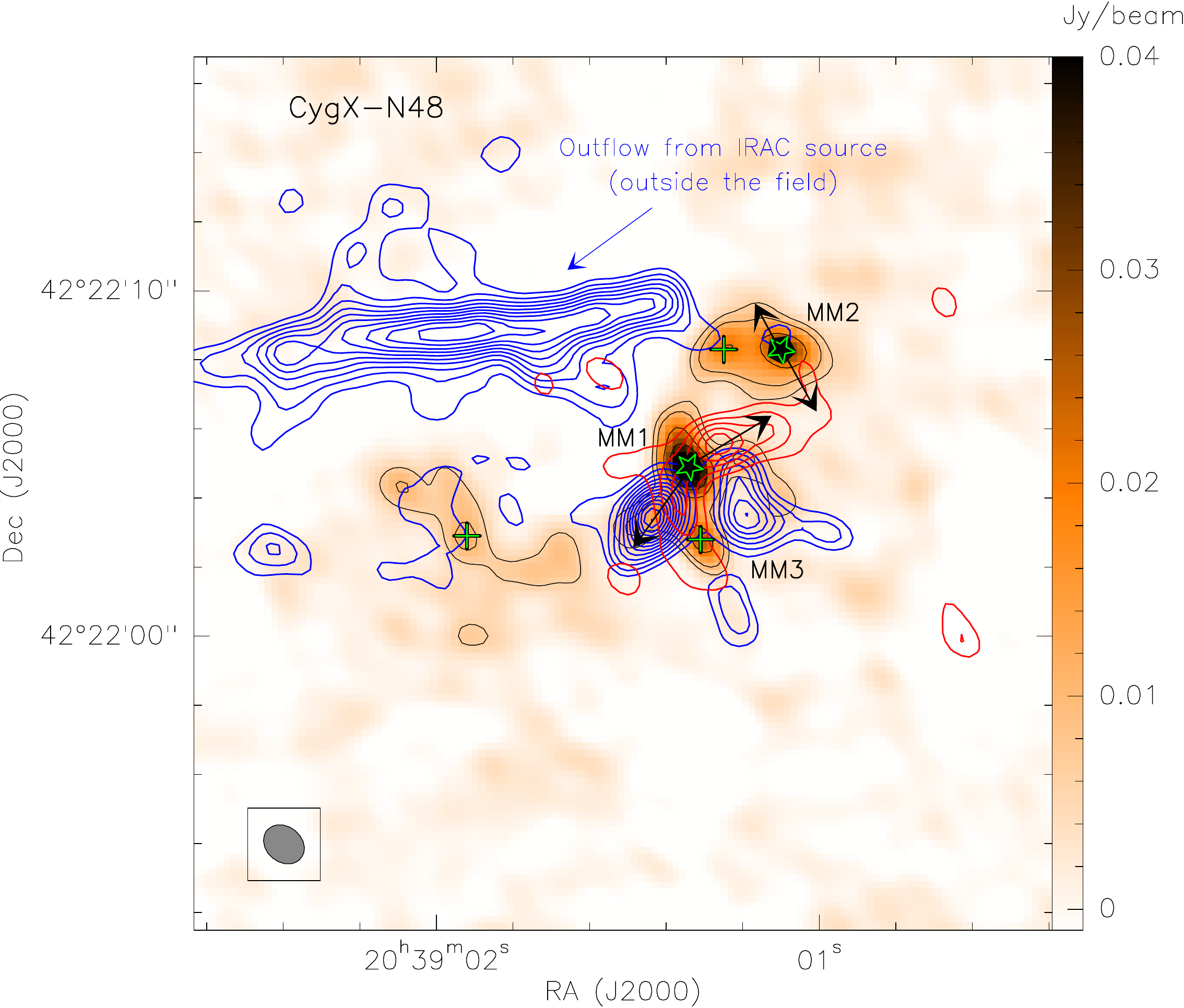}\\
	\vspace{0.3cm}
	\includegraphics[width=0.4\textwidth]{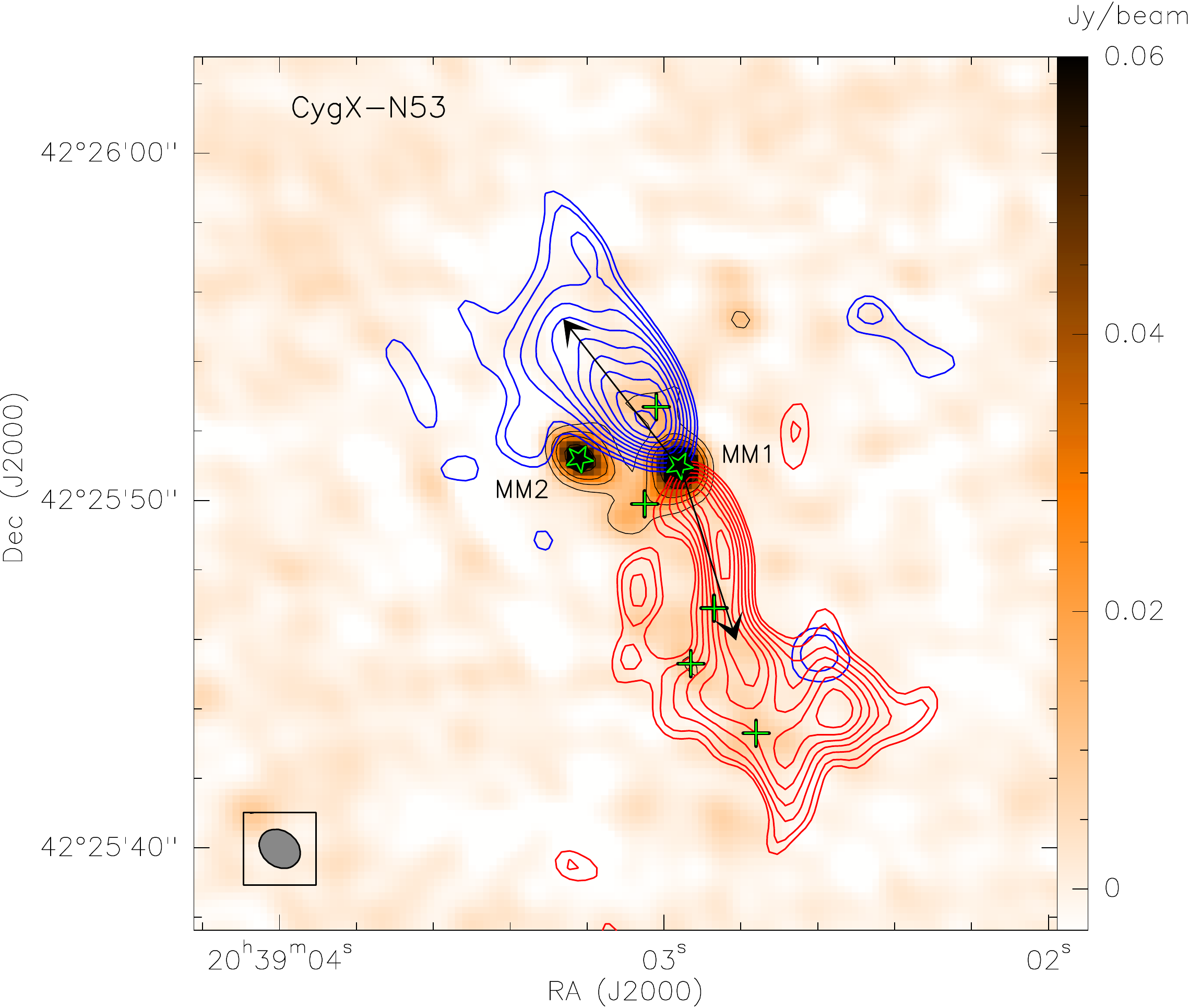}
	\hspace{1cm}
	\includegraphics[width=0.4\textwidth]{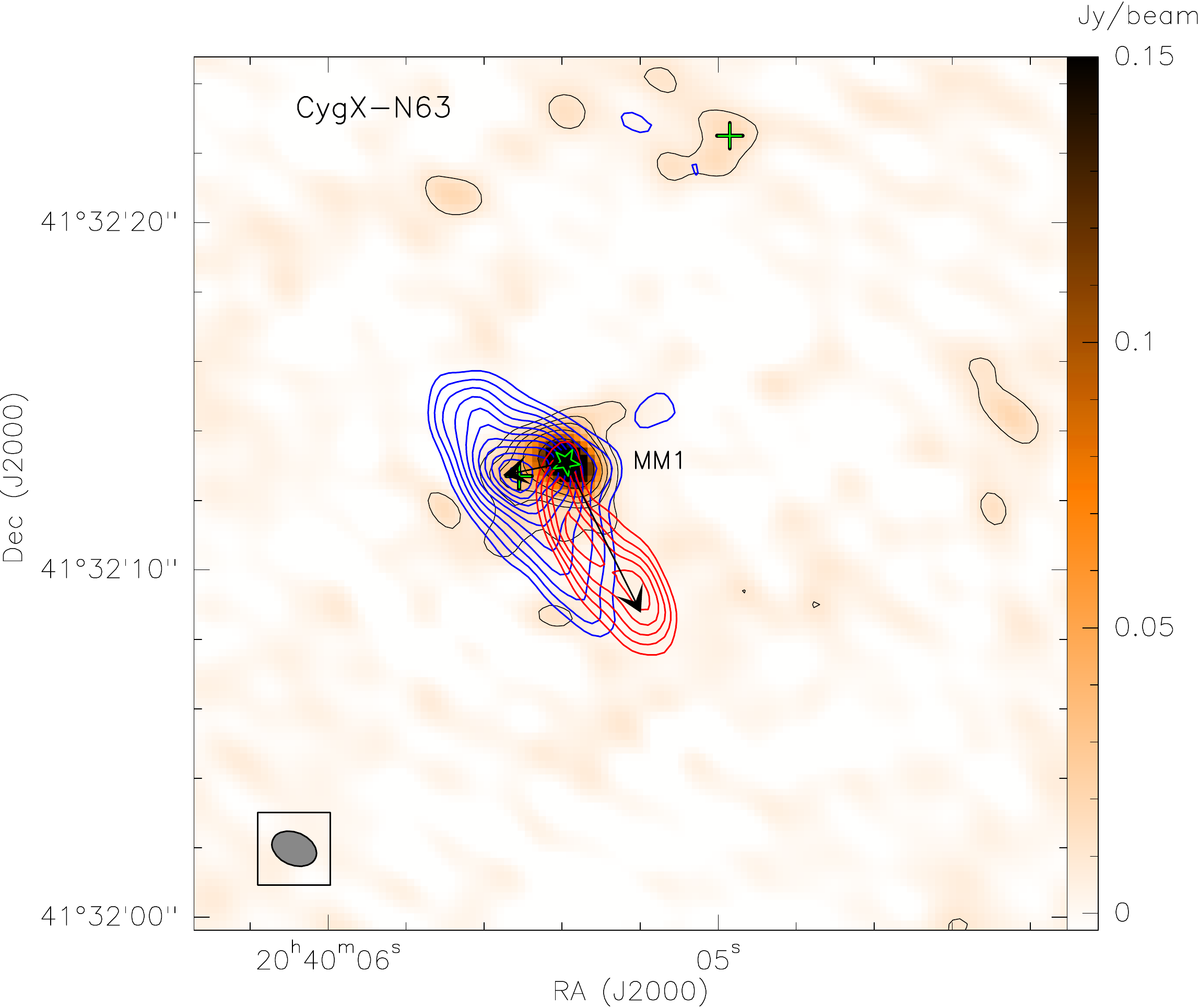}\\
	\caption[]{\small{PdBI 1.2mm continuum emission in colour scale, overplotted with the blue and red contours from the CO emission of N3 (top-left), N12 (top-right), N40 (middle-left), N48 (middle-right), N53 (bottom-left), N63 (bottom-right). The beam sizes are represented in the lower-left corner of each panel. The 9 high-mass cores identified in \citet[][]{2010A&A...524A..18B} are marked as green stars, and the possible low mass fragments identified are marked as green crosses. The sources discussed here are labelled. The arrows show the directions of the outflows identified in this paper, for which we estimated the energetics.}}
	\label{fig:outflows_id}}
\end{figure*}

\subsection{Outflow maps}
\label{sec:outflow_maps}

To identify the individual outflows and their respective driving source, we integrated the $^{12}$CO high-velocity wing emission. The detected $^{12}$CO outflow wings are bright and extend up to high velocities (e.g. CygX-N53 MM1 and N63 MM1 reach velocities of 45\,km\,s$^{-1}$ offset from ambient cloud velocities, see Table~\ref{tab:energetics}) making their interpretation as powerful outflow emission unambiguous, as expected for young high-mass protostars (see Sect.~\ref{sec:scale_up}). An example of a line profile is shown in Fig.~\ref{fig:n53_spectra} (see more examples of line profiles in Appendix ~\ref{areas}). To choose the convenient velocity ranges (i.e. outside the systemic velocities of the cloud where the $^{12}$CO emission is optically thick), we have taken the average spectra over each field and fitted a number of Gaussians designed to fit: a broad outflow component, the bulk of the cloud, and the self absorption feature. This procedure meant to get the best estimate of the systemic velocity and velocity dispersion of the bulk of the cloud. In Table~\ref{tab:summary_velo_energetics} we summarise the systemic velocity of the cloud as from N$_{2}$H$^{+}$ ($v_{o}$), and the result from the Gaussian fitting of the CO ambient cloud component ($v_{peak}$ and $\sigma_{v}$). We then consider that the outflow emission is all $^{12}$CO emission outside 2.5\,$\sigma_{v}$ from the systemic velocities (see Sect.~\ref{sec:outflow_fco} and Appendix ~\ref{areas} for more details, and Table~\ref{tab:energetics} in particular, for a full layout of the velocity ranges used). The resulting outflow maps are shown as contours in Fig.~\ref{fig:outflows_id} for all six fields, overplotted on the PdBI 1.3mm dust continuum from  \citet[][]{2010A&A...524A..18B}.

\subsection{Strong CO outflows from the nine high-mass protostars}
\label{sec:outflow_hmpo}

Our $^{12}$CO observations show clear, strong outflow emission in all six fields. The driving sources of such outflows were identified with the help of the PdBI 1.3\,mm continuum \citep[][]{2010A&A...524A..18B}. Even though \citet[][]{2010A&A...524A..18B} have found several low-mass cores in the continuum of each field, the emission that we detect here is dominated by the outflows powered by the most massive protostars of each MDC (Fig.~\ref{fig:outflows_id}), with the exception being the CygX-N48 field where there is a strong outflow from a source outside our coverage. 

From the nine high-mass protostars, eight are driving a clear strong individual outflow (for CygX-N53 MM2, we only have a tentative detection). These are generally collimated outflows, similar to those of low-mass Class 0 protostars. A detailed description of the outflows and their driving sources is described in the next section.

\subsection{Detailed outflow identification}
\label{sec:outflow_id}

In CygX-N3, MM1 powers a collimated bipolar outflow, perpendicular to the filamentary shape of the continuum. The blue lobe reaches the edges of our image coverage (to the east), and the red is embedded in a more diffuse extended emission to the west. MM2 shows overlapping blue and red weak emission towards the south, which is indicative of an outflow close to the plane of the sky. No clear outflow emission is detected from the other weaker sources.

In CygX-N12 , the identification of the outflows is complicated by the apparent alignment of the outflows with the dust continuum filament.  We consider that the detected outflow emission arises from the two most massive cores, MM1 and MM2, and that the two weaker sources that lie along the dust filament have a negligible outflow emission.
The large spatial extent of the SiO emission observed with the PdBI towards this region (with a wider field of view, Duarte-Cabral et al. in prep.) is in fact indicative of an outflow close to the plane of the sky, with the red outflow lobe becoming blue-shifted further away from the driving source(s), and with significant SiO emission arising from systemic velocities. This is also consistent with the fact that the CO outflow wings do not attain very high velocities. Nevertheless, we tentatively separate the emission of the outflows from MM1 and MM2, by the realisation that MM1 is driving a more collimated outflow than MM2. We stress that if the outflows are indeed close to the plane of the sky, the momentum flux estimates will be highly underestimated. We do not, however, have enough indications to be able to constrain the inclination angle, so we refrain from doing so.

In CygX-N40, the continuum emission is not centrally condensed, but there is a low-mass protostellar candidate, MM1, for which we see a weak CO outflow emission. We can also see very weak CO wing emission to the NE (red) and NW (blue) of MM1, but it is not clear if these high-velocity lobes trace back to MM1 or if they are part of a different outflow (from an undetected low-mass protostar). We have estimated the properties of the outflow from MM1 using the CO emission closer to the source. 

CygX-N48 is the most clustered region of the sample. Firstly, it is worth noting that the blue elongated lobe in the north of the map, extending to the east, does not arise from any of the millimetre sources in our field. In fact, with the help of the SiO emission observed with the PdBI towards this region (with a wider field of view, Duarte-Cabral et al. in prep.), we suspect that this high-velocity outflow arises from a source outside our field (at RA: 20:39:02.927 and Dec.: 42:22:07.32) detected and resolved in IRAC bands at 3$\mu$m and 4$\mu$m, and unresolved at 8$\mu$m and onwards \citep[a source to the north of ERO-2 from][]{2004ApJS..154..333M,2007MNRAS.374...29D}. This source is defined as a more evolved object by Hennemann et al. (in prep). Within our field there are three sources that are from intermediate- to high-mass protostars. For MM1, despite not being a strong outflow emission, we clearly see an outflow oriented SE-NW. Close to MM1, however, there is some emission that is either still associated with MM1's outflow (as an open outflow) or is powered by the less massive MM3 (in the opposite direction of the MM2 outflow). For simplicity, we assume that all the outflow emission we detect around MM1 is from itself (given its higher mass), even though this is likely to include some contamination from the lower mass MM3. If this is the case, the configuration of the outflow lobes would suggest that this outflow is along a direction that is close to the line of sight. 

CygX-N53 embeds two massive cores, MM1 and MM2. MM1 is powering a strong and nicelyV-shaped outflow, but we cannot detect a clear CO outflow emanating from MM2. At lower levels, we can detect some CO high-velocity weak emission near this source, and we use this as our estimate of the outflow momentum flux for MM2. This is merely an upper limit since the CO emission in this field is dominated by the MM1 outflow, and MM2 is therefore likely contaminated by this.

\begin{figure*}[!t]
	\centering
	{\renewcommand{\baselinestretch}{1.1}
	\vspace{0.5cm}\hbox{\hspace{-0.4cm}\includegraphics[width=0.3\textwidth]{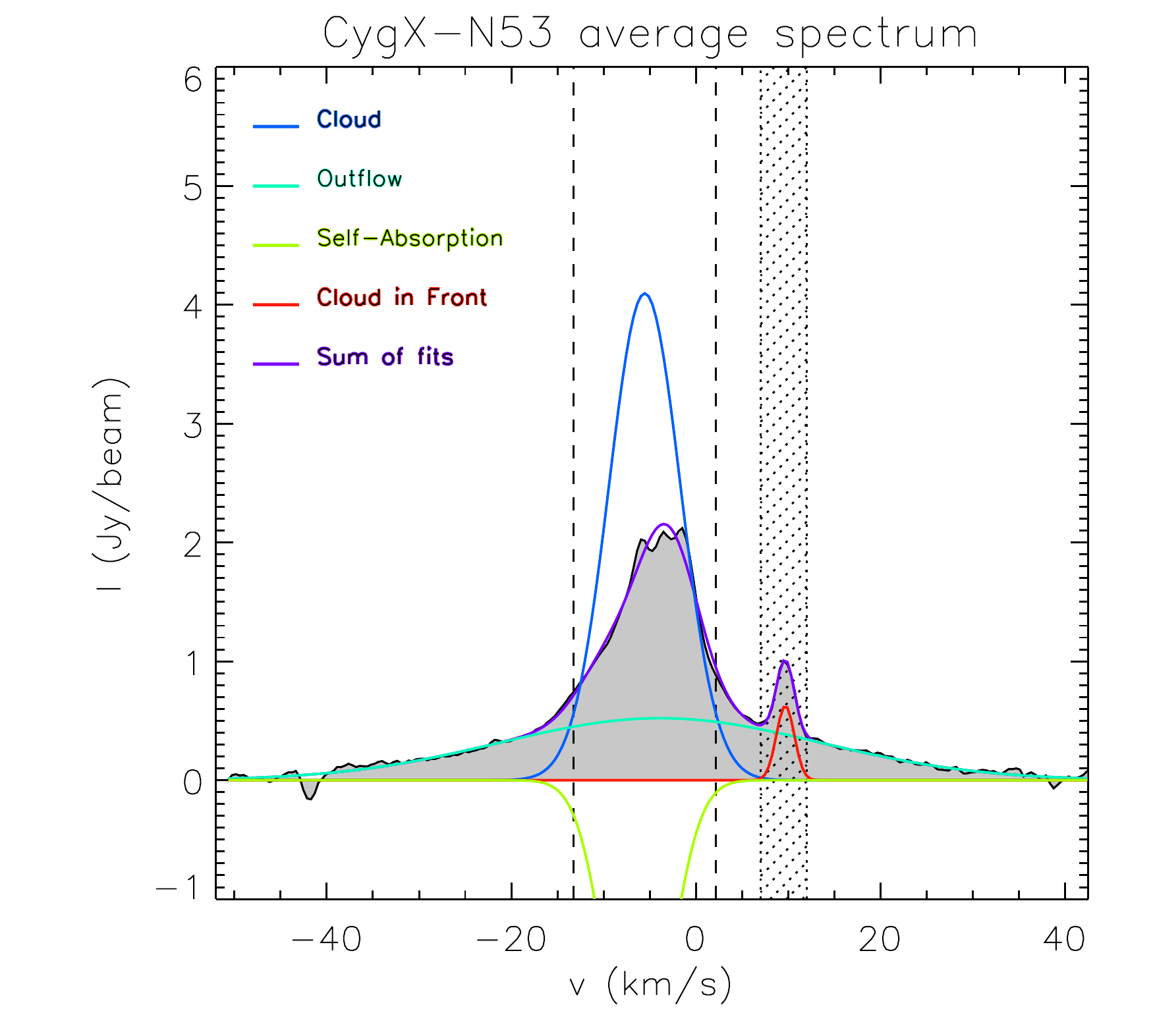}}
	\vspace{-5cm}\hbox{\hspace{4.6cm}\includegraphics[width=0.3\textwidth]{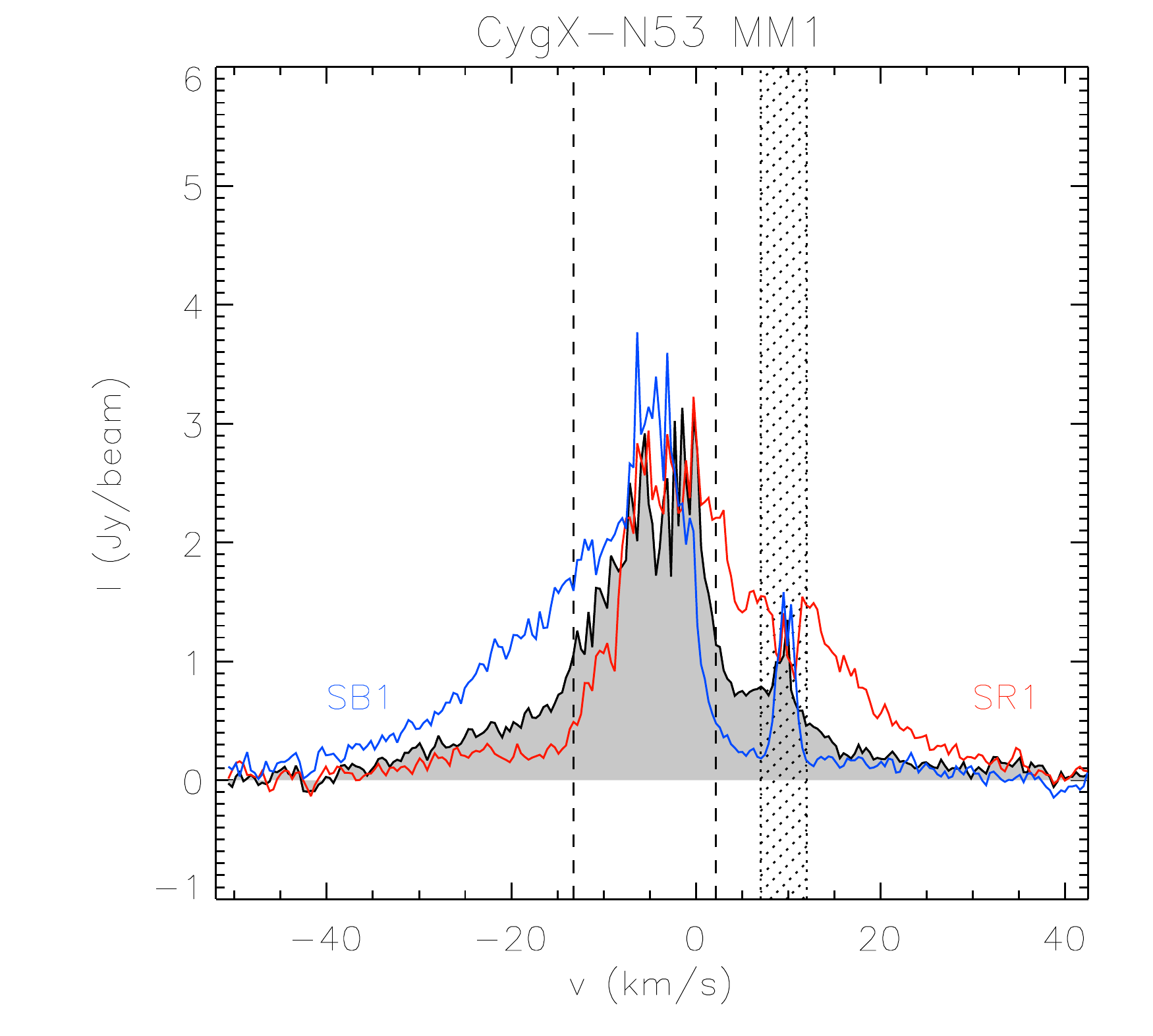}}
	\vspace{-5.3cm}\hbox{\hspace{10.0cm}\includegraphics[width=0.46\textwidth]{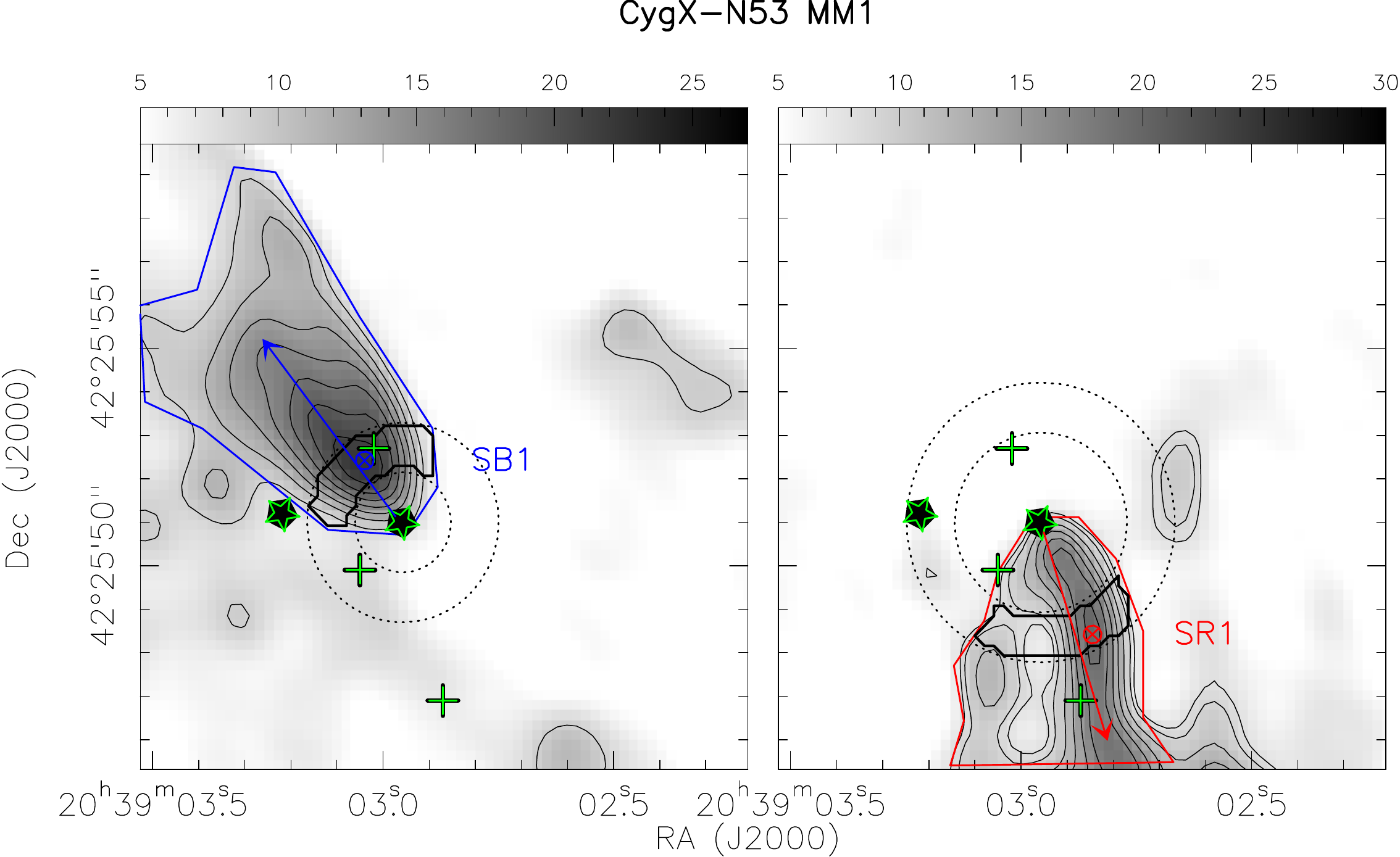}}	
	\caption[]{\small{{\textit{Left}: Average $^{12}$CO (2-1) spectrum in CygX-N53. The Gaussian fits made to the data are shown with curves in different colours: ambient cloud in dark blue, broad outflow emission in light blue, self-absorption in olive-green, emission from another cloud in red, and final model spectra in purple. The vertical dashed lines constrain the systemic velocities of the cloud excluded for the momentum flux calculations. The shadowed area shows the velocity range affected by a cloud in front, also excluded for momentum flux estimates. \textit{Centre}: $^{12}$CO (2-1) spectra at the position of CygX-N53 MM1 in grey, with the spectra at the peak of the blue and red emission (SB1 and SR1 positions marked as blue and red crosses on the right panels). \textit{Right}: Example of the areas used for measuring the CO outflow momentum flux for CygX-N53 MM1, with the blue and red CO emission in grey scale and contours. The intersection between the polygons and rings are the areas taken to measure the respective momentum flux for each wing. }}
	\label{fig:n53mm1_rings}
	\label{fig:n53_spectra}}}
\end{figure*} 

Finally, CygX-N63 is the only single object that we have in this sample, and is the most massive one. It powers a clear clean outflow that extends to its left-hand side, perhaps simply due to the existence of more material towards the east which gets shocked and entrained by the outflowing gas. We neglect the possible outflow emission from the weaker millimetre continuum source detected by \citet[][]{2010A&A...524A..18B} to the east of MM1, since it is not a well-defined millimetre peak, and it may in fact be part of the elongated envelope of MM1. 

\subsection{Outflow momentum flux}
\label{sec:outflow_fco}

To estimate the energetics of the outflows from this sample of sources, we have used an approach similar to what \citet[][]{1996A&A...311..858B} used for low-mass protostellar objects. This method consists of estimating the momentum flux of the outflows on a ring centred on the driving source. It assumes that the momentum flux is conserved along the outflow direction, and it is a particularly useful method for small maps where the total extent of the outflow may not be covered. A detailed description of this method can be found in Appendix~\ref{areas}. In particular, an example of the rings and areas used for CygX-N53 MM1 is shown in Figs.~\ref{fig:n53mm1_rings}, and those used for our entire sample of objects are shown in Fig~\ref{fig:n3_rings} to \ref{fig:n63_rings}. The velocity ranges used to integrate the blue and red emission, as well as the momentum flux estimates, are detailed in Table~\ref{tab:energetics}.

In Table~\ref{tab:summary_velo_energetics} we summarise the systemic velocities of each region from N$_{2}$H$^{+}$, $v_{o}$, and the parameters of the ambient cloud component as from the Gaussian fittings of the CO emission ($v_{peak}$ and $\sigma_v$). The velocities used to integrate the momentum fluxes are those more than 2.5$\sigma_v$ away from $v_{peak}$. This choice of velocity ranges was supported by a visual inspection of the individual datacubes, so as to neither include the extended (optically thick) emission nor miss out the compact outflow emission close to the systemic velocities. From the average spectra of the regions (e.g. Fig.~\ref{fig:n53_spectra}), this choice also corresponds to where the broad outflow component starts dominating the ambient cloud emission. The last column of Table~\ref{tab:summary_velo_energetics} shows the total  $F_{\rm{co}}$ for the nine high-mass protostars corrected by an average inclination angle (factor 2.9), estimated as twice the value for the wing with the highest momentum flux, as a consequence of assuming that the outflows are symmetrically bipolar. If using different velocity ranges, the $F_{\rm{co}}$ values would change (e.g. by 5-10\% if using 2\,$\sigma_{v}$ or 3\,$\sigma_{v}$, instead of 2.5\,$\sigma_{v}$). However, this is negligible compared to the remaining uncertainties, in particular, those associated with the assumptions on optical depth, symmetry, and inclination angles, giving $F_{\rm{co}}$ an uncertainty that can easily be as high as a factor 2 (or even 4). Nevertheless, we adopt a factor 2 as the uncertainty of our momentum flux estimates in the remainder of the article.

From a quick comparison of our sample of high-mass protostars with the low-mass Class 0 VLA1623 (last row of Table~\ref{tab:luminosities}), we can see that the outflow momentum flux of our sample is typically an order of magnitude stronger.

\subsection{Morphology of the outflows}

For low-mass protostars \citep[e.g.,][]{2006ApJ...646.1070A} it has been found that young protostars (Class 0) have opening angles below $\sim$55$^{\circ}$, while it can reach more than $\sim$75$^{\circ}$ at the Class I stage. To investigate whether the protostellar outflows we find in our sample of high-mass protostars have similar properties to those of young low-mass protostars, we measured the opening angles based on the CO high velocity wing emission alone. However, for CygX-N53 there are some indices of a wide-angled low-velocity CO emission that could still be associated with the outflows and that are not taken into account. We found that the high-velocity outflow opening angles range between 15 and 35$^{\circ}$. The only exception is CygX-N48 MM1 whose outflow has an apparent opening angle of $\sim$100$^{\circ}$. However, as discussed in Sect.~\ref{sec:outflow_id}, this outflow has a complex morphology, and it either is a combination of MM2 and MM3 outflow emission or, if it is indeed a single outflow, then it has overlapping blue and red emission. The small spatial extent of such emission and the relatively high velocities it reaches in both the blue and red wings (nearly 30~km\,s$^{-1}$ offset from the systemic velocities), can in fact be consistent with an outflow direction close to the line of sight. In this case, the opening angle cannot be determined accurately, and the true value will likely lie much below $\sim$100$^{\circ}$. Our estimates of the opening angles are, therefore, consistent with those found for low-mass Class 0 protostars.

\subsection{Outflows and the large-scale dynamics}

The different large-scale structure and dynamics associated with our sample of clumps may play a role in the small scale conditions. Because outflows are thought to be centrifugally launched perpendicular to the forming protostellar discs with Keplerian rotation, the outflow direction provides information on the rotational motion on small scales  ($\sim$1\,AU). A comparison of the outflow orientations with the filamentary structures shows that the most massive objects (CygX-N53 and N63) have an outflow perpendicular to the elongated continuum structures (or the alignment of the two sources in the case of N53). These two regions also present a rotation \citep[][]{2011A&A...527A.135C} whose axis is approximately parallel to that of the outflows. Therefore, the global rotation of the clump having the same rotation axis as the disc suggests that the angular momentum of the MDC is transferred down to small scales in CygX-N53 and N63.

For the highly filamentary CygX-N3 region, the outflows also appear perpendicular to the filamentary structure. However, the N3 filament lies along the tip of a pillar in DR17 and experiences a velocity shear from material being compressed and converging perpendicularly to the filament \citep[][]{2011A&A...527A.135C}. If this large-scale infall motion were transferred directly onto a rotational motion of the envelopes, the expected outflow directions would be parallel to the filament, rather than perpendicular. That this is not the case implies that the apparent bulk motion of the infalling gas in N3 is not being directly transferred into the individual rotation of the protostellar envelopes. 

Finally, even though CygX-N12 has outflows aligned with its filamentary structure, the larger scale motions \citep[][]{2011A&A...527A.135C} are too complex to find any correlation with the outflows. The more cluster-like CygX-N48 does not show any clear large-scale gradient and does not have a clear filamentary structure, making it hard to correlate the outflow directions with the dust emission or large-scale dynamics. Despite the complex velocity fields, \citet[][]{2011A&A...527A.135C,2011ApJ...740L...5C} have discussed the possible existence of small-scale converging flows in these two fields. 

Our analysis shows that, even though low-velocity converging flows could be an important provider of mass to the forming protostars and crucial to the formation of the surrounding structures, by the time the gas is accreted from the small-scale envelopes and discs onto the protostars, the momentum of the gas no longer shows any clear remnants of the initial velocity of the converging flows themselves.

\section{Analysis}
\label{sec:analysis}

\subsection{Nature of the nine high-mass cores}

The systematic detection of high-velocity outflows towards all the massive cores of our sample, is evidence of ongoing accretion of mass. That these are mostly single and unconfused outflows also indicates that these cores are not fragmented further into several protostars\footnote{We note that if fragmentation was to occur on smaller scales within the disc, it would not necessarily change the outflow properties \citep[][]{2012MNRAS.422..347S}. This means that close binaries cannot be ruled out because we see a single well-defined outflow.}. This together with the position of these high-mass cores in the $M_{\rm  env}-L_{\rm  bol }$ diagram (Fig.~\ref{fig:mass_lum_nomodel}), confirms the nature of our sample as individual high-mass Class 0-like protostars. 	

One exception may be CygX-N53 MM2. This source is among the three most massive of our sample, and yet we cannot detect any outflow emission being clearly powered by it. As mentioned before, the values we provide are a tentative detection of a possible outflow, and they only represent an upper limit. The similar mass to its companion, CygX-N53 MM1, the non-counterpart of CygX-N53 MM2 in 24$\mu$m emission (and tentative detection at 70$\mu$m) and their differences in the outflow power seem to indicate that CygX-N53 MM2 is at a younger stage than CygX-N53 MM1. Perhaps this source still needs to achieve the high accretion rates needed to power a stronger outflow, meaning that it could well be either a true high-mass prestellar core or a source at the transition stage between prestellar and protostellar.

\subsection{Scaling up of outflow power}
\label{sec:scale_up}

\begin{figure*}[!t]
	\centering
	{\renewcommand{\baselinestretch}{1.1}
	\includegraphics[width=0.46\textwidth]{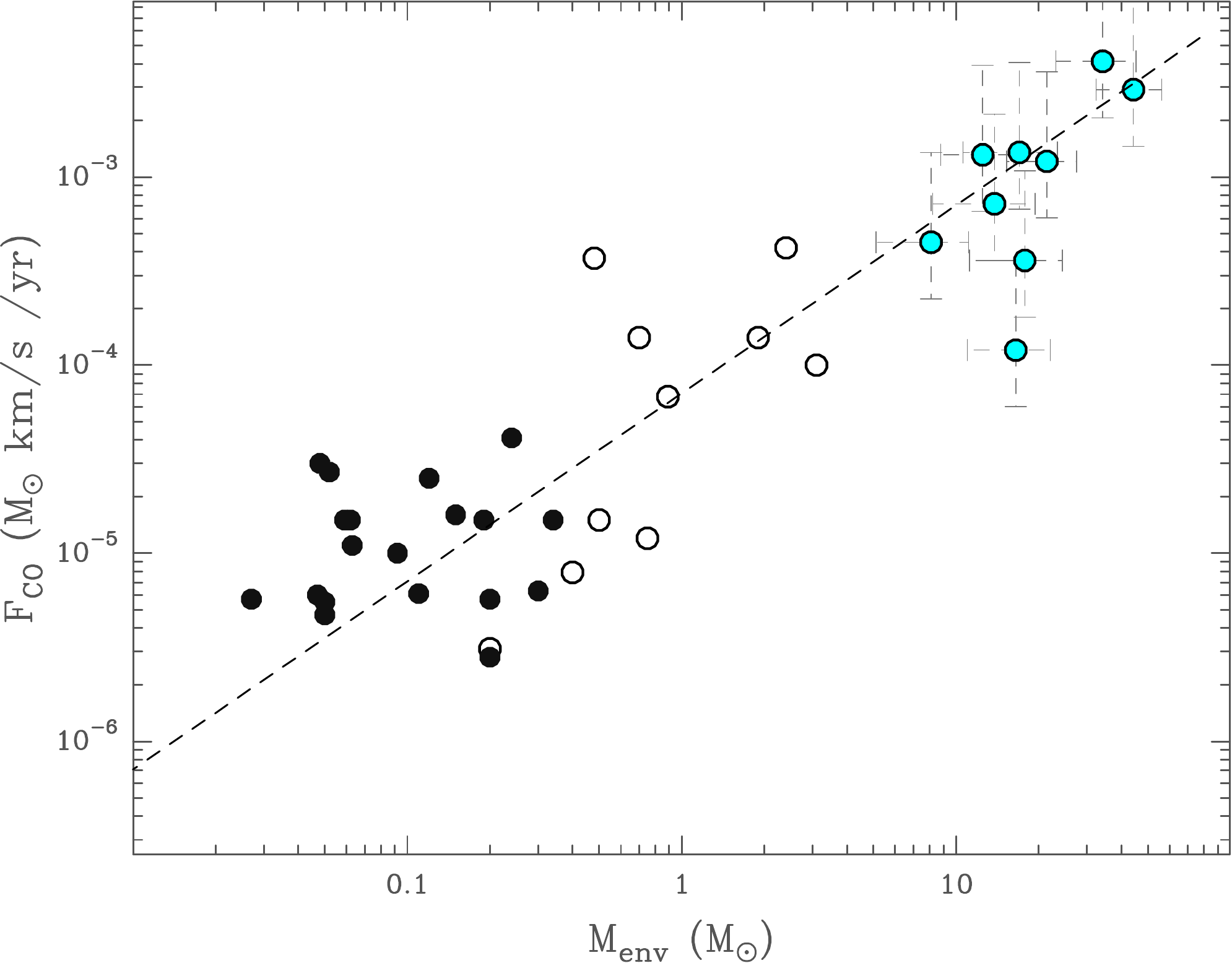}
	\hfill
	\includegraphics[width=0.46\textwidth]{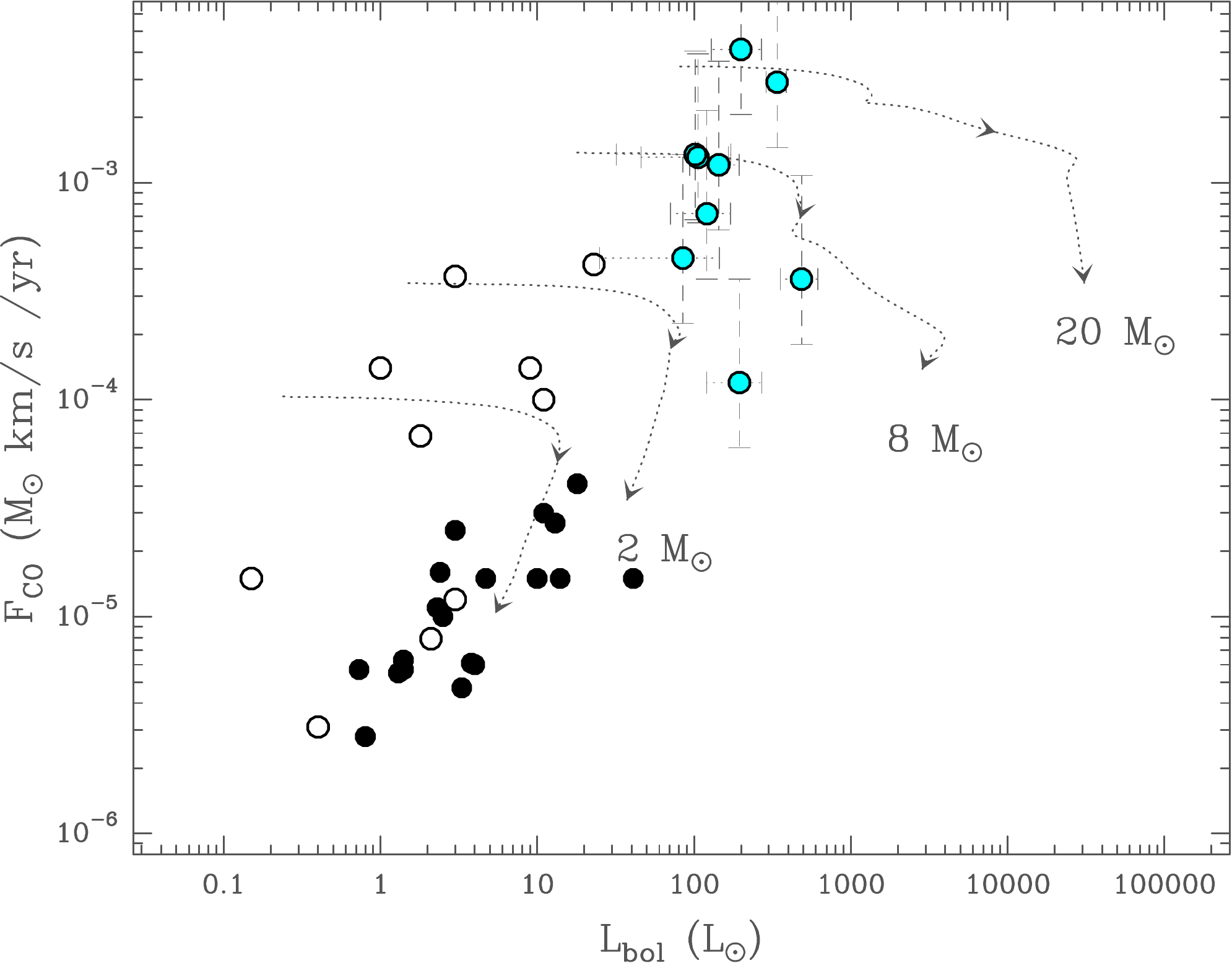}
	\caption[]{\small{Outflow momentum flux correlation with envelope mass (left) and with bolometric luminosity (right). The black filled and open circles are the low-mass Class I and Class 0 sources from \citet[][]{1996A&A...311..858B}. The light blue circles are the massive protostars from our sample. In the left panel, the black dashed line is the observational linear relation found by  \citet[][]{1996A&A...311..858B}, showing that it also holds for the massive individual protostars. On the right, we show the evolutionary tracks corresponding to those of Fig.~\ref{fig:mass_lum_nomodel}.}}
	\label{fig:fco_mass}}
\end{figure*} 

Our sample of nine true individual high-mass protostars allows us to study the scaling of outflow power with the final stellar mass with an unprecedented accuracy. The comparison with low-mass protostellar populations is particularly relevant here since we were able to recognise and measure properties of the high-mass protostars at the same physical scale than their low-mass counterparts. 

In Table~\ref{tab:luminosities} we have compiled the basic properties of these nine high-mass star precursors, including envelope mass ($M_{\rm  env}$), bolometric luminosity ($L_{\rm  bol}$), and the total CO outflow momentum flux ($F_{\rm{co}}$). 
Figure~\ref{fig:fco_mass} shows the correlation of $F_{\rm{co}}$ with $M_{\rm  env}$ (left) and $L_{\rm  bol}$ (right) for the low-mass sample of \citet[][]{1996A&A...311..858B} and our high-mass protostellar sample. It is clear that the outflows of high-mass protostars are, as expected, much more powerful than their low-mass counterparts. The left-hand panel of Fig.~\ref{fig:fco_mass} shows that the high-mass Class 0 objects extend the correlation of $F_{\rm  co}$ with  $M_{\rm  env}$ up to the $20-50\,$M$_\odot$ regime. The linear relation for low mass stars found by \citet[][]{1996A&A...311..858B} 
is also shown. This correlation now extending over 2.5 orders of magnitude and including both Class 0 and Class I protostars (for the low-mass part) suggests there is a fundamental relationship between mass reservoir ($M_{\rm  env}$) and outflow/accretion activity ($F_{\rm  co}$).

\begin{table}[!t]
\caption{Timescales and accretion rates}
\begin{tabular}{l | c | c c c}
\hline 
\hline 
\multirow{3}{1cm}{Source}   	&	\multirow{2}{*}{$4 t_{\rm ff}$}  	& \multirow{2}{*}{{$<\dot{M}_{\rm acc}>^{\rm 4t_{ff}}$}} 		& \multirow{2}{*}{{$<\dot{M}_{\rm acc}>^{\rm 0.3Myr}$} }			& \multirow{2}{*}{{$\dot{M}_{\rm acc}^{\rm obs}$}} \\
    			&	\multirow{2}{*}{(10$^{5}$ yr) }  		&  \multicolumn{3}{c}{\multirow{2}{*}{($10^{-5}\,$M$_{\odot}$\,yr$^{-1}$)}} 	\\
		        &		&            &            &           \\
\hline
{N3 MM1}		&    0.19    &    67.    &    4.1  &    5.2  \\
{N3 MM2	}	&    0.18    &    77.    &    4.6  &    2.9   \\
{N12 MM1}	&    0.16    &    113.  &    5.9  &    1.4   \\
{N12 MM2}	&    0.16    &    101.  &    5.5  &    0.48   \\
{N48 MM1}	&    0.16    &    106.  &    5.7  &    5.4   \\
{N48 MM2}	&    0.23    &    35.    &    2.7  &    1.8   \\
{N53 MM1}	&    0.11    &    302.  &    11.4  &    16.5   \\
{N53 MM2}	&    0.14    &    150.   &    7.1  &    $<$ 4.8   \\
{N63 MM1}	&    0.09    &  	445.  &  	14.8  &    11.6   \\
		        &		  &            &            &           \\
{VLA1623}	&   {0.79$^{(a)}$}    &    {0.88}  &  0.23  &    0.56  \\
\end{tabular}
\\
 {$^{(a)}$ We stress that this $4 t_{\rm ff}$ is not the measured collapse timescale, but the one estimated under the assumption that all cores started their collapse from identical initial volumes (in this case 4000\,AU FHWM). In reality, VLA1623 is currently twice more compact than the size we used for this estimate, which could be a consequence of the collapse of VLA1623, with its envelope getting more compact with time.} 
 \label{tab:accrates}
\end{table}

From the right-hand panel of Fig.~\ref{fig:fco_mass} we can see that the high-mass objects  
have luminosities that are 10 to 100 times higher than low-mass Class 0s and that they cluster at the location expected for young precursors of stars of $\sim 5$ to 20 M$_\odot$. The spread in $F_{\rm  co}$ for a certain $L_{\rm  bol}$ (or $M_{\rm  env}$) is similar to that found in the low-mass regime, about a factor of 20.

\subsection{Accretion timescales for high-mass stars}
\label{sec:acc_rates}

The linear relation $F_{\rm  co} = 7.5\times10^{-5} M_{\rm  env}$  (Fig.~\ref{fig:fco_mass}, left), found between low-mass Class 0s and Class Is, had been interpreted as due to evolution alone ($\dot{M}_{\rm  acc}$ decreasing with decreasing $M_{\rm  env}$, with $M_{\rm  env}$ being an age indicator) in \citet[][]{1996A&A...311..858B}. That this relation also holds between low- and high-mass protostars points to a fundamental relationship between $M_{\rm  env}$ and $F_{\rm  co}$, i.e. between mass reservoir and outflow activity. Young protostars have most of the mass in the collapsing envelope, therefore $M_{\rm  env}$ measures the amount of mass in the local gravitational well. The infall of material and respective accretion rate could scale with $M_{\rm  env}$ as a consequence of a larger gravitation field. Since outflow power should trace accretion rates, this direct effect of self-gravitation could explain this fundamental relation.

In fact, accretion rates are basically expressed as  $\dot{M}_{\rm acc} = \epsilon_{\rm acc} M_{\rm  core} / t_{\rm  acc}$, where $\epsilon_{\rm acc}$  is the efficiency of the collapse to gather mass from the core into the final star, $M_{\rm  core}$ is the mass of the original collapsing core\footnote{We assumed a monolithic collapse in which the initial collapsing core contains most/all of the mass that will end up in the star. Alternatively, the stellar material may come from larger distances with a longer effective free-fall time (see discussion in Sect.~\ref{sec:common_time}).} (expected to be well traced by $M_{\rm  env}$ for earliest stages Class 0 protostars), and  $t_{\rm  acc}$ the timescale for main accretion, i.e. roughly the collapse time. If gravity dominates the collapse, $t_{\rm  acc}$ should be directly linked to the free-fall time $t_{\rm  ff}$ of the original core. Before the accretion phase, in models of monolithic collapse, cores have a central flat density profile (up to a few thousand AU), and a $\rho \propto r^{-2}$ at larger radii \citep[e.g.,][]{1994MNRAS.268..276W,2001A&A...365..440M}. In the low-mass regime, the timescales for the collapse of cores have been found to be typically about four to five times larger than the free fall-time estimated from the original flat density region \citep[][]{1997A&A...323..549H,1993ApJ...416..303F}.

Since the fragmentation scales tend to be similar from the low- to the high-mass regime \citep[of $\sim$4000\,AU here, and of 6000\,AU in $\rho$-Ophiuchi from][]{1998A&A...336..150M}, one could expect that, in a scenario of quasi-static monolithic collapse, the size of the initial prestellar cores is also similar for all mass ranges - and set by the fragmentation scale itself. In this case, the density would be proportional to $M_{\rm  core}$ (traced here by $M_{\rm  env}$). Since $t_{\rm  ff} \propto {n_{\rm  H_2}}^{-1/2}$, this hypothesis would predict the $t_{\rm  ff}$ (and $t_{\rm  acc}$) scaling as $M_{\rm  env}^{-1/2}$, resulting in shorter $t_{\rm  acc}$ for higher mass cores\footnote{From a completely independent approach, \citet[][]{2003ApJ...585..850M} obtained $t_{\rm  acc} \propto M_{*}^{1/4} \Sigma^{-3/4}$, which also implies shorter $t_{\rm  acc}$ for high mass cores, since the surface densities $\Sigma$ of the regions are much higher for high-mass star-forming regions, which lead to higher densities in the high-mass collapsing cores.}. This leads to $\dot{M}_{\rm  acc}  \propto  M_{\rm  env}^{3/2}$, i.e. $F_{\rm  co}  \propto  M_{\rm  env}^{3/2}$. It is then reasonable to expect this relation between $\dot{M}_{\rm  acc}$ and $M_{\rm  env}$, set at the beginning of the collapse, to hold along the main accretion phase of the protostars, especially for the youngest protostars (the Class 0s). However, we observe a pure linear correlation between $F_{\rm{co}}$ (i.e. $\dot{M}_{\rm  acc}$) and $M_{\rm  env}$. Such pure proportionality actually implies that, surprisingly, $t_{\rm  acc}$ does not depend on $M_{\rm  env}$.

To further emphasise the discrepancy between the observed accretion rates and those expected from a scenario of monolithic core collapse where all cores have similar initial sizes, we estimated the expected collapse timescale as $4 t_{\rm  ff}$ for our sample of high-mass cores as well as for the example low-mass protostar, VLA1623, by assuming that the initial cores enclosed the current envelope mass in a diameter of 4000\,AU (set by the fragmentation scale),\footnote{Even though we use a fixed value for the size of cores for this exercise, there is probably a dispersion on the initial sizes that could vary from 2000\,-\,6000\,AU. This alone could be partly responsible for the observational scatter of accretion rates ($\sim$ factor 2\,-\,3).} but with an initial flat density profile.
These collapse times and their corresponding accretion rates {$<\dot{M}_{\rm acc}>^{\rm 4t_{ff}}$} are given in Table~\,\ref{tab:accrates}, for the nine high-mass protostars and the low-mass Class 0. For comparison, we also show the observed accretion rates, $\dot{M}_{\rm  acc}^{\rm obs}$, from the $F_{\rm{co}}$ measurements (using $\dot{M}_{\rm  acc}^{\rm obs}= F_{\rm  co} / 20\,$km\,s$^{-1}$, see Sect.~\ref{sec:model} below), as well as the accretion rates in case of an accretion time identical for all masses and equal to $3\times 10 ^5$\,yr, $<\dot{M}_{\rm  acc}>^{\rm 0.3Myr}$ \citep[e.g.,][]{1993ApJ...402..635M}. While the accretion timescales assuming 4$t_{\rm ff}$ for low-mass protostars is within one order of magnitude of the measured timescales obtained from global statistics of young stellar objects in nearby star-forming regions \citep[e.g.,][]{greene1994, kenyon&hartmann1995, evans2009,2011A&A...535A..77M}, the high-mass Class 0s should have timescales that are one order of magnitude smaller, leading to accretion rates that are 100 times higher than their low-mass counterparts. The observed values calculated directly from $F_{\rm  co}$, are only of the order of 10 times greater than for low-mass protostars and are compatible with the values obtained assuming the same accretion time for protostars of all masses.

We conclude that the observed power of outflows in the Cygnus X high-mass protostars suggests a constant accretion time for all stars, pointing to a protostellar time of $\sim 3 \times 10 ^5\,$yr, identical for both low- and high-mass stars. The reason $t_{\rm  acc}$ does not decrease with increasing mass is unclear, but we explore different possibilities in Sect.~\ref{discussion}.

\section{Evolutionary models}
\label{sec:model}

The few observed quantities ($M_{\rm  env}$, $L_{\rm  bol}$, and $F_{\rm  co}$) that characterise protostars are indirect proxies for the true basic parameters of the protostars (ages, accretion rates, stellar masses). In the previous section, we derived an estimate of the timescale for the accretion phase for high-mass protostars by comparing them with low-mass protostars. To go one step further, we can model all observed quantities to build a coherent picture that includes all observed parameters for low- and high-mass protostars.

\subsection{Mass reservoir and collapse/accretion efficiency}

The material stored in the collapsing envelope is usually assumed to correspond to the reservoir of mass available to build up the star. A fraction of this material will be lost, however, during the accretion phase due to outflows and radiative/ionisation pressures. The envelope mass $M_{\rm  env}$ is therefore expected to decrease as the star grows and as the outflow blows away part of the envelope. As such, the evolution of $M_{\rm  env}$ can be expressed as 

\begin{equation}
 {\rm d} {M_\star}(t) =  \dot{M}_{\rm acc} (t) {\rm d}t = \epsilon_{\rm acc }\, {\rm d} (M_{\rm  core} - M_{\rm  env} (t)) = - \epsilon_{\rm acc }\,  {\rm d}M_{\rm  env} (t)
\label{eq:menv}
\end{equation}

\noindent
where $\epsilon_{\rm acc}$ is the accretion efficiency on the scale of the envelope, and $M_{\rm  core}$ the mass of the  core at the beginning of its collapse. If some competitive accretion plays a role, $M_{\rm  env}$ could increase with time with the infall of material from larger scales increasing the efficiency $\epsilon_{\rm acc}$, which could become greater than 1.

\subsection{Accretion and stellar luminosities}

 The bolometric luminosity $L_{\rm  bol}$ is the total luminosity of the protostar, which is the sum of the accretion luminosity, $L_{\rm  acc} (t)$, radiated at the surface of the stellar embryo at a radius $R_\star$(t) and of the stellar luminosity, $L_\star (t)$, as
 
\begin{equation}
L_{\rm  bol} (t) =  L_{\rm  acc} (t) + L_{\star} (t)
\label{eq:lbol}
\end{equation}
with
\begin{equation}
L_{\rm  acc} (t) =  G \,  M_{\star} (t) \,  \dot{M}_{\rm acc} (t) / R_{\star} (t).
\label{eq:lacc}
\end{equation}

\noindent
The stellar luminosity and radius $L_\star$(t) and $R_\star$(t) are calculated from the Hosokawa tracks provided for high accretion rates, which are adequate for high-mass protostars \citep{2009ApJ...691..823H}.

\begin{figure*}[!t]
	\centering
	{\renewcommand{\baselinestretch}{1.1}
	\vspace{-1.5cm}
	\includegraphics[width=0.4\textwidth]{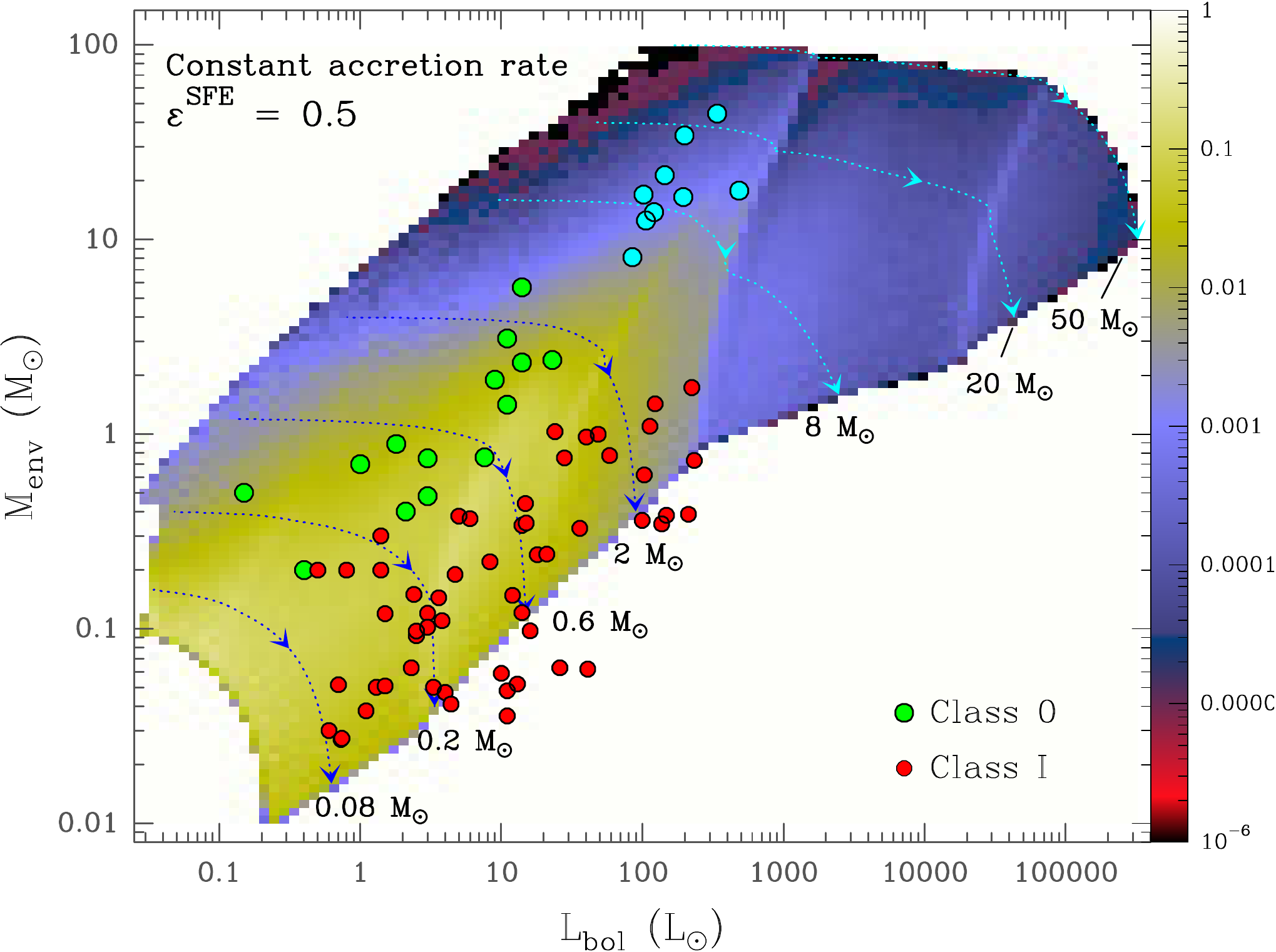}
	\hspace{1cm}
	\includegraphics[width=0.4\textwidth]{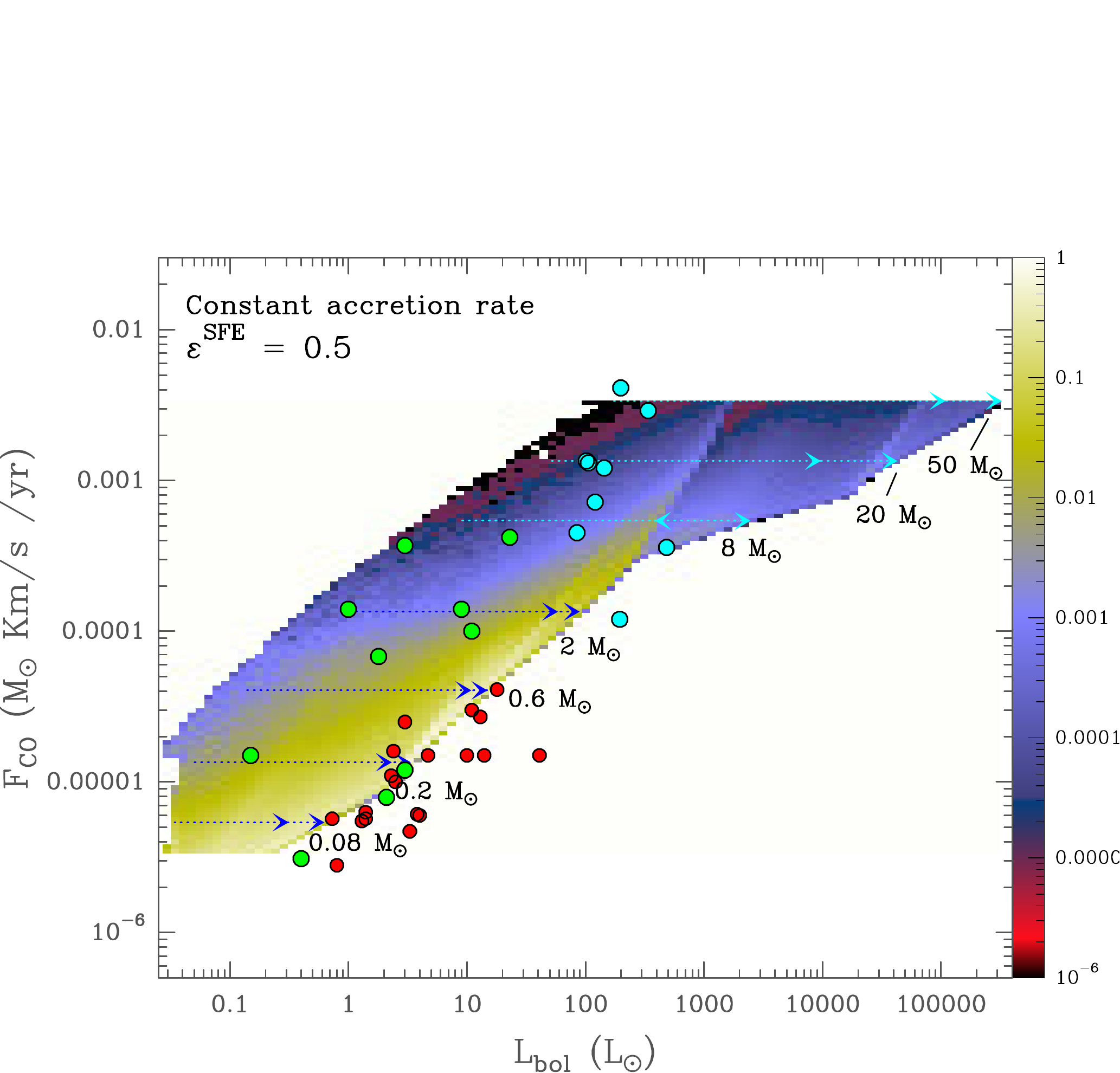}
	\includegraphics[width=0.4\textwidth]{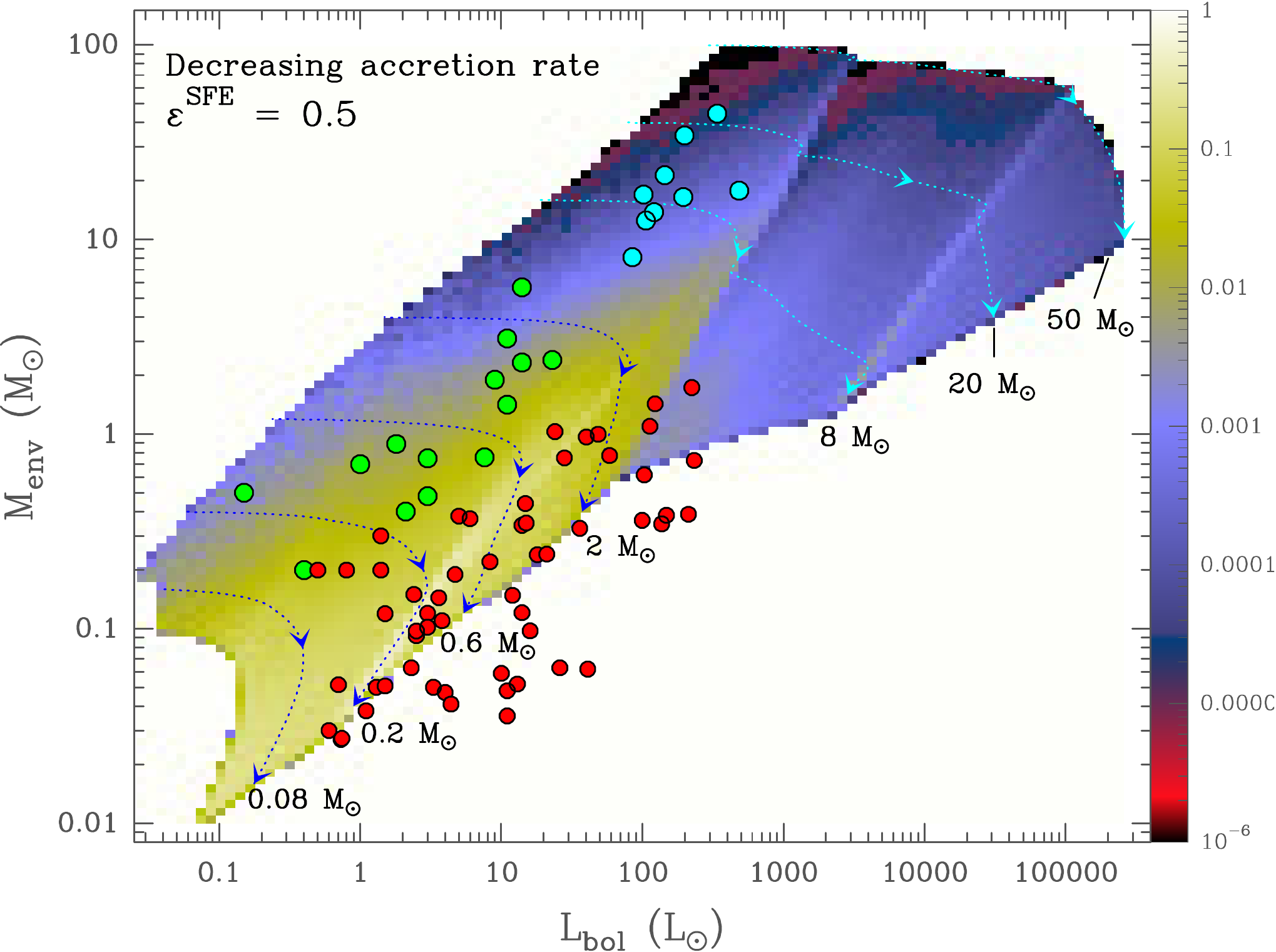}
	\hspace{1cm}
	\includegraphics[width=0.4\textwidth]{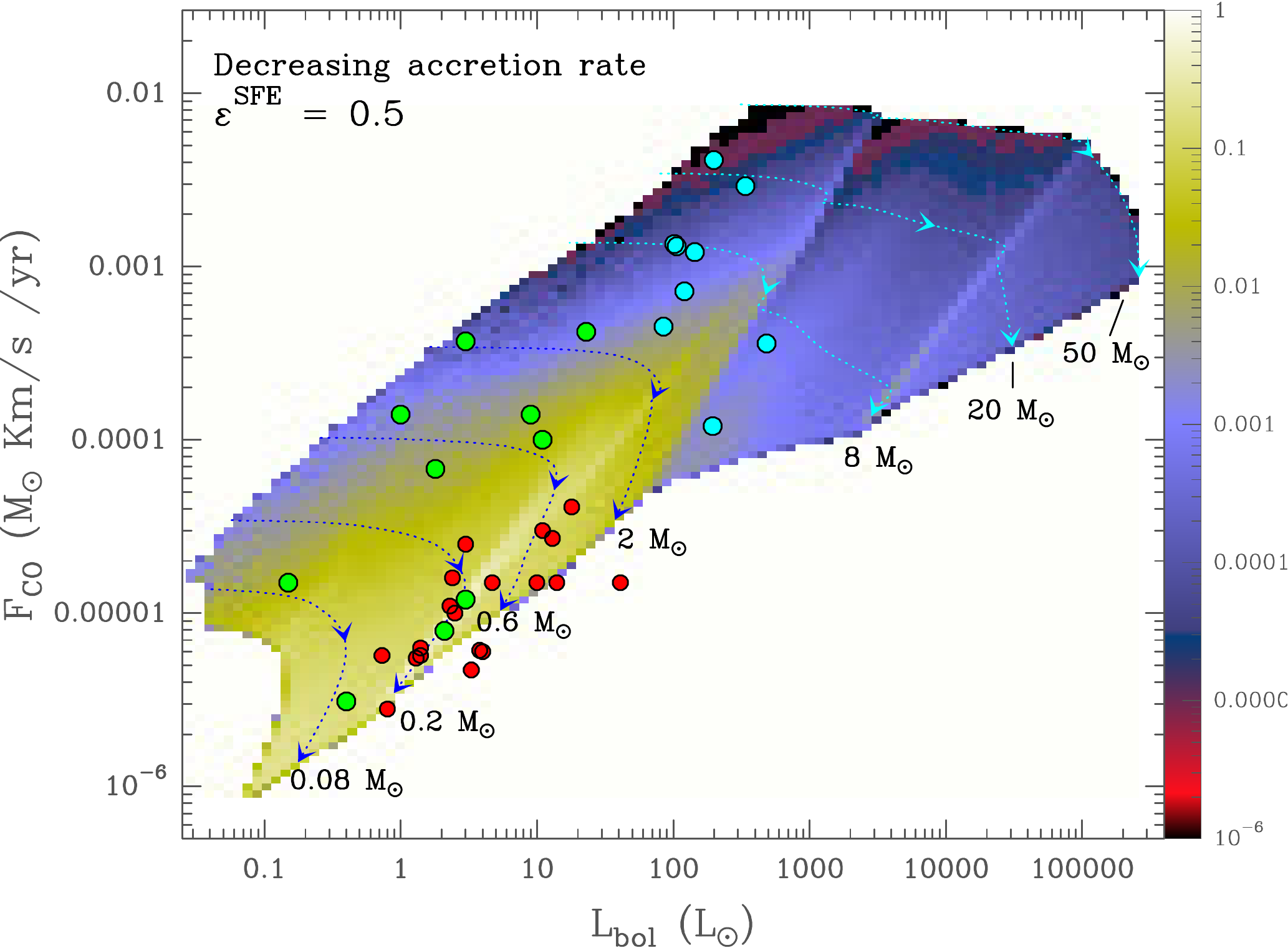}
	\includegraphics[width=0.4\textwidth]{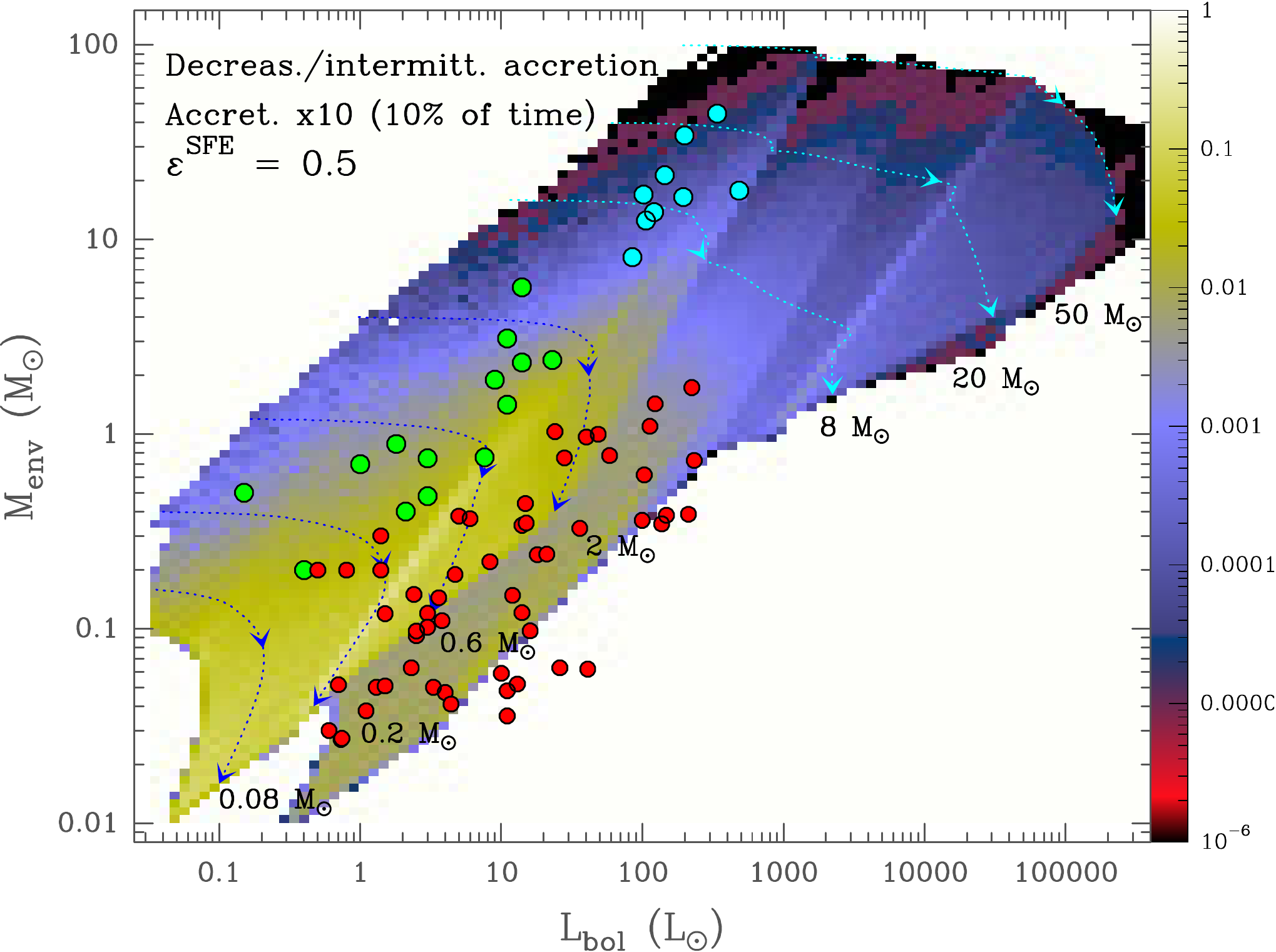}
	\hspace{1cm}
	\includegraphics[width=0.4\textwidth]{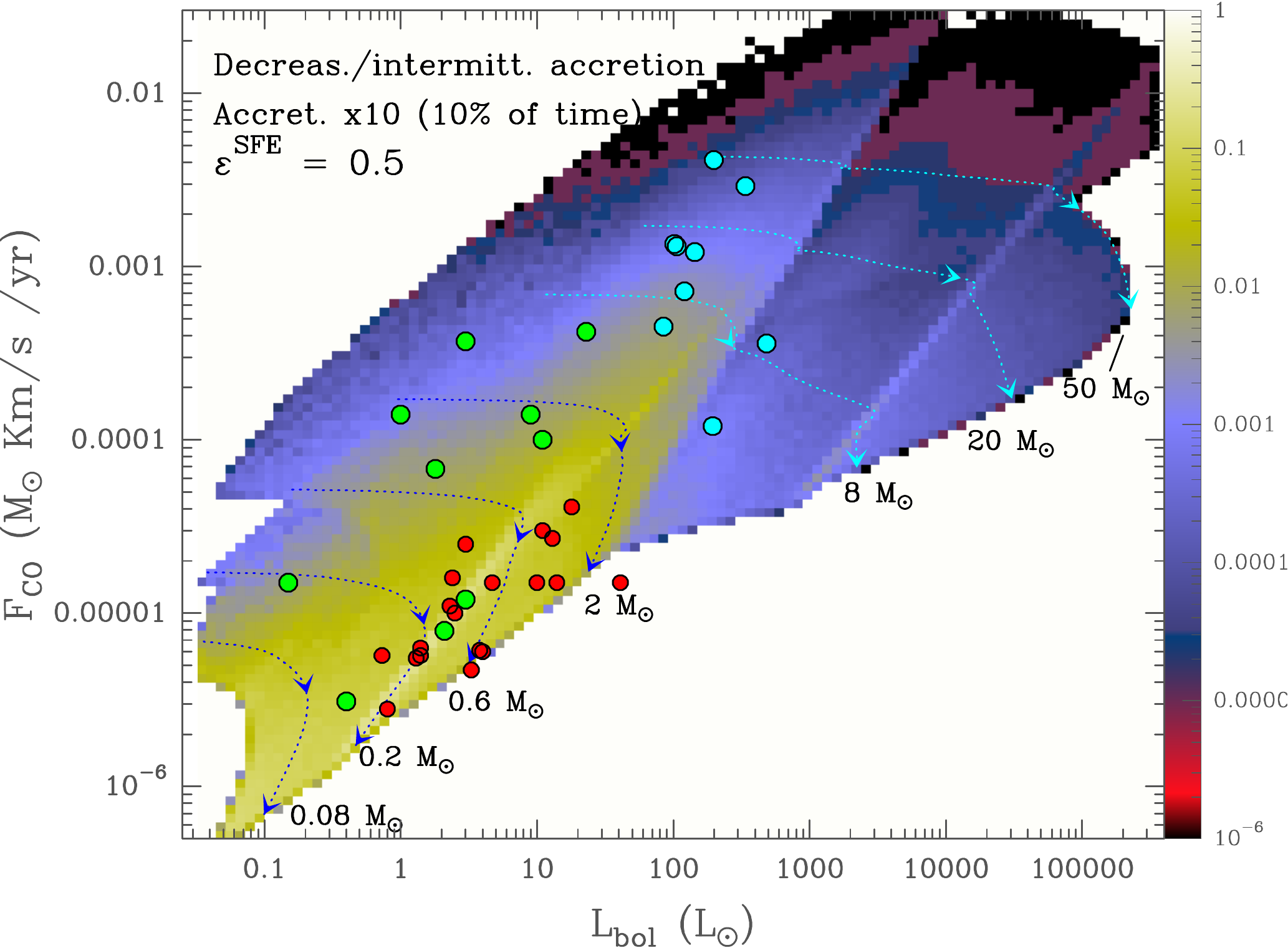}
	\caption[]{\small{\textit{Left column:} Envelope mass with respect to the bolometric luminosity. \textit{Right column:} F$_{\rm co}$ with respect to bolometric luminosity. The top row shows the model with constant accretion rate over time, the middle row shows the model with decreasing accretion rate, and the bottom panels present the model with decreasing accretion rates and intermittent accretion. For all panels, the dotted curves show the evolutionary tracks for 0.08, 0.2, 0.6, 2, 8, 20, and 50~M$_{\odot}$, and the arrows show the positions in each track where 50\% and 90\% of the envelope mass has been accreted onto the protostar. The coloured area represents a normalised surface density of the number of sources predicted to be at a given position, by taking the IMF distribution and the timescales of the evolutionary tracks into account. The massive protostars studied here are in light blue circles. The sources from \citet[][]{1996A&A...311..858B} and \citet[][]{2000prpl.conf...59A} are plotted as green- and red-filled circles (for Class 0 and Class I, respectively).}}
	\label{fig:mass_lum}}
\end{figure*} 

\subsection{Relation between accretion and ejection}

The outflow momentum flux, $F_{\rm  co}$, can be expressed as a function of the accretion rate $\dot{M}_{\rm  acc}$ as

\begin{equation}
F_{\rm  co} = f_{\rm  ent} \frac{\dot{M}_{\rm  w}}{\dot{M}_{\rm  acc}}  v_{\rm  w}  \dot{M}_{\rm  acc} 
\label{eq:outflowmomfl}
\end{equation}

\noindent
where $f_{\rm  ent}$ is the entrainment efficiency, $\dot{M}_{\rm  w}$ is the mass ejection rate and $v_{\rm  w}$ the wind speed at the ejection. Outflows are believed to extract the angular momentum of the inner regions of the accretion discs to allow/accelerate accretion at an effective radius $R_{\rm  eff}$. Thanks to conservation of momentum between the inner disc and the jet, the value of the expression $\frac{\dot{M}_{\rm  w}}{\dot{M}_{\rm  acc}}  v_{\rm  w}$, which is a velocity, tends to be constant if the ejection occurs in the same region of the disc. In the case of the X-celerator \citep[e.g.,][]{1994ApJ...429..781S} its value varies between 50 and 70 km\,s$^{-1}$. In the disc wind models, it is more of the order of 20 km\,s$^{-1}$ at a radius of $\sim 1$\,AU for a 1 M$_\odot$ stellar mass \citep[e.g.,][]{1992ApJ...394..117P,1993ApJ...410..218W,1995A&A...295..807F}. We adopt a value of 40 km\,s$^{-1}$ for this velocity\footnote{Since the outflow is expected to efficiently slow the disc material down, this effective velocity of 40 km\,s$^{-1}$ should represent a significant fraction of the Keplerian velocity of the inner disc, at the effective radius of the ejection. For a 4 M$_\odot$ stellar embryo, the Keplerian velocity is equal to 40 km\,s$^{-1}$ at $R_{\rm  eff} = 0.28$\,AU.}. For a fraction of ejected mass of 15~\% ($\dot{M}_{\rm  w}/\dot{M}_{\rm  acc}=0.15$, e.g. \citealp{2007IAUS..243..203C}), this corresponds to a realistic velocity of the jet of 300 km\,s$^{-1}$. If adopting, in addition, an entrainment efficiency of 50~\% ($f_{\rm  ent} =0.5$), then $F_{\rm  co} / \dot{M}_{\rm  acc} \approx 20$ km\,s$^{-1}$.

\subsection{Monte Carlo modelling of a typical protostellar population}

The accretion rate is expected to be constant over time only in the case of the singular isothermal sphere \citep{1977ApJ...214..488S}. In more general cases, the accretion rates may decrease over time \citep{1969MNRAS.145..271L,1969MNRAS.144..425P,1985MNRAS.214....1W,1997A&A...323..549H}. We adopt two extreme cases: the constant accretion rate and a decreasing rate using the toy model described in \citet{1996A&A...311..858B}, i.e., with ${M}_{\rm  env} (t) = M_{\rm  core} e^{-t/\tau}$, and therefore $\dot{M}_{\rm  acc} (t) = \epsilon_{\rm acc } (M_{\rm  core} / \tau) e^{-t/\tau}$, where $\tau$ is a characteristic time. To characterise $\tau$, we assume that the duration of the accretion phase is $\sim 3 \times 10 ^5\,$yr and that 90\% of the final mass of the star has been accreted at this stage.

To compare with observations we calculated the distribution of properties expected for a population of protostars formed from a constant star formation rate and with a final stellar mass distribution reproducing a normal initial mass function \citep[IMF,][]{1993MNRAS.262..545K}. For this we used a Monte Carlo code to sample time and final stellar mass. For each modelled protostar of a certain final mass and at a certain time, we calculated the current $M_\star$ and $\dot{M}_{\rm  acc}$ from the assumed accretion history. Using the work of \cite{2009ApJ...691..823H}, we could then predict the current stellar radius $R_\star$ and stellar luminosity $L_\star$ to calculate $L_{\rm  bol}$, $M_{\rm  env}$, and $F_{\rm  co}$ from the above formulas. For the case of decreasing accretion rates, we interpolated the curves of \cite{2009ApJ...691..823H} since they are only for constant accretion rates. These models are shown in Fig.~\ref{fig:mass_lum}.

\subsection{Accretion efficiency of 50~\%}

To be able to reproduce the location of the protostars in the $M_{\rm env} - L_{\rm bol}$ diagrams, we adjusted $\epsilon_{\rm acc}$ to 0.5. 
The constant accretion rate scenario (top panels of Fig.~\ref{fig:mass_lum}) does not predict a decrease in luminosity for the Class I stage, while the decreasing accretion rate scenario does (middle-row panels of Fig.~\ref{fig:mass_lum}). Therefore, when taking the spread in $M_{\rm env}$ for the Class I's $M_{\rm env} - L_{\rm bol}$ diagram into account, the constant accretion rate scenario provides a better fit. However, it is the opposite in the $F_{\rm co} - L_{\rm bol}$ diagram where the constant accretion rate scenario does not reproduce the location of protostars well, with the Class I's observed outflow momentum flux lower than that expected by the model. 

Furthermore, in a constant accretion rate scenario, the two diagrams ($M_{\rm env} - L_{\rm bol}$ and $F_{\rm co} - L_{\rm bol}$) are not compatible. In the $M_{\rm env} - L_{\rm bol}$ diagram the low-mass Class 0s actually evolve into low-mass Class Is, while this is not the case in the $F_{\rm co} - L_{\rm bol}$ diagram. In the latter, there would be no young low-mass protostars, and no evolved intermediate-mass protostars in the nearby molecular clouds. In contrast, the decreasing accretion rate scenario naturally explains the evolution from Class 0s to Class Is in both diagrams, and the outflow diagram is reproduced acceptably well in this scenario. 

Even though we fail to reproduce the range of spread of the protostars in the diagrams, we conclude that the best scenario is one with a decreasing accretion rate, and a collapse/accretion efficiency of the order of  50~\%.

\subsection{Intermittent accretion}

The only property that is not reproduced well by the decreasing accretion rate scenario is the spread in the diagrams. On the other hand, accretion is not expected to be smooth over time. The accreting material may transit in an inner disc region where material could be stored for a while until an episodic large accretion event allows for a burst of accretion and subsequent ejection \citep[e.g ][]{2012ApJ...747...52D,2007A&A...463.1017B}. Intermittence could also be due to external accretion of mass into the envelope \citep[e.g.,][]{2009MNRAS.400.1775S}, with a delay between the induced changes in the envelope mass and the respective change in accretion rates. Intermittent accretion is highly probable and is clearly observed in the T Tauri star phase. Such a variability in $\dot{M}_{\rm  acc}$ may explain a significant spread in $F_{\rm co}$ as a direct tracer of $\dot{M}_{\rm  acc}$ and in  $L_{\rm bol}$ due to changes in  $L_{\rm acc}$.

To illustrate the impact of having bursts of accretion in the distribution of sources in these diagrams, we implemented a simple intermittent accretion history in addition to the decreasing average accretion rate by adding random bursts of accretion over the whole evolution of the protostars. In the lower panels of Fig.~\ref{fig:mass_lum} we display the evolutionary diagrams for bursts of accretion occurring 10~\% of the time, with accretion rates ten times higher than at quiescent state for the high accretion state. The displayed tracks corresponds to a smooth (decreasing) accretion history, with an equivalent average rate which is roughly two times higher than the quiescent accretion rate (and five times lower than the accretion rate in the high accretion state).

With this model, which is purely meant to illustrate intermittency, we can reproduce the observed scatter better in both the $M_{\rm env} - L_{\rm bol}$ and $F_{\rm co} - L_{\rm bol}$ diagrams. In fact, this is most likely closer to a realistic situation, where accretion is not a simple function of the mass available, but it naturally suffers from bursts and fluctuations.



\section{Discussion}
\label{discussion}

\subsection{The existence of individual high-mass Class 0 protostars}

The systematic detection of powerful outflows, collimated and driven by each of the high-mass cores recognised by \citet[][]{2010A&A...524A..18B} (except CygX-N53 MM2 which could be a true prestellar core), clearly shows that these objects are the high-mass analogues of low-mass Class 0 protostars, both from their position in $M_{\rm env} - L_{\rm bol}$ diagram and their outflow properties. 

The existence of high-mass Class 0 protostars strongly suggests a similar formation process for low- and high-mass stars, meaning a monolithic collapse from a prestellar core with a finite reservoir of mass. Using the \emph{Herschel} data, we could derive precise dust temperatures of the collapsing envelopes with values ranging from $\sim16$ to 20$\,$K, i.e. very close to the value of 20~K used in \citet[][]{2010A&A...524A..18B}. It confirms that the envelopes are massive (typically 20 times more massive than low-mass Class 0s) ranging from $\sim10$ to $50\,$M$_\odot$. These masses are significantly higher than the Jeans masses in the host clumps, which range from 0.5 to 1$\,$M$_\odot$ (from the clump properties listed in Table~3 in \citealp[][]{2010A&A...524A..18B}, scaled to the new distance of Cygnus X of 1.4\,kpc). Since the turbulent support in the Cygnus X MDCs is known to presently be too weak to explain much higher effective Jeans masses than the above-quoted thermal Jeans masses \citep[see][]{2011A&A...527A.135C}, the origin of such super-Jeans cores is unclear and appears as one the main questions to be addressed for understanding the origin of high-mass Class 0 protostars. 

\subsection{Origin of a common collapse timescale}
\label{sec:common_time}

The inferred sizes of the Cygnus X high-mass cores are very similar to the sizes of Class 0 protostars in clustered low-mass star-forming regions (e.g. $\rho\,$Ophiuchi by \citet[][]{1998A&A...336..150M}). More precisely, it is the fragmentation scale, i.e. the typical separation between objects, which is important for limiting the sizes of cores. It has been estimated to be of the order of 4000\,AU for the high-mass cores of Cygnus X (3000$-$5000\,AU in \citealp[][]{2010A&A...524A..18B}), which is roughly identical to the fragmentation scale of 4000-6000\,AU reported for $\rho\,$Ophiuchi by \citet[][]{1997A&A...323..549H} and \citet[][]{1998A&A...336..150M}. 

In a quasi-static view, such a similar fragmentation scale would indicate that cores at birth, i.e. just at the beginning of the collapse, should have had similar sizes for all masses\footnote{In Cygnus-X, only CygX-N63 has no fragmentation scale since it is single. Even though its current size is limited by the flux retrieved by the single dish, the initial core before collapse of N63 could have then been significantly larger than 4000\,AU.}.
The existence of such pre-collapse cores spanning a wide range of masses, but within identical volumes, brings important implications on the collapse history of these cores. Providing gravity plays the dominant role, the collapse should be ruled, at least indirectly, by the free-fall timescales. 
In \citet[][]{1997A&A...323..549H}, it has been shown that the total protostellar time of low-mass objects (Class 0s and Class Is), presumably corresponding to the total collapse time, is four to five times the free-fall time of the inner 4000\,AU regions of the collapsing cores at birth. 
From low to high masses, an identical size of cores at birth implies that the density in these cores scales up and that the free-fall time (and collapse time) should decrease, leading to much higher accretion rates for high-mass cores. Though less dramatically, the magnetised turbulent core model of \citet[][]{2003ApJ...585..850M} also predicts shorter collapse/formation times for high-mass stars as a result of the higher densities at birth (due to high external pressure).

But in reality, despite providing reasonable estimates for low-mass objects, the accretion rates for massive protostars do not increase as much as expected from a purely monolithic collapse on 4000\,AU scales. The observed linear dependence of accretion rates with envelope mass actually implies that we have a common collapse timescale for all cores (see Sect.~\ref{sec:acc_rates}). The origin of such a common timescale is unclear, but it could either foresee that cores accrete from larger areas or that there is a stronger inner support for the individual massive cores. We explore the different possibilities in the following sections.

\subsubsection{Collapse of magnetically supported or turbulent high-mass cores?}
\label{sec:magnetic}

In a quasi-static view, the most obvious explanation for a similar collapse timescale for all masses, despite differences in initial densities, would be that there is additional pressure from turbulence or magnetic field, which would slow down the collapse for the highest mass protostars (e.g. in line with the TNT model of  \citealt[][]{1993ApJ...402..635M}, or the turbulent core model of \citealt[][]{2003ApJ...585..850M}).
However, turbulent support has been found to be too weak on the scale of clumps ($\sim 0.02 - 0.1\,$pc) to increase the effective Jeans masses to the high-mass regime, and therefore to significantly affect the collapse \citep{2011A&A...527A.135C}. Only a strong additional source of turbulence on the scale of the high-mass cores ($\sim 4000$\,AU, i.e. $0.02\,$pc) could slow down the collapses as required. This source of turbulence still needs to be discovered.
Alternatively, the magnetic field could play the role of providing this additional support.

\subsubsection{Dynamical collapse from large scales?}
\label{sec:prestellar}

In a dynamical view, the effective collapse timescale could be set by the size and density of the regions from which the gas comes, and this can go beyond the original fragmentation seeds. The origin of a similar timescale of collapse for high- and low-mass stars could be the consequence of a large-scale collapse starting at roughly the same typical density. High-mass cores are at least 20 times more massive than low-mass star precursors. To get the same density, the volume has to be 20 times greater, which corresponds to a size 2.7 times larger. The gas of the 4000\,AU high-mass cores should therefore have started its global collapse from a region of $\sim10000$\,AU, i.e. almost the size of the host MDCs (0.1 pc, i.e. 20000\,AU). The fragmentation on the lower scales of 3000-5000\,AU would then result from a dynamical fragmentation during the collapse of the entire clump. This dynamical view is in line with observations of globally collapsing high-mass star-forming clouds \citep[e.g.,][]{2010A&A...520A..49S, 2011A&A...527A.135C, 2011ApJ...740L...5C,2013A&A...555A.112P}.

\subsection{Competitive accretion VS turbulent core collapse}

In models of quasi-static monolithic collapse, we expect the formation of high-mass prestellar cores, which will be the precursors of high-mass stars. These prestellar cores should be observed since their lifetime should be longer than the Class 0 protostellar lifetime. From our sample, only one core out of nine could be a candidate for a prestellar, CygX-N53 MM2, leading to a statistical lifetime that is eight times shorter than for Class 0s \citep[which have lifetimes of $4-9\times10^4\,$yr according to][]{2011A&A...535A..77M}. As such, high-mass prestellar cores would have a lifetime of 0.5 to $1\times 10^4\,$yr. With a typical velocity dispersion of 2 km/s inside the larger scale MDCs, this lifetime corresponds to only one crossing time for the typical size of 4000\,AU. These prestellar cores can therefore not be formed through a quasi-static process, so they would need to be formed by dynamical processes. The statistical base for Cygnus X is still low, so that the true timescale of such a prestellar phase stays uncertain. But our estimate is in line with other recent estimates of high-mass prestellar phase by \citet{2007A&A...476.1243M} and \citet{2012A&A...538A.142R} derived for larger scales, suggesting that the prestellar phase of high-mass stars has to be short. One could also wonder whether the selection is biased and whether we may have missed larger, colder cores that would represent the prestellar phase for high-mass star formation. We believe that this is unlikely to be the case, since the selection is based on the \citet{2007A&A...476.1243M} survey of the entire Cygnus X complex, and no MDCs on the scale of 0.1 - 0.2 pc ($\sim 8$ times larger than 4000\,AU) were found to be in a pre-stellar phase. 
We can therefore conclude that despite witnessing monolithic collapses on the scale of 4000\,AU in a population of high-mass Class 0 objects, they cannot have been formed through a quasi-static evolution of turbulent or magnetically supported MDCs. 

Alternatively, if massive stars were to be formed in a competitive accretion scenario, the precursor of a massive star would in fact be a low-mass protostar, as a result of the original fragmentation of the cloud, accreting strongly from the environment, according to their position within the potential well. Indeed this would  agree with the existence of higher accretion rates for the precursors of massive stars, but it does not explain the higher envelope mass in a similarly sized volume as those of low-mass protostars. Observing high masses in compact envelopes for the precursors of massive stars implies that such massive protostellar envelopes have an early origin. Our sample of young Class 0 protostars (and one candidate prestellar core) already contain enough mass to form high-mass stars, and the accretion rates already reflect the current amount of mass available, pointing to, at least at first order, a view of a monolithic collapse.

Altogether, our favoured scenario is therefore a combination of these two extreme views. A global collapse on the scale of the MDCs sets on, followed by some degree of turbulent fragmentation on scales of a few thousand AU, and the possible formation of short-lived high-mass prestellar cores (such as in the turbulent core model, but driven by a high level of dynamics, with dynamical times of $\sim10^4\,$yr). The cores then collapse in a monolithic way. An overlap of the collapsing phase and of the dynamical accretion of mass from global collapse is likely to happen during the Class 0 phase, since the dynamical timescales are similar to the typical lifetimes of Class 0 protostars. 


\section{Summary and conclusions}
\label{concl}

We have studied the properties of nine high-mass cores in the Cygnus X complex, which are candidates for young accreting protostellar objects. From our estimates of their masses, bolometric luminosities, and outflow momentum flux, we conclude that eight out of nine sources of our sample are clearly true equivalents of Class 0 protostars, in the high-mass regime, and that one core could be a rare example of a high-mass prestellar core on the verge of collapsing to form a single (or a close) binary high-mass star(s). 

The main result of our study lies in the fact that the momentum flux of high-mass objects scales linearly with the reservoir of mass in the envelope, as a true scale-up of the relations previously found for low-mass protostars. Assuming that outflow momentum flux is an indicator of the accretion rates, the observed linear relation suggests that the accretion rates are proportional to the envelope mass, which points to a fundamental relationship between accretion and the mass of the core. Such a linear relation also implies that the timescales for accretion have to be similar for all masses, and are therefore close to what is found for low-mass star formation, namely $\sim 3 \times 10^{5}\,$yr.

The existence of a common duration of collapse for all stellar masses could arise from having an original free-fall timescale similar for all objects, thanks to similar initial densities. This would have the underlying consequence of more massive objects accreting mass from larger regions than low-mass ones, favouring the models of large-scale global collapse of dense clumps. This is in line with recent results for the observed short timescales for IR-quiet dense clumps \citep{2007A&A...476.1243M,2012A&A...538A.142R} and for the dynamics at early stages in high-mass star-forming clumps \citep[e.g.,][]{2010A&A...520A..49S, 2011A&A...527A.135C, 2011ApJ...740L...5C}, and more recently by \citet[][]{2013A&A...555A.112P}.


\begin{acknowledgements}

We thank the referee, J. Tan, for his important remarks that made the article clearer and stronger. We are grateful to the HOBYS key programme and the specialist astronomy group dedicated to star formation within the SPIRE Consortium (SAG 3) for letting us use a preliminary version of their \emph{Herschel} photometric results in advance of publication. A first-generation paper including the full \emph{Herschel} sources catalogue in the Cygnus-X complex will be published shortly (Hennemann et al., in prep.).
ADC is supported by the project PROBeS, and NS is supported by the project STARFICH, both funded by the French National Research Agency (ANR). IRAM is supported by INSU/CNRS (France), MPG (Germany), and IGN (Spain). The data reduction and analysis made use of the GILDAS software (http://www.iram.fr/IRAMFR/GILDAS).  
      
\end{acknowledgements}

\bibliographystyle{aa}	
\bibliography{references}		

\onecolumn

\begin{appendix}

\section{Flux extraction and SED fittings}
\label{ap:SEDs}

To estimate the mass, bolometric luminosity, and dust temperatures of our sample of nine high-mass fragments, we have constructed SEDs using the emission from 24\,$\mu$m to 3.5mm. Since our sample of sources consists of IR-quiet cores, the flux bellow 8\,$\mu$m is in fact negligible, and integrating the SEDs down to 24\,$\mu$m is enough to get a good estimate of the bolometric luminosity. We used the 24\,$\mu$m from Spitzer MIPS (6'' resolution), the 70, 160, 250, 350, and 500\,$\mu$m from \emph{Herschel} PACS and SPIRE (8.4'', 13.5", 18.2'', 24.9'', and 36.3'' resolution, respectively) observed as part of the HOBYS programme (\citealp{2012A&A...543L...3H}; in prep), the 1.2\,mm from MAMBO \citep[11'' resolution,][]{2007A&A...476.1243M}, and the 1.3\,mm and 3.5\,mm emission from PdBI \citep[1'' and 2.4'' resolution respectively,][]{2010A&A...524A..18B}.

\begin{table*}[!b]
\small
\caption{SED results}
\begin{tabular}{l | c c | c c | c | c c c | c c }
\hline 
\hline
\multirow{4}{*}{Source}  	& \multicolumn{5}{c |}{Best SED fit}		&  \multicolumn{3}{c |}{\multirow{2}{*}{SED with fixed M$_{env}$ and T$_{dust}^{(a)}$}}  &	\multicolumn{2}{c}{\multirow{2}{*}{Using datapoints}} \\
				 	& \multicolumn{2}{c |}{Cold component} & \multicolumn{2}{c |}{Warm component} & Total	&  		& &				& 			&   \\
					& M$_{env}$ 	 & T$_{dust}$		& M$_{env}$ 		& T$_{dust}$ 	& L$_{bol}$	&  \multirow{1}{1cm}{\,\,\,M$_{env}$}     & 	T$_{dust}$	& L$_{bol}$	& L$_{max}$ 	& L$_{min}$ \\ 
					& (M$_{\odot}$) & (K)			& (M$_{\odot}$)		& (K)			& (L$_{\odot}$) & (M$_{\odot}$)	 &	 (K) 			& (L$_{\odot}$) 	&   \multicolumn{2}{c}{(L$_{\odot}$)} \\
\hline
CygX-N3 MM1			& $12.5\pm3.7$ & $19.0\pm1.3$	& 			-	& 	-		  & 	106		& 	9.4		& 		20		&	109		&	185	   & 48 \\ 
CygX-N3 MM2			& $13.8\pm5.6$ & $18.0\pm3.0$ 	& $0.02\pm0.03$	& $46.8\pm7.0$  & 	121		& 	7.4		& 		20		&	122		&	197	   & 68  \\ 
\hline
CygX-N12 MM1		& $17.7\pm6.9$ & $20.4\pm5.2$	& $0.54\pm0.78$	& $36.9\pm3.5$  & 	481		& 	15.6		& 		20		&	432		&	816	   & 350 \\ 
CygX-N12 MM2		& $16.4\pm5.8$ & $19.1\pm3.3$	& $0.04\pm0.11$	& $43.0\pm9.0$ & 	185		& 	12.5		& 		20		&	187		&	309	   & 113 \\ 
\hline	
CygX-N48 MM1		& $17.0\pm6.4$ & $17.9\pm4.0$	& -				& 	-		  & 	102		& 	11.8		&		 20		&	136		&	470	   & 5 \\ 
CygX-N48 MM2		& $8.1\pm3.0$  & $19.7\pm4.1$	& -				& 	-		  & 	85		& 	8.2		&		 20		&	95		&	483	   & 10 \\ 
\hline
CygX-N53 MM1		& $34.2\pm11.1$ & $15.5\pm2.2$	& $0.20\pm0.12$	& $39.2\pm3.0$  &   199		& 	26.6		& 		20		&	422		&	343	   & 130 \\ 
CygX-N53 MM2		& $21.4\pm6.1$  & $18.3\pm1.8$	& -				& -			  & 	144		& 	16.9		& 		20		&	196		&	285	   & 95 \\ 
\hline
CygX-N63 MM1		& $44.3\pm11.9$ & $16.6\pm1.6$	& $0.49\pm0.30$	& $35.2\pm1.6$  & 	339		& 	38.6		&		 20		&	621		&	537	   & 290 \\ 
\end{tabular}
\\
$^{(a)}$ from \citet[][]{2010A&A...524A..18B}, with masses corrected for the new distance of 1.4\,kpc.
\label{tab:sed_results}
\end{table*}

\begin{figure*}[!b]
\centering
	\hbox{\includegraphics[width=0.69\textwidth]{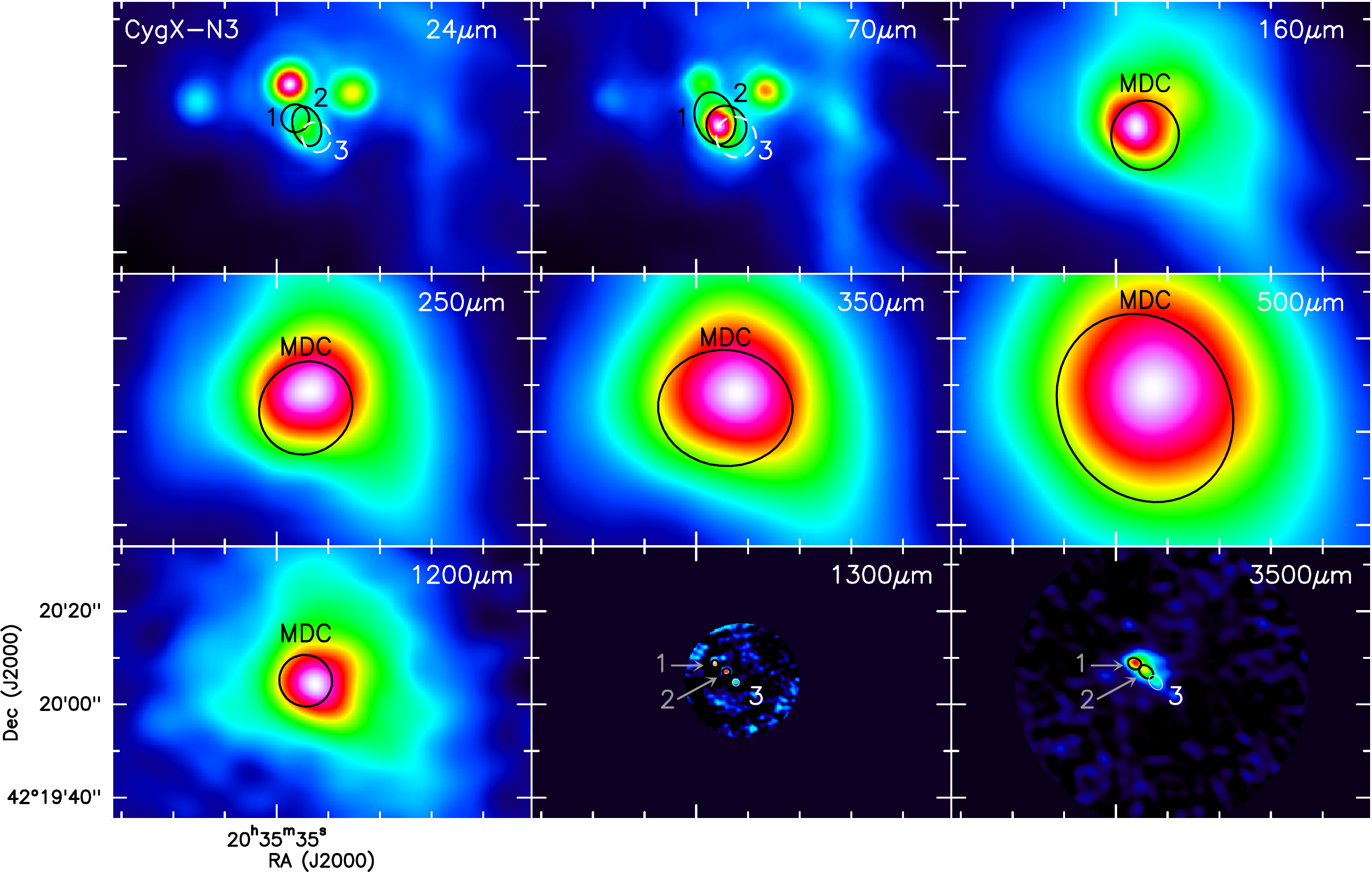}\hspace{-6.3cm}
	\vbox{\includegraphics[width=0.3\textwidth]{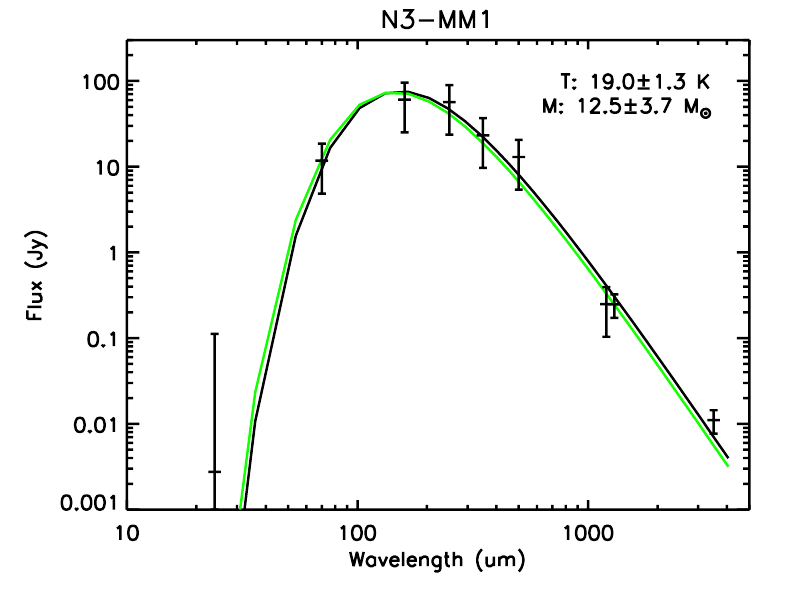}\\
	\includegraphics[width=0.3\textwidth]{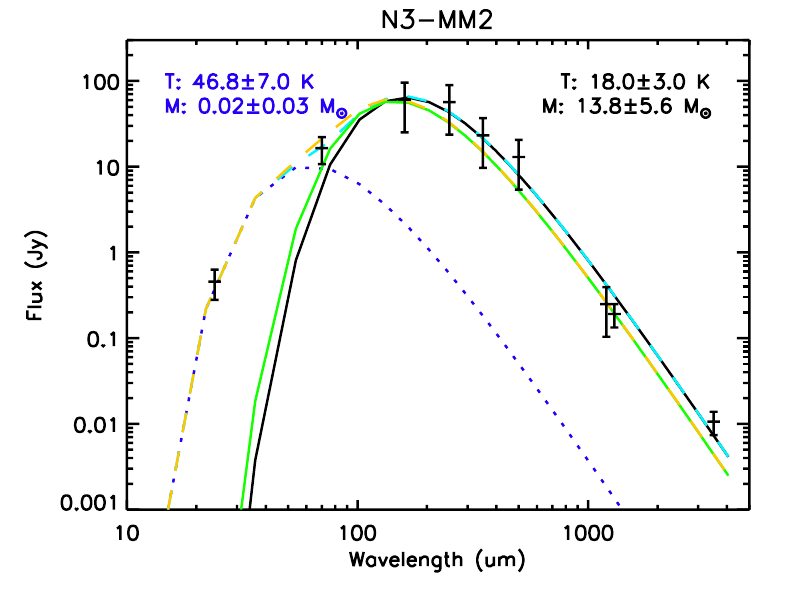}}}
	\caption{\small{\textit{Left:} Results of the source extraction by {\it getsources} in CygX-N3. All images cover the same angular scale. The detections of {\it getsources} at 1.3mm are used to extract the fluxes at 24\,$\mu$m, 70\,$\mu$m, 1.3\,mm, and 3.5\,mm (even though we study only sources 1 and 2, in black). The {\it getsources} detections at 24\,$\mu$m are used to retrieve the flux of the MDC in the remaining bands (MDC shown in black). These MDC fluxes are divided by the 3 fragments detected at 1.3mm. \textit{Right:} SEDs of CygX-N3 MM1 (top) and MM2 (bottom). The black curves (and top-right parameters) are the grey-body fits to the cold component. The green line shows the grey-body curve of the cold component assuming fixed parameters \citep[from][]{2010A&A...524A..18B}. When applicable, the best SED fit of a warm component is shown as a dotted blue curve and top-left parameters. The sum of the cold and warm components is plotted with dashed lines (light blue for the sum of best SED fit, and yellow for the sum of the warm component with the fixed-parameters SED).}}
	\label{fig:sed_n3}
\end{figure*}

The extractions of the fluxes were made using the {\it getsources} code \citep[v1.120126,][]{2012A&A...542A..81M}. Because the PdBI 1.3mm emission filters out the more extended emission, we cannot derive precise envelope/core sizes. However, since the mean separation between the high-mass fragments detected by \citet[][]{2010A&A...524A..18B} varies between 3000\,AU and 5000\,AU, we take 4000\,AU (i.e. $\sim$3'' resolution) as the size for estimating the properties of the individual high-mass collapsing cores. Therefore, owing to the different resolutions of the datasets, two separate extractions were made. 

\begin{figure*}[!t]
\centering
	\hbox{\includegraphics[width=0.69\textwidth]{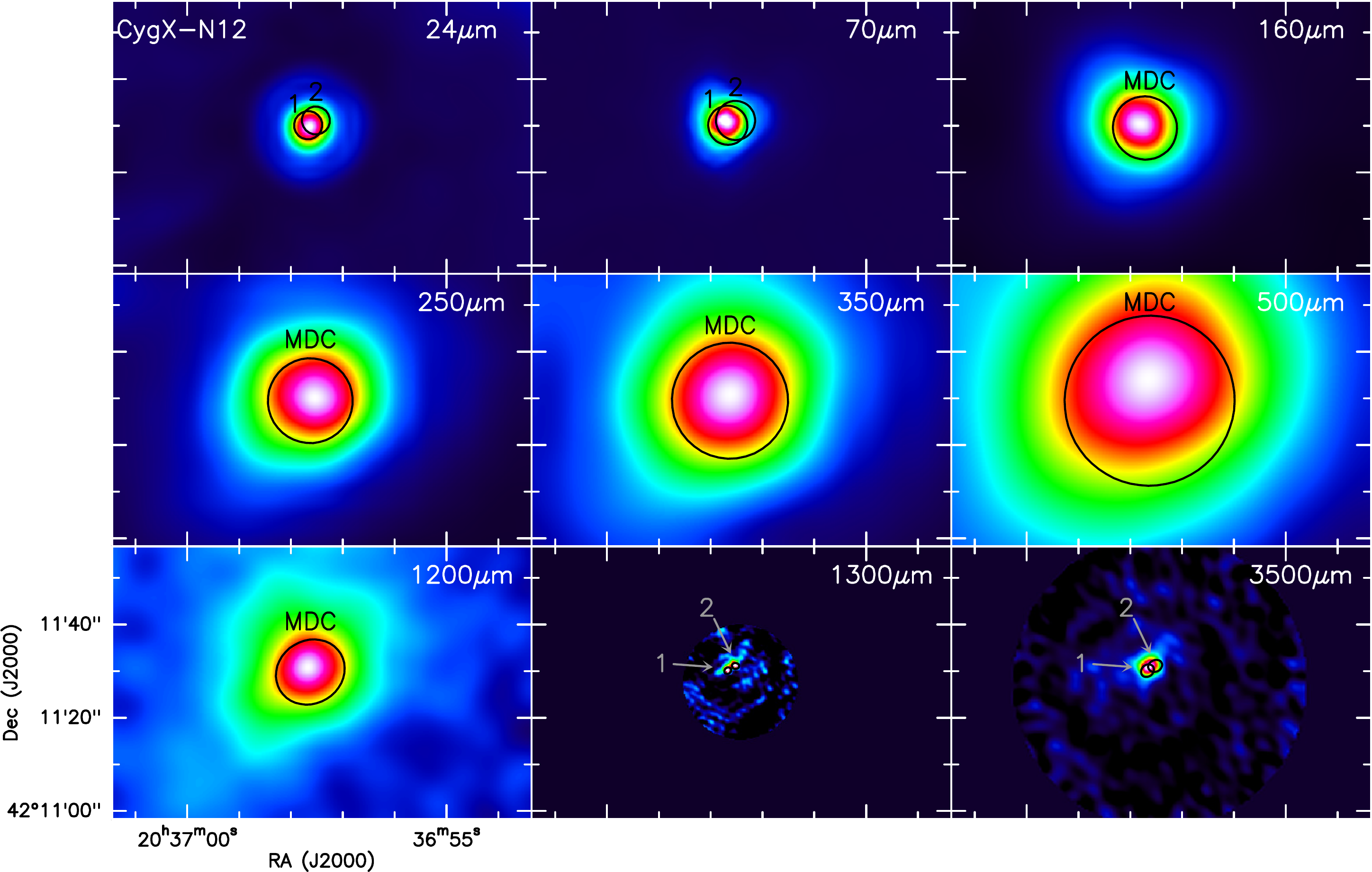}\hspace{-6.3cm}
	\vbox{\includegraphics[width=0.3\textwidth]{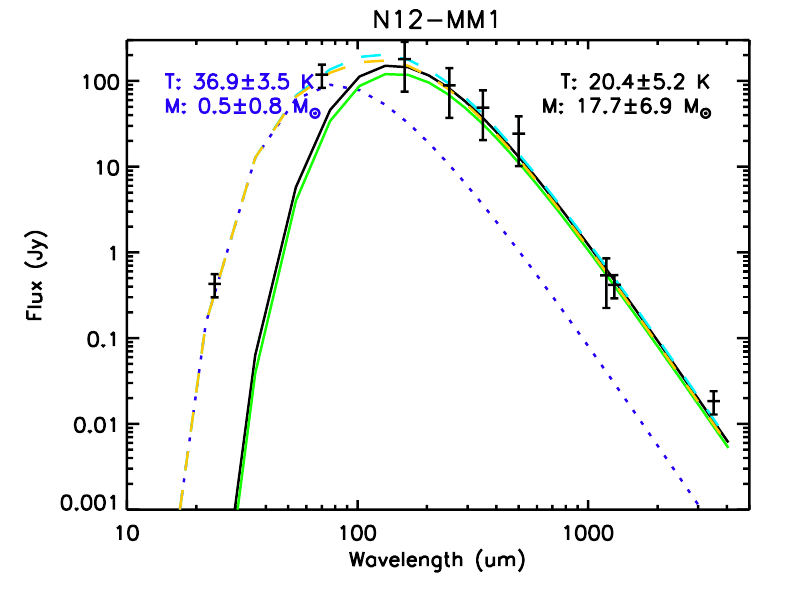}\\
	\includegraphics[width=0.3\textwidth]{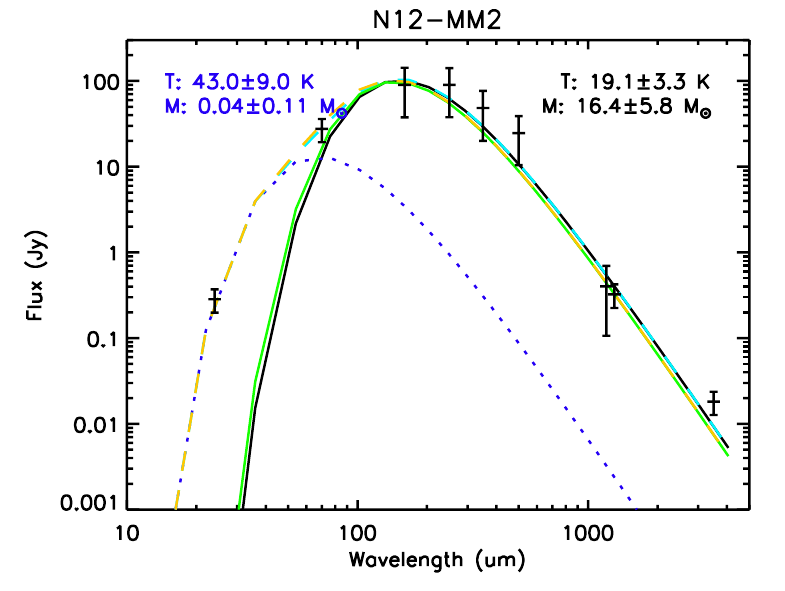}}}
	\caption{\small{\textit{Left:} Results of the source extraction by {\it getsources} in CygX-N12. The MDC fluxes are divided by the 2 fragments detected at 1.3mm. \textit{Right:} SEDs of CygX-N12 MM1 (top) and MM2 (bottom). Curves are the same as in Fig.~\ref{fig:sed_n3}. }}
	\label{fig:sed_n12}
\end{figure*}

\begin{figure*}[!t]
\centering
	\hbox{\includegraphics[width=0.69\textwidth]{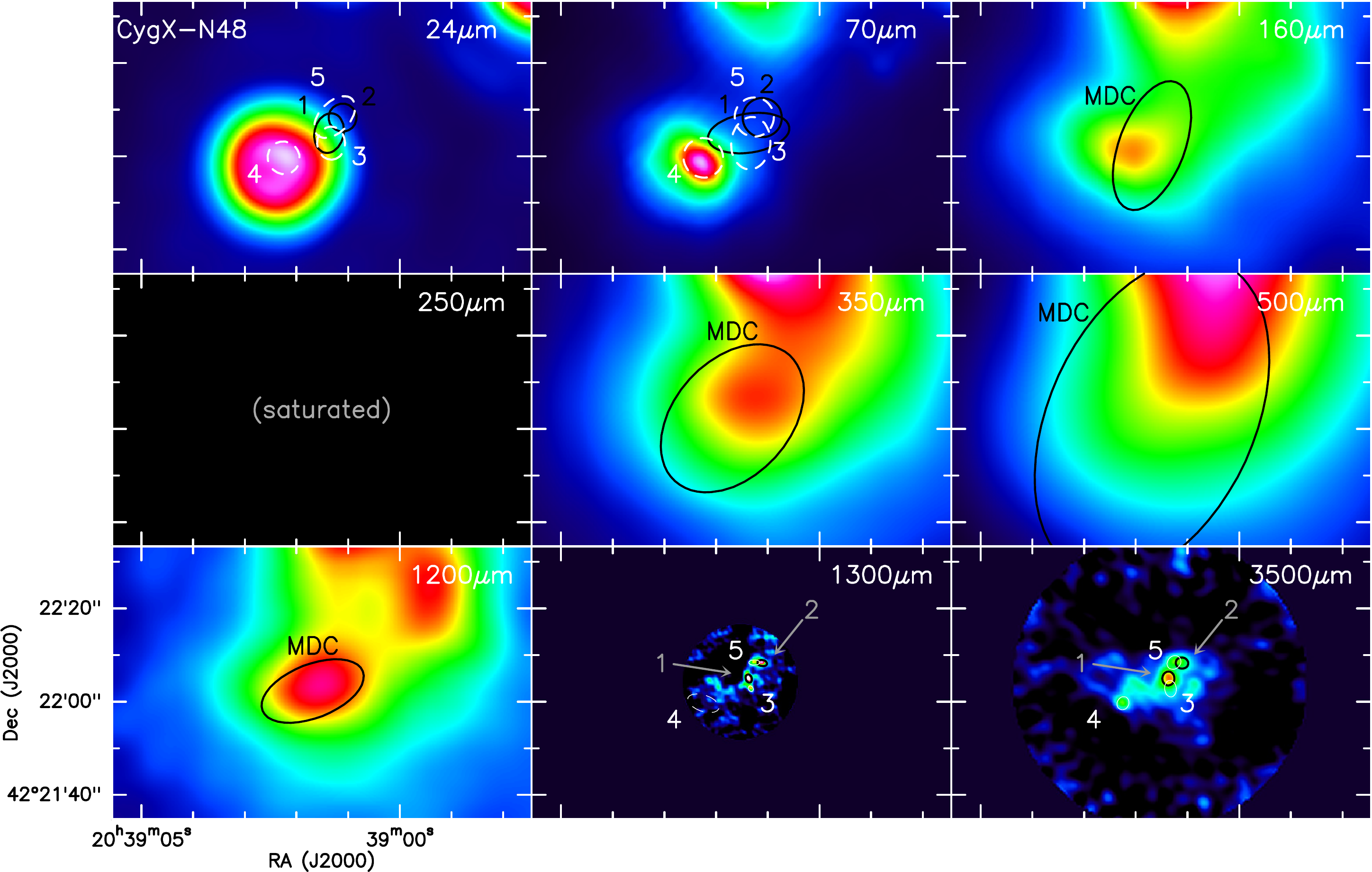}\hspace{-6.3cm}
	\vbox{\includegraphics[width=0.3\textwidth]{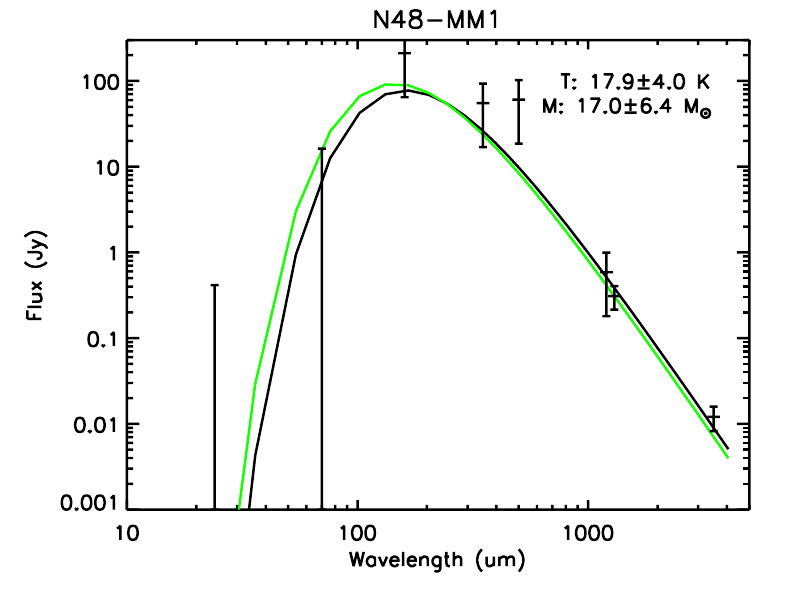}\\
	\includegraphics[width=0.3\textwidth]{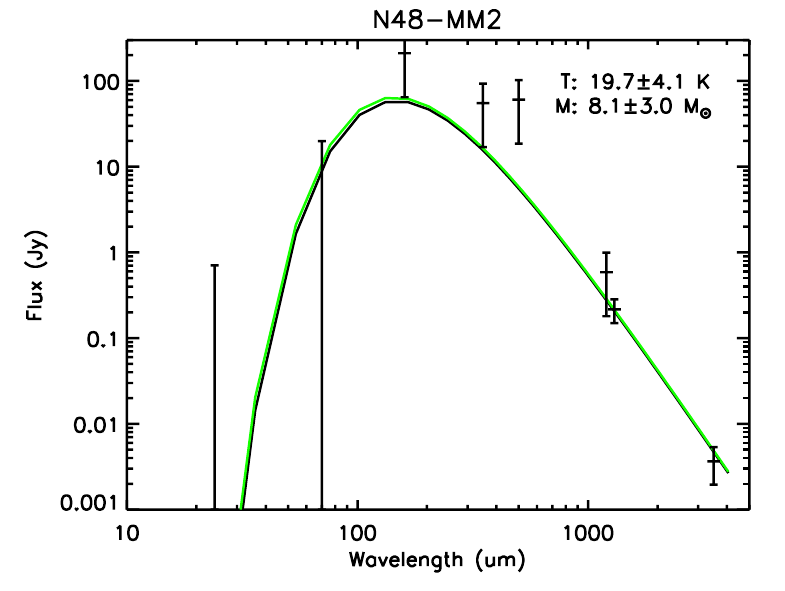}}}
	\caption{\small{\textit{Left:} Results of the source extraction by {\it getsources} in CygX-N48. The 250\,$\mu$m image is unusable since it is saturated. In this case, because the 24\,$\mu$m emission does not trace the millimetre sources, we have used the combined detections at 350\,$\mu$m and 1.2\,mm for estimating the emission from the MDC, whose flux is then divided by 5 fragments. The flux extraction at 500\,$\mu$m is not very well constrained to the MDC only, and therefore the flux measurement at 500\,$\mu$m is likely overestimated. \textit{Right:} SEDs of CygX-N48 MM1 (top) and MM2 (bottom). Curves are the same as in Fig.~\ref{fig:sed_n3}. }}
	\label{fig:sed_n48}
\end{figure*}

\begin{figure*}[!t]
\centering
	\hbox{\includegraphics[width=0.69\textwidth]{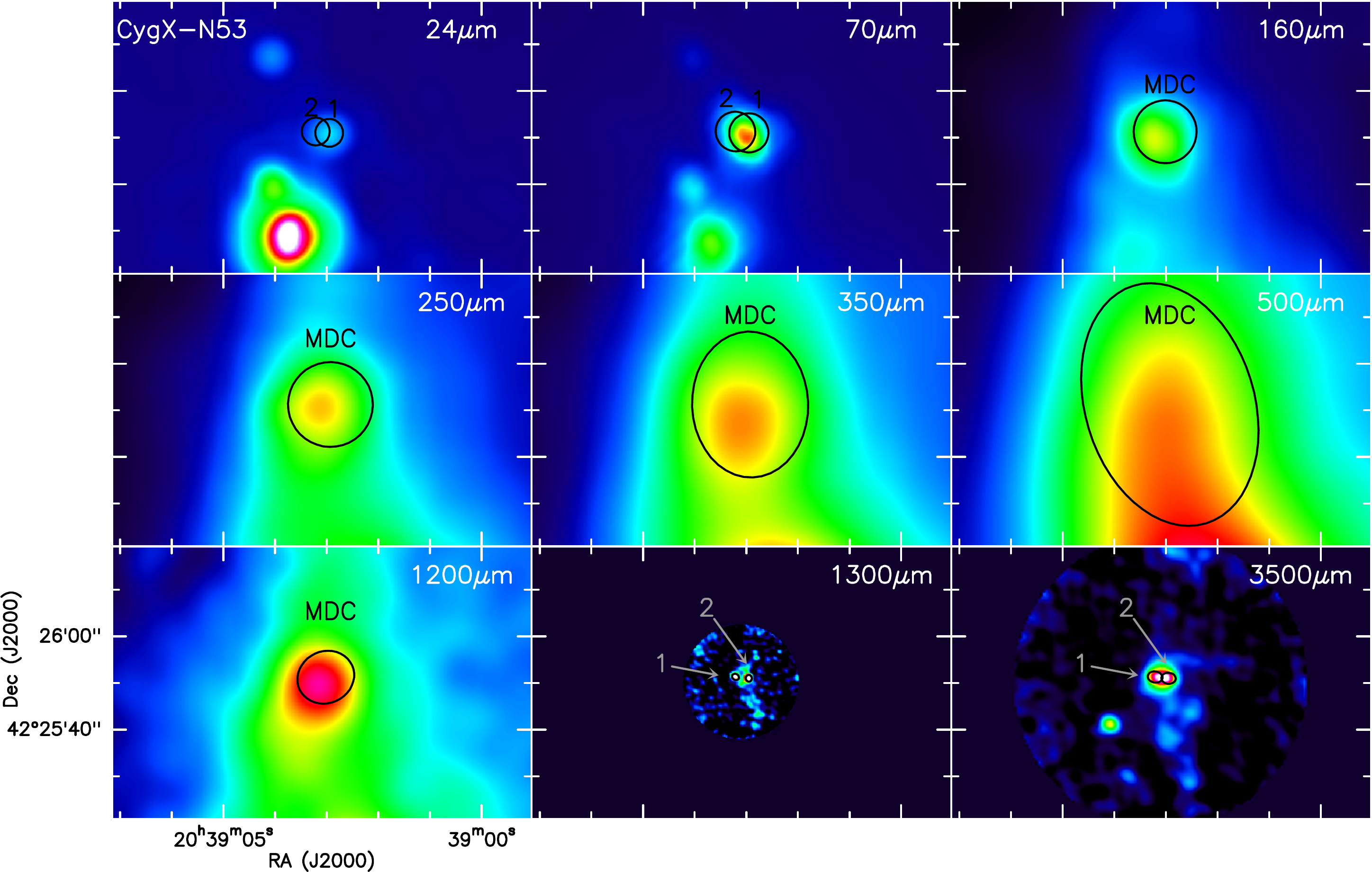}\hspace{-6.3cm}
	\vbox{\includegraphics[width=0.3\textwidth]{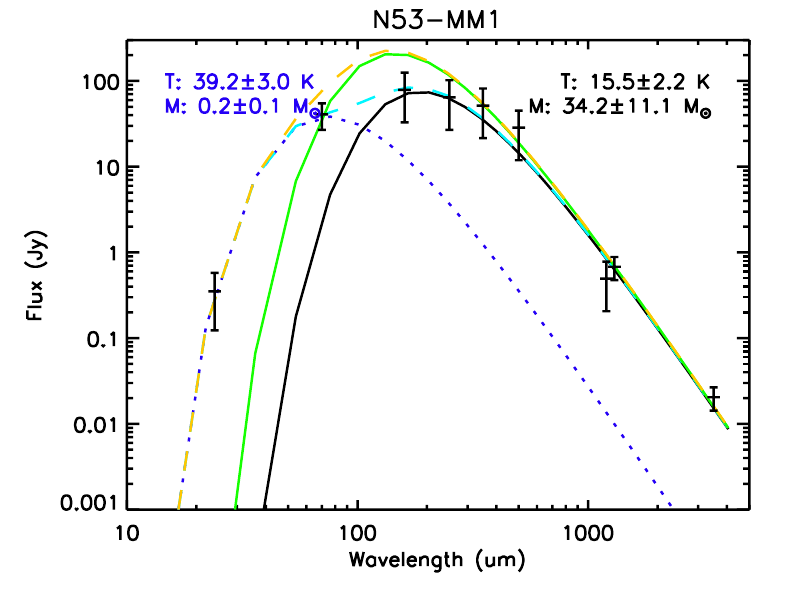}\\
	\includegraphics[width=0.3\textwidth]{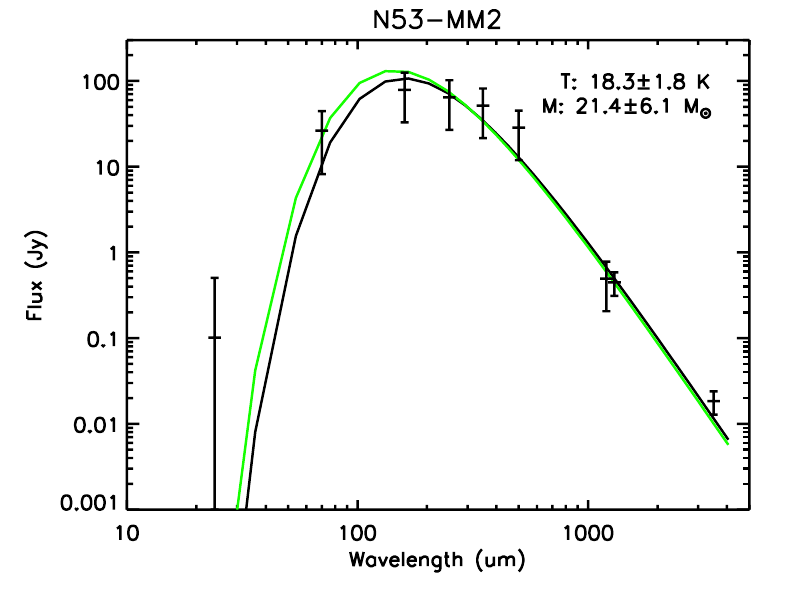}}}
	\caption{\small{\textit{Left:} Results of the source extraction by {\it getsources} in CygX-N53. The emission from the MDC is then divided by 2 fragments. \textit{Right:} SED of CygX-N53 MM1 and MM2. Curves are the same as in Fig.~\ref{fig:sed_n3}. The extraction from {\it getsources} assigned a significant flux to MM2 at 70\,$\mu$m even though this emission is mostly centred on MM1. Therefore, we believe that the 70\,$\mu$m flux measurement is slightly overestimated for MM2. For MM1 we can see that a temperature of 20K (green curves), as assumed by \citet[][]{2010A&A...524A..18B}, greatly overestimates the emission around the peak of the SED.}}
	\label{fig:sed_n53}
\end{figure*}

\begin{figure*}[!t]
\centering
	\hbox{\includegraphics[width=0.69\textwidth]{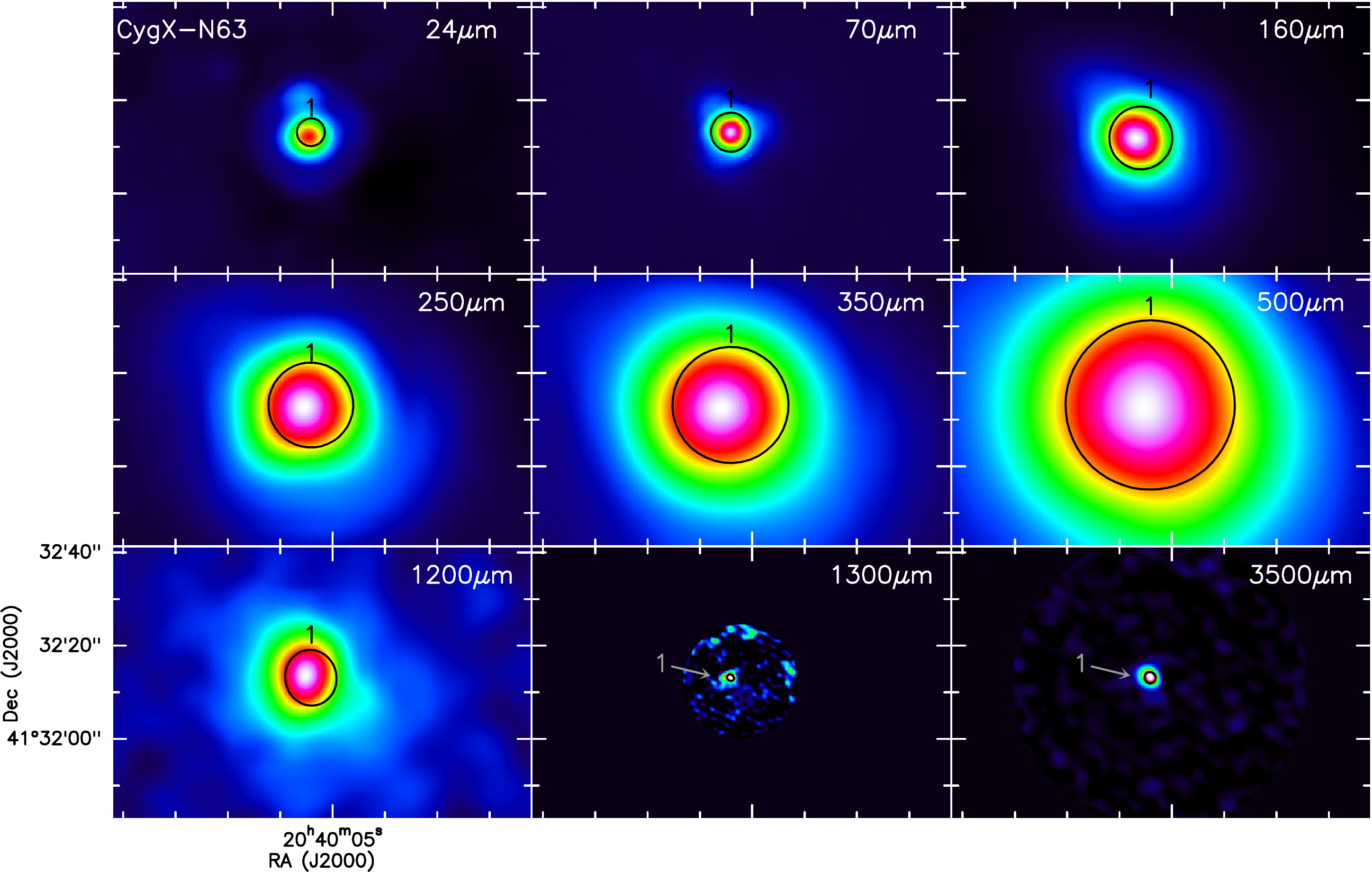}\hspace{-6.3cm}
	\vbox{\includegraphics[width=0.3\textwidth]{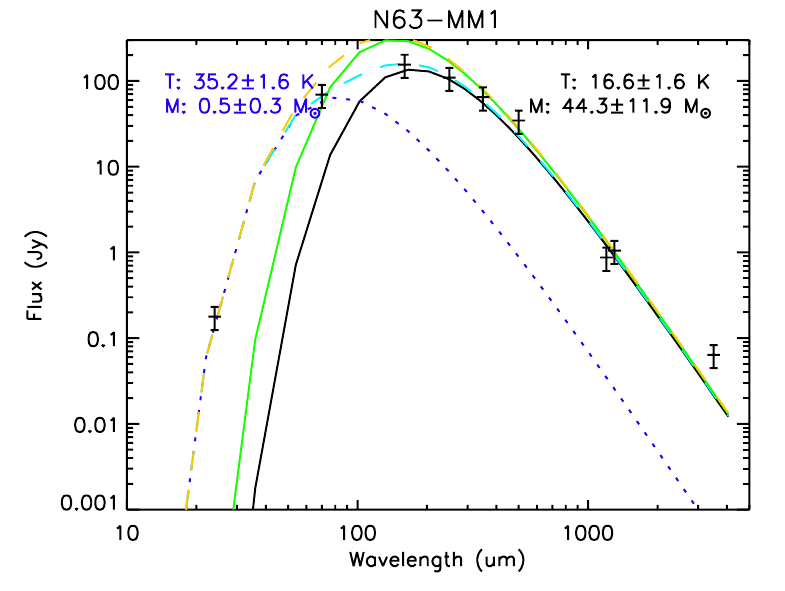}}}
	\caption{\small{\textit{Left:} Results of the source extraction by {\it getsources} in CygX-N63, where all the emission arises from MM1 (not sub-fragmented). \textit{Right:} SED of CygX-N53 MM1. Curves are the same as in Fig.~\ref{fig:sed_n3}. Similar to CygX-N53 MM1, the temperature of 20K (green curves) for CygX-N63 MM1, as assumed by \citet[][]{2010A&A...524A..18B}, overestimates the emission around the peak of the SED.}}
	\label{fig:sed_n63}
\end{figure*}

For resolutions worse than 10'' (i.e. physical scales greater than 14\,000\,AU), which is the case for the 160\,$\mu$m, 250\,$\mu$m, 350\,$\mu$m, 500\,$\mu$m, and 1.2\,mm images, we are no longer able to separate the emission from the different fragments, and therefore we have estimated the flux correspondent to the MDC as a whole. For these five wavelengths, we used the 24\,$\mu$m emission as the detection wavelength, in order to give an idea of the position of the MDC (black ellipses, labelled with ``MDC" on Figs.~\ref{fig:sed_n3} to \ref{fig:sed_n63}), plus the existence of other nearby 24\,$\mu$m sources that could be blended within the coarser resolutions. 
Because {\it getsources} uses the positional information coming from the detection wavelength to make the best deblended flux estimates at 
the other wavelengths \citep[see][for details on how the deblending is made]{2012A&A...542A..81M}, the flux we retrieve should represent the flux of the entire MDC, deblended for any nearby sources. We then assume that the contribution of each individual fragment to the continuum is similar, and therefore we divided the MDC flux by the number of PdBI 1.3\,mm fragments as an estimate of their individual fluxes. Since this method only provides a rough estimate of the individual fluxes, we increased the uncertainties associated with these measurements accordingly. (We assume uncertainties of $\sim$60\,-\,70\% on the individual fluxes.) The lower weight we associate to these points on the SEDs implies that the SED fittings do not strongtly rely on the actual fluxes from 160\,$\mu$m-1.2\,mm, and these only provide a trend in the behaviour of the peak of the SEDs, useful to help constrain the temperature.

The most reliable measurements are those whose resolution is enough to separate the individual fragments. That is the case for the 24\,$\mu$m, 70\,$\mu$m, 1.3\,mm, and 3.5\,mm. For these four wavelengths, we extracted the fluxes using {\it getsources} detections at 1.3mm. Even though at 24\,$\mu$m and 70\,$\mu$m the sources are still blended, in most cases the peak is centred well on one of the mm sources, allowing {\it getsources} to properly assign the flux arising from the individual fragments. Since the emission at 24\,$\mu$m (and 70\,$\mu$m) arises from the inner warm regions around a protostar, they are expected to correspond to sizes of $<$7000\,AU. Therefore, even though the area where the flux is measured is larger than this, we need no rescaling of the fluxes because all the inner emission is recovered in the beam. For the 3.5mm emission we also do not need to rescale, since {\it getsources} extracts the flux from fragments using the beam size, i.e 3", which corresponds to regions of $\sim$4000\,AU size. For the 1.3mm, however, with a resolution of 1'', the PdBI filters out all extended emission and is only sensitive to the inner $\sim$1000\,-\,2000\,AU. Therefore, to construct the SEDs we used the PdBI 1.3mm fluxes as measured by \citet[][]{2010A&A...524A..18B} that correspond to deconvolved sizes in the range of 800-1500\,AU (for a distance of 1.4kpc), and rescaled these to more realistic envelope sizes by assuming a density profile as $\rho \propto r^{-2}$ outside the central $\sim$1000\,AU \citep[as in][]{2010A&A...524A..18B}. Density profiles close to $\rho \propto r^{-2}$ have been observed for individual protostars \citep[e.g.,][]{2001A&A...365..440M,2002ApJS..143..469M}, and even though studies of starless/prestellar cores have suggested shallower slopes \citep[e.g.,][]{2012ApJ...754....5B}, these studies also suggest that steepening of these profiles could be expected with the evolution and contraction of such cores. In fact, if using a shallower density profile, we would very quickly recover all the single-dish emission. For instance, for the isolated CygX-N63 MM1, the most massive core of our sample, all the single dish 1.2\,mm emission is in fact recovered by rescaling the PdBI 1.3\,mm flux to simply $\sim 2500$\,AU using a steep $r^{-2}$ density profile. Using a shallower profile would imply that this core contained all its mass within a mere thousand AU, and it would require a sharp edge after that. Therefore, we have taken a profile as $r^{-2}$, because our sample of cores already show important outflow activity, indicative of on-going star formation, and we rescaled the PdBI 1.3\,mm emission to envelope sizes of $\sim$\,4000\,AU FWHM for all sources, except for CygX-N63 MM1 where we used 2500\,AU. Despite the small errors retrieved by the flux extraction by {\it getsources}, we assumed conservative uncertainties of the flux measurements for these four wavelengths,  as being 30\% of the flux extracted. This is meant to account for biases linked to the flux extraction (using different parameters for the {\it getsources} extraction), the assumptions on the need for rescaling and observational uncertainties.

Using these flux measurements, we constructed the SEDs for the nine high-mass fragments of our sample. Each SED was fit with a grey-body curve, assuming an opacity law as in \citet[][]{1983QJRAS..24..267H} with $\beta = 2$, assuming a dust emissivity of 1.0\,cm\,g$^{-1}$ at 1.3mm \citep[][]{1994A&A...291..943O} and a dust-to-gas ratio of 100. For sources with 24\,$\mu$m emission, two-temperature grey-body fits were made with both a cold and a warm component. The SEDs and their grey-body fittings can be seen in Figs.~\ref{fig:sed_n3} to \ref{fig:sed_n63}. We have estimated the bolometric luminosities using three methods:  \textit{1)} integrating the best SED curve fitting; \textit{2)} integrating the SED fitting retrieved from assuming the mass from \citet[][]{2010A&A...524A..18B} estimated using a dust temperature of 20K for all sources, and corrected for the new distance of Cygnus-X; \textit{3)} integrating the SED points (without any fitting), and estimating the uncertainties on the bolometric luminosity by using the maximum and minimum points from the error bars. The results from the three methods agree nicely, and therefore we take the masses and bolometric luminosities estimated from the best SED fitting. The uncertainties on the bolometric luminosity were estimated as the difference between methods \textit{1)} and the minimum estimate of method \textit{3)}. It is worth noting that despite the rough method for determining the fluxes for the unresolved wavelengths, the bolometric luminosity is most sensitive to the flux estimates at 24 and 70\,$\mu$m. Since the flux extraction for these two wavelengths is relatively well constrained, we consider that the errors we provide are adequate. The results from these SED fittings are summarised in Table~\ref{tab:sed_results}. 

\end{appendix}

\begin{appendix}
\section{Areas for momentum flux estimates}
\label{areas}

\begin{figure*}[!t]
	\centering
	{\renewcommand{\baselinestretch}{1.1}
	\hbox{\vspace{-6cm} \includegraphics[width=0.6\textwidth]{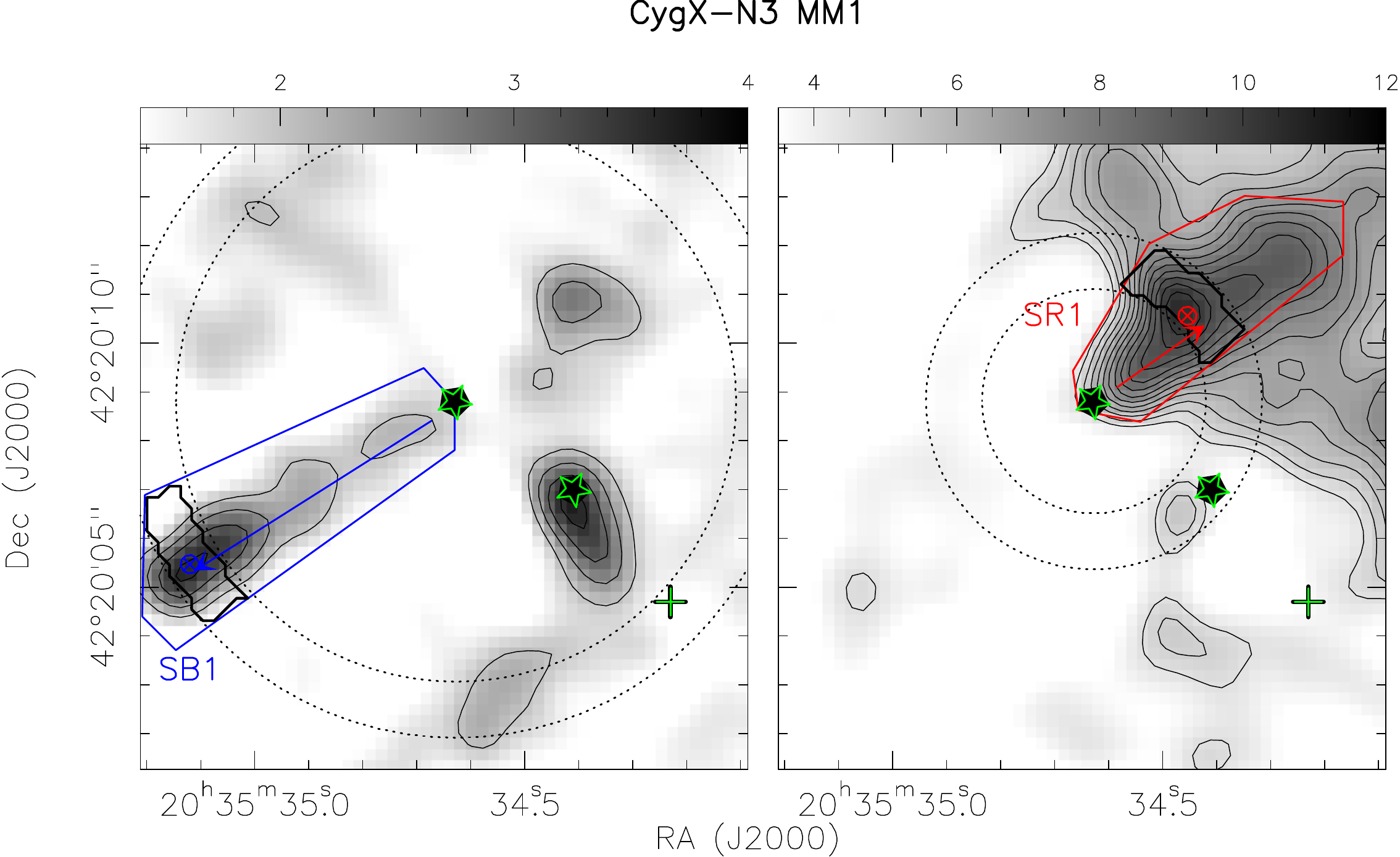}}	
	\hspace{11cm} \hbox{\includegraphics[width=0.37\textwidth]{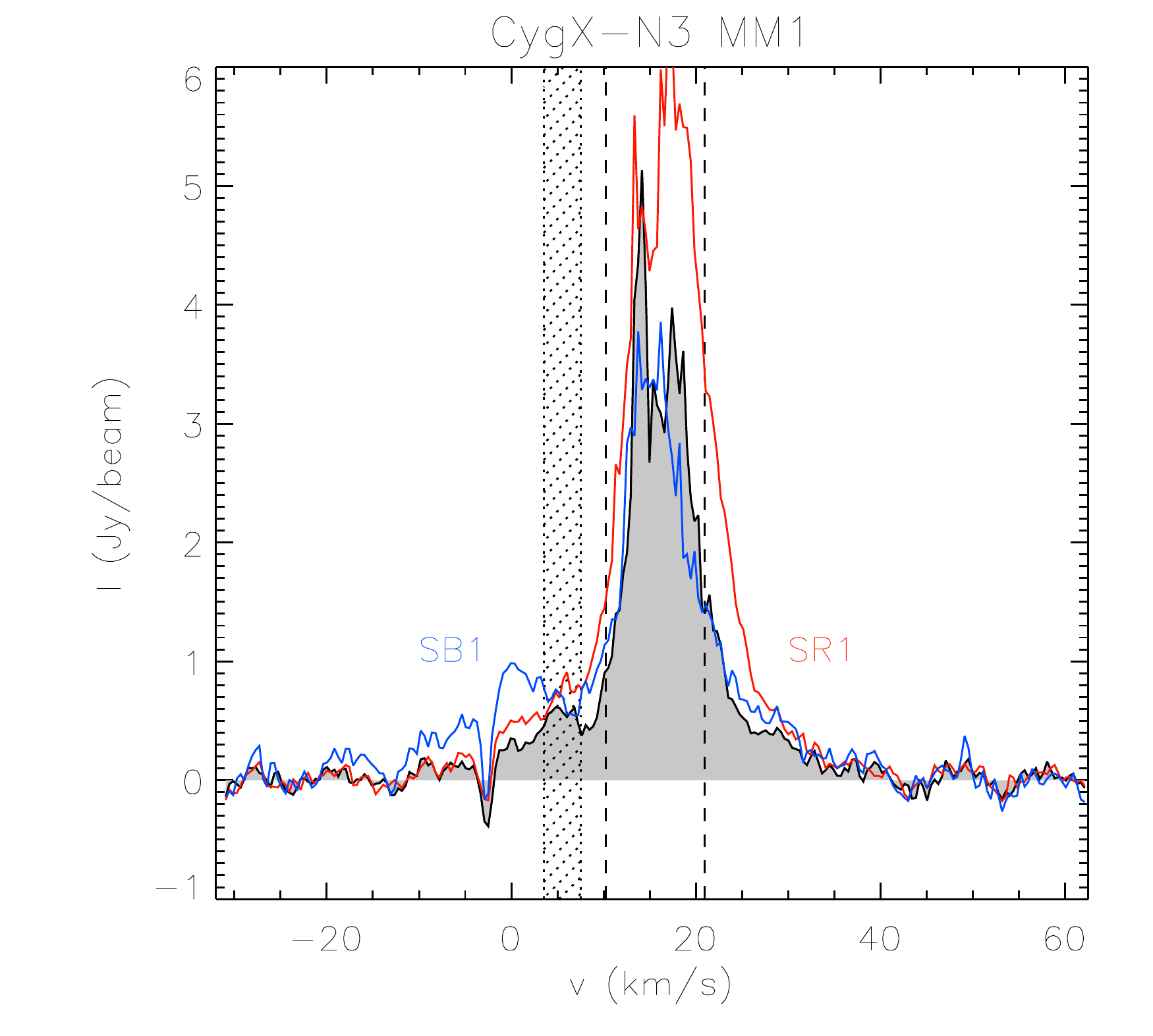}}\\
	\vspace{0.5cm} 
	\hbox{\vspace{-6cm} \includegraphics[width=0.6\textwidth]{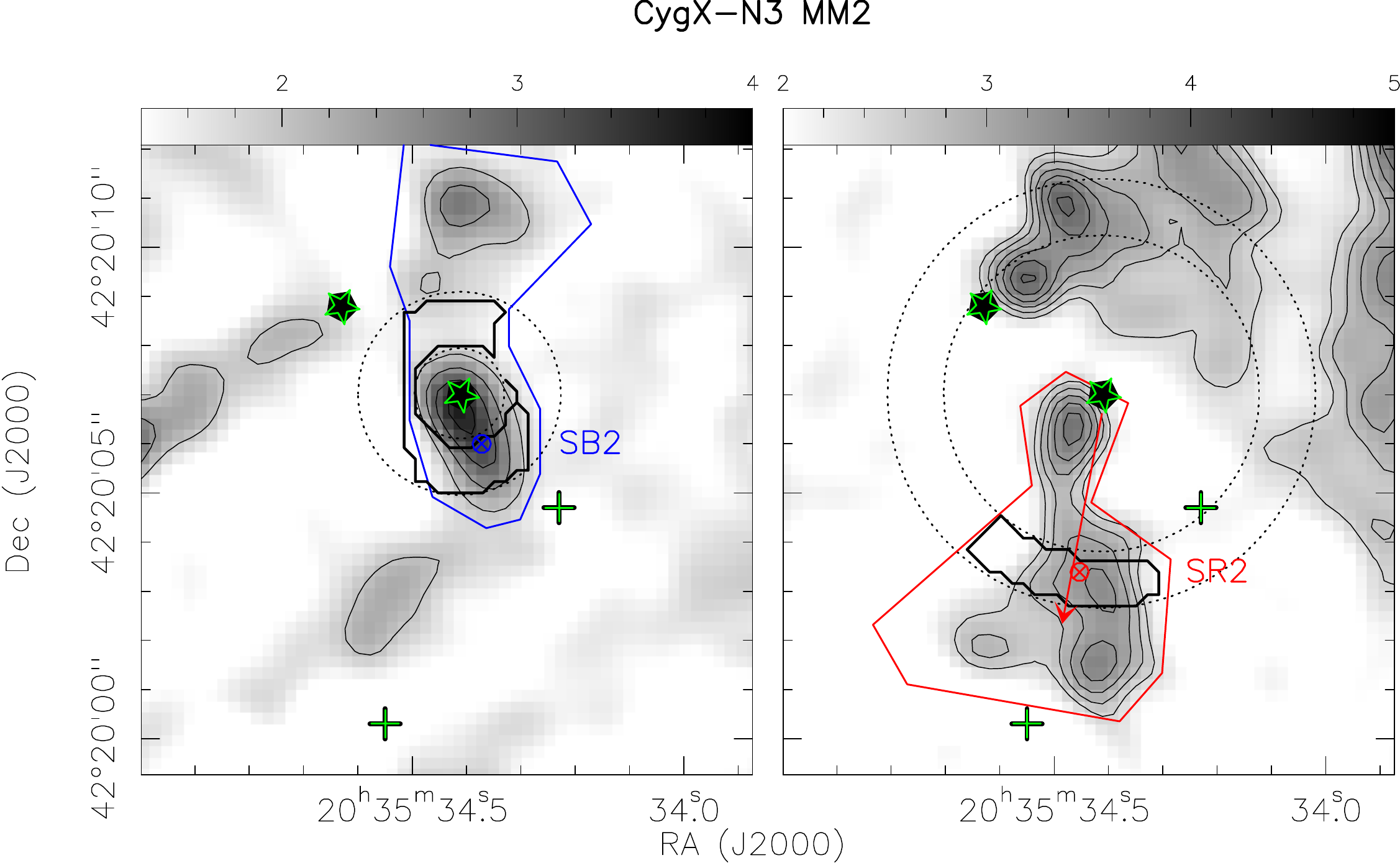}}	
	\hspace{11cm} \hbox{\includegraphics[width=0.37\textwidth]{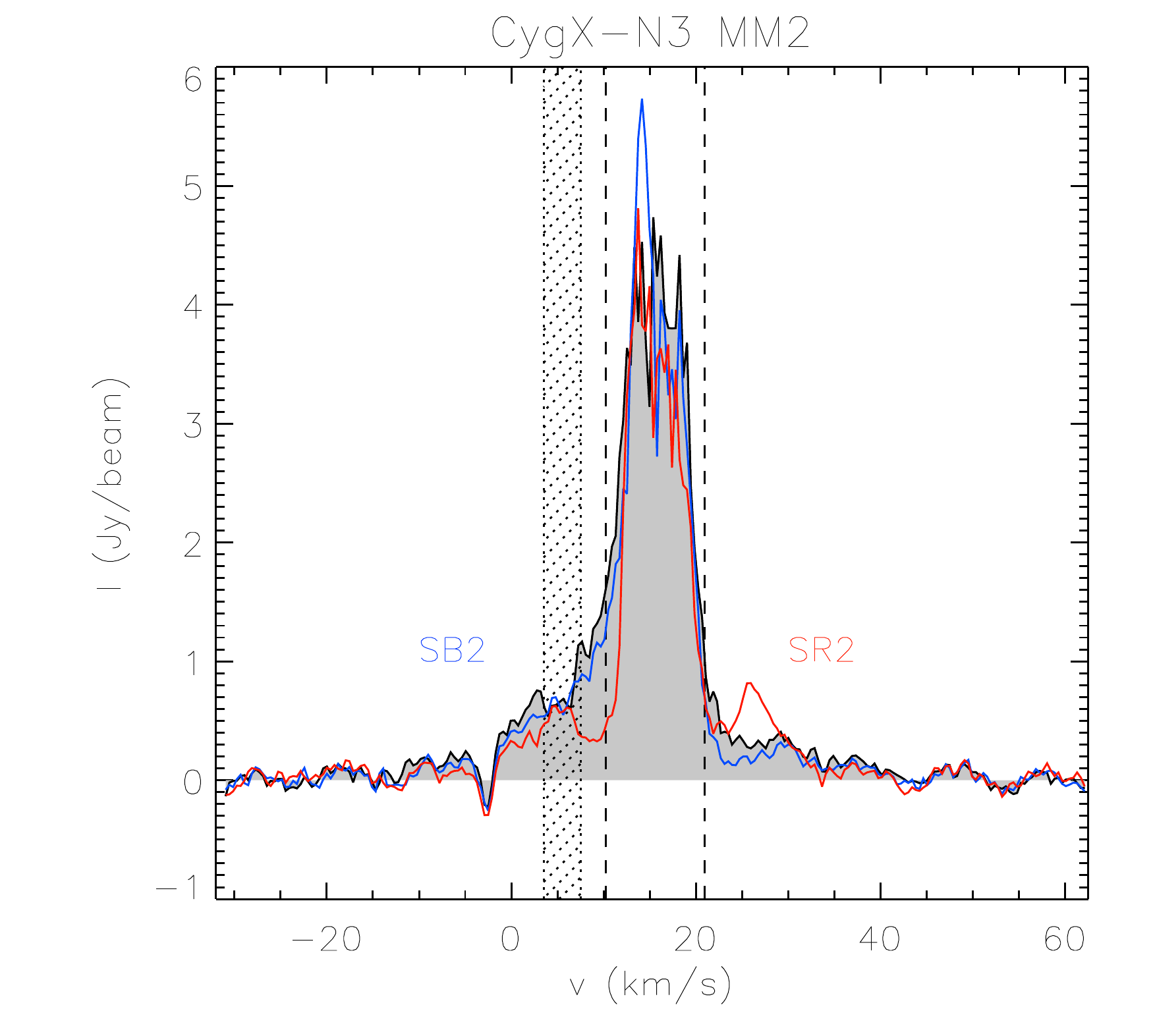}}
	\caption[]{\small{\textit{Left}: Blue and red CO emission in grey scale and contours for CygX-N3 MM1 (top) and MM2 (bottom). For clarity purposes only, the shown integrated intensity for MM2 red emission covers a narrower velocity range than used for the actual estimation. The intersection between the polygons and rings are the areas taken to measure the respective momentum flux for each wing. The blue and red crosses show the positions of the spectra shown on the right panel. \textit{Right}: Spectra at the position of the source in grey (CygX-N3 MM1 in the top panel and MM2 in the lower), with the spectra at the peak of the blue and red emission (SB1 and SR1 for MM1, SB2 and SR2 for MM2). The vertical dashed lines constrain the systemic velocities of the cloud excluded for the momentum flux calculations. The shaded area shows the velocity range affected by a cloud in front, also excluded.}}
	\label{fig:n3_rings}}
\end{figure*}

\begin{figure*}[!t]
	\centering
	{\renewcommand{\baselinestretch}{1.1}
	\hbox{\vspace{-6cm} \includegraphics[width=0.6\textwidth]{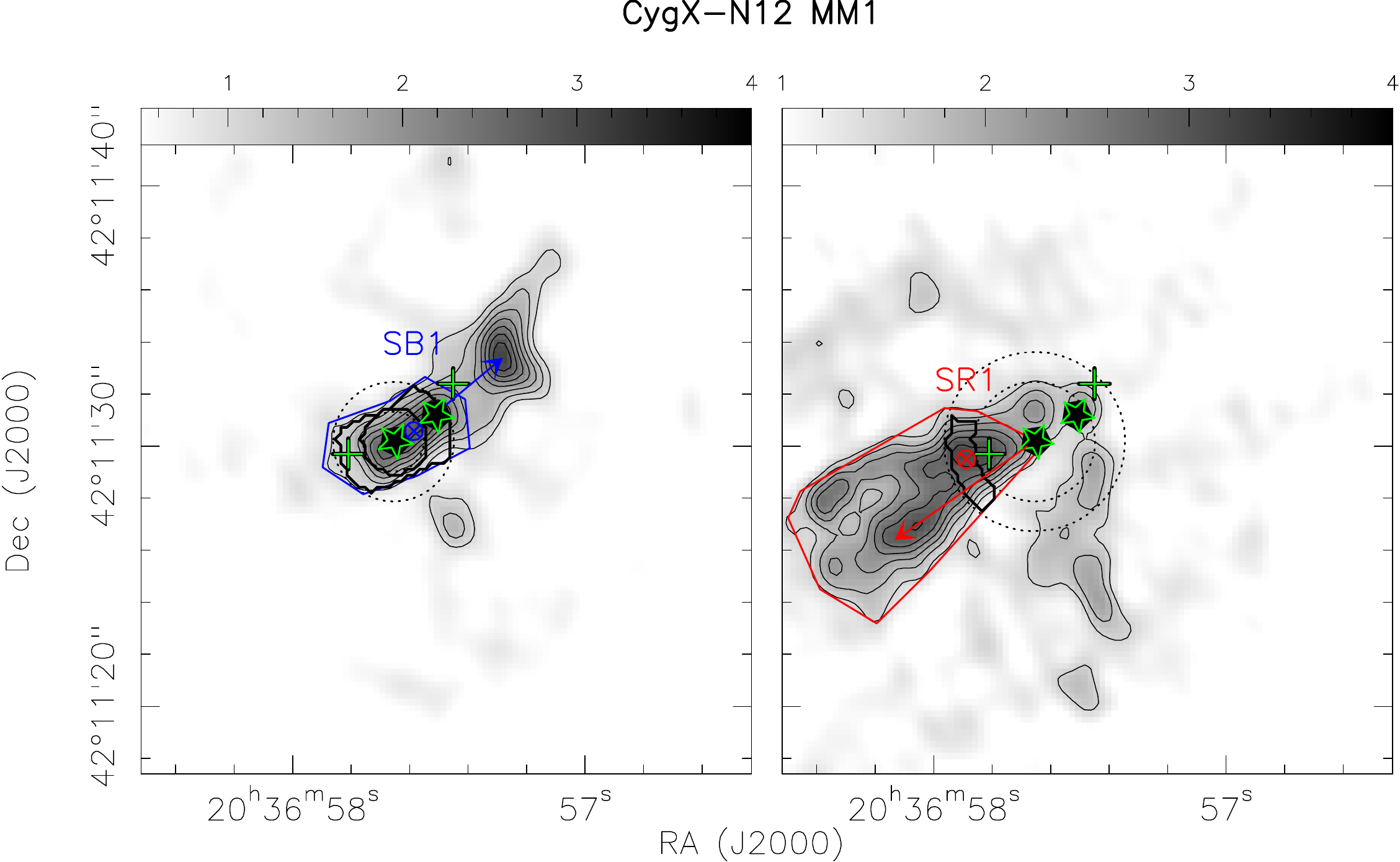}}	
	\hspace{11cm} \hbox{\includegraphics[width=0.37\textwidth]{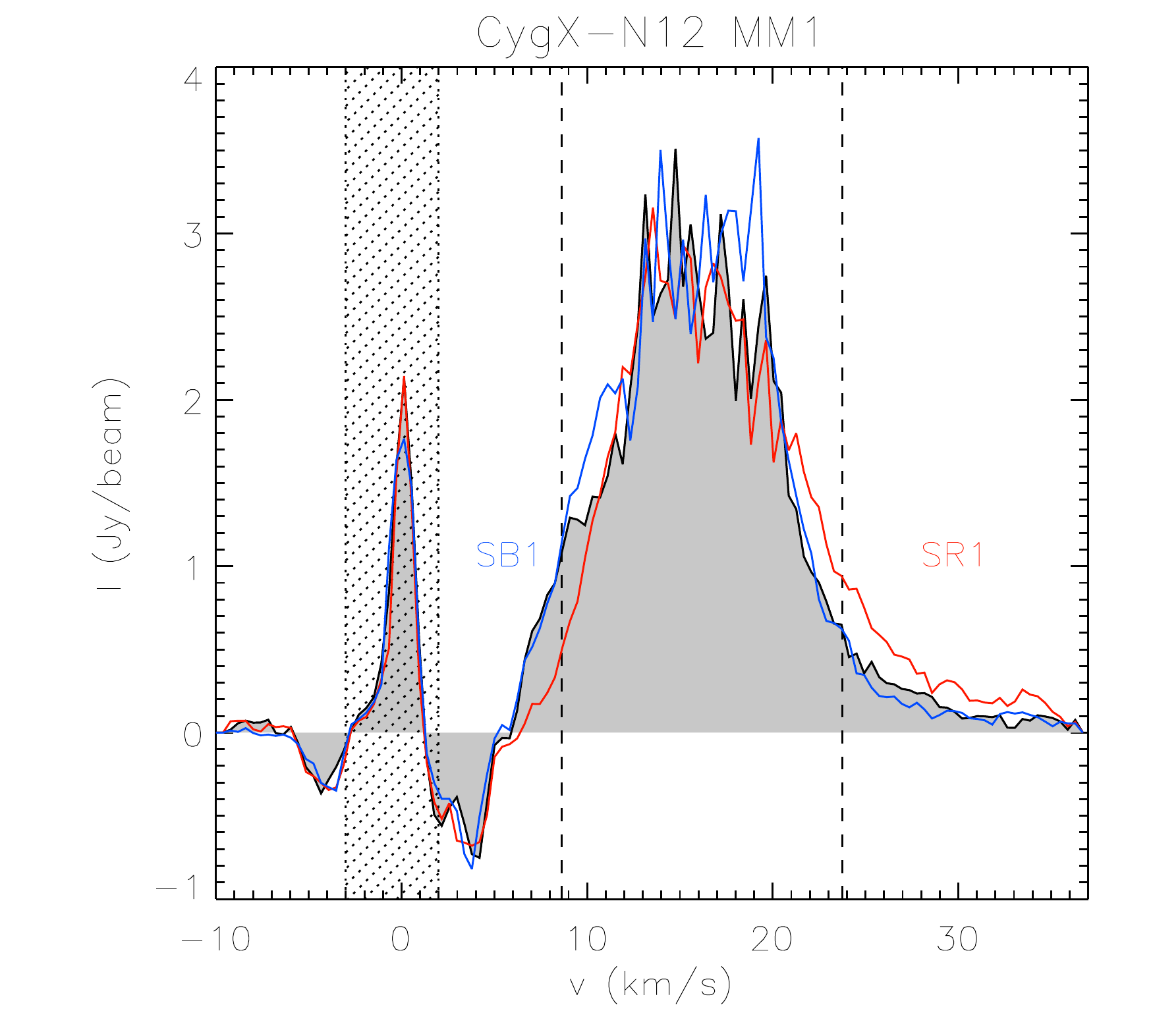}}\\
	\vspace{0.5cm} 
	\hbox{\vspace{-6cm} \includegraphics[width=0.6\textwidth]{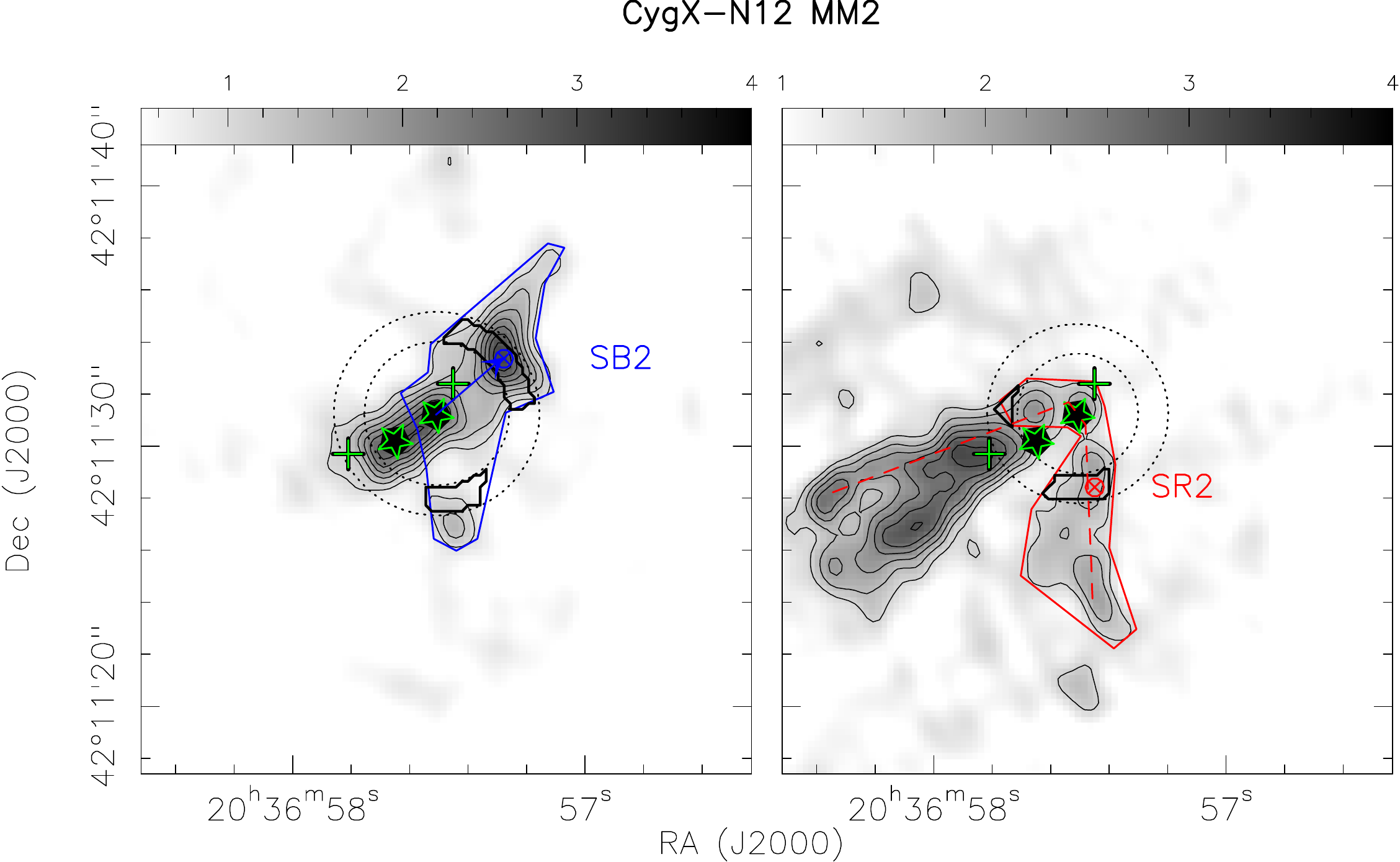}}	
	\hspace{11cm} \hbox{\includegraphics[width=0.37\textwidth]{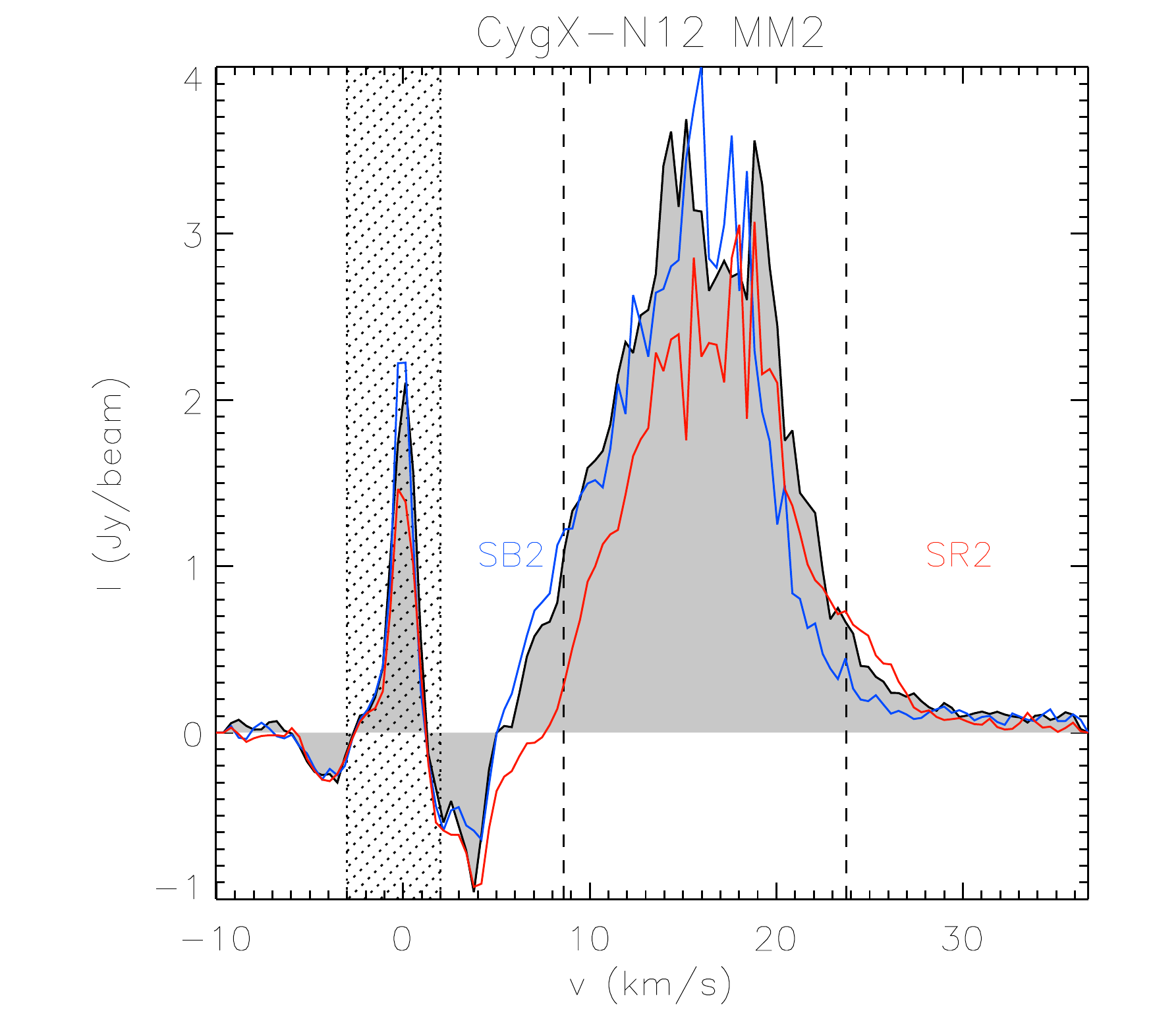}}
	\caption[]{\small{Same as Fig.~\ref{fig:n3_rings}, for CygX-N12 MM1 (top) and MM2 (bottom).}}
	\label{fig:n12_rings}}
\end{figure*}

\begin{figure*}[!t]
	\centering
	{\renewcommand{\baselinestretch}{1.1}
	\hbox{\vspace{-6cm} \includegraphics[width=0.6\textwidth]{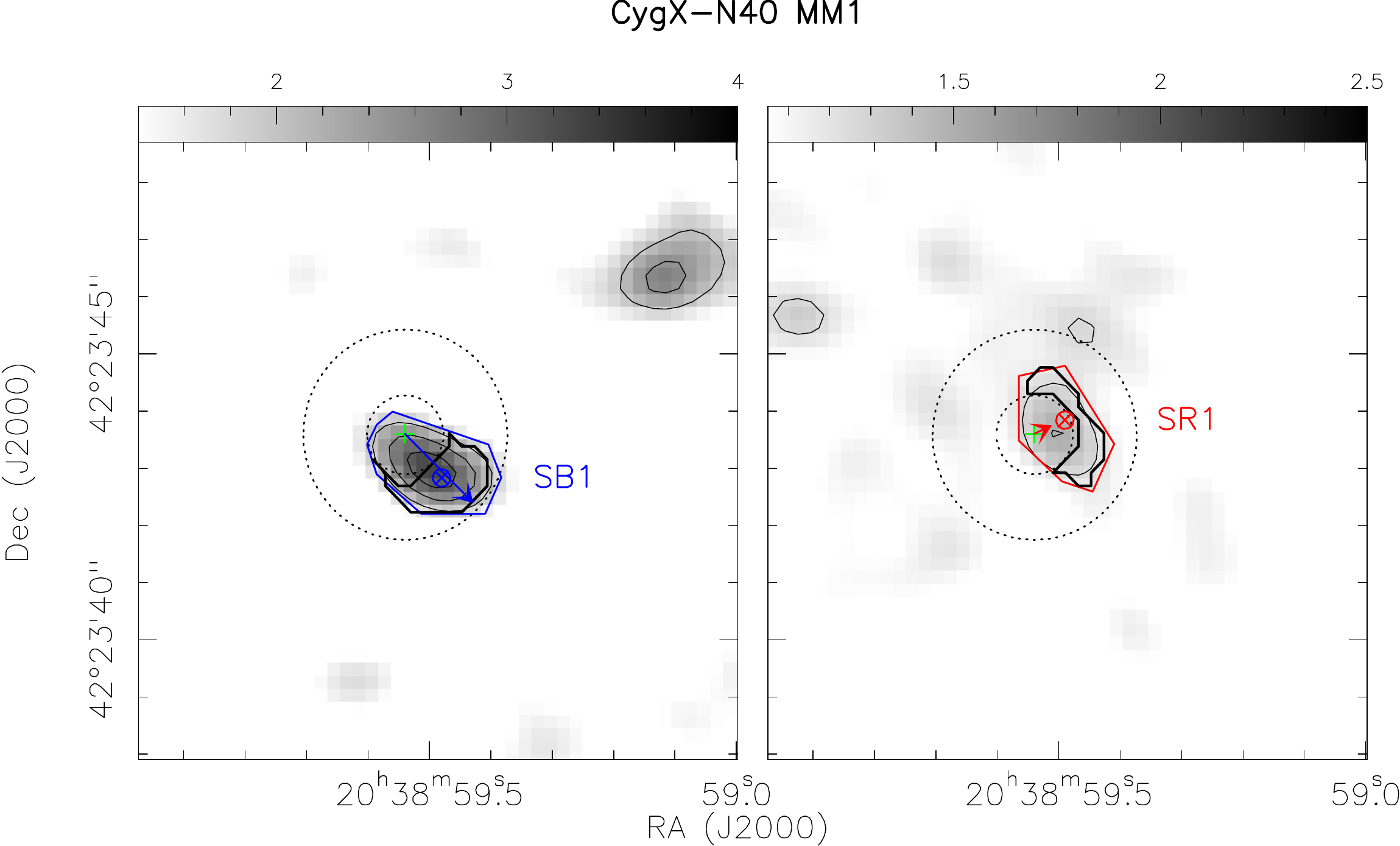}}	
	\hspace{11cm} \hbox{\includegraphics[width=0.37\textwidth]{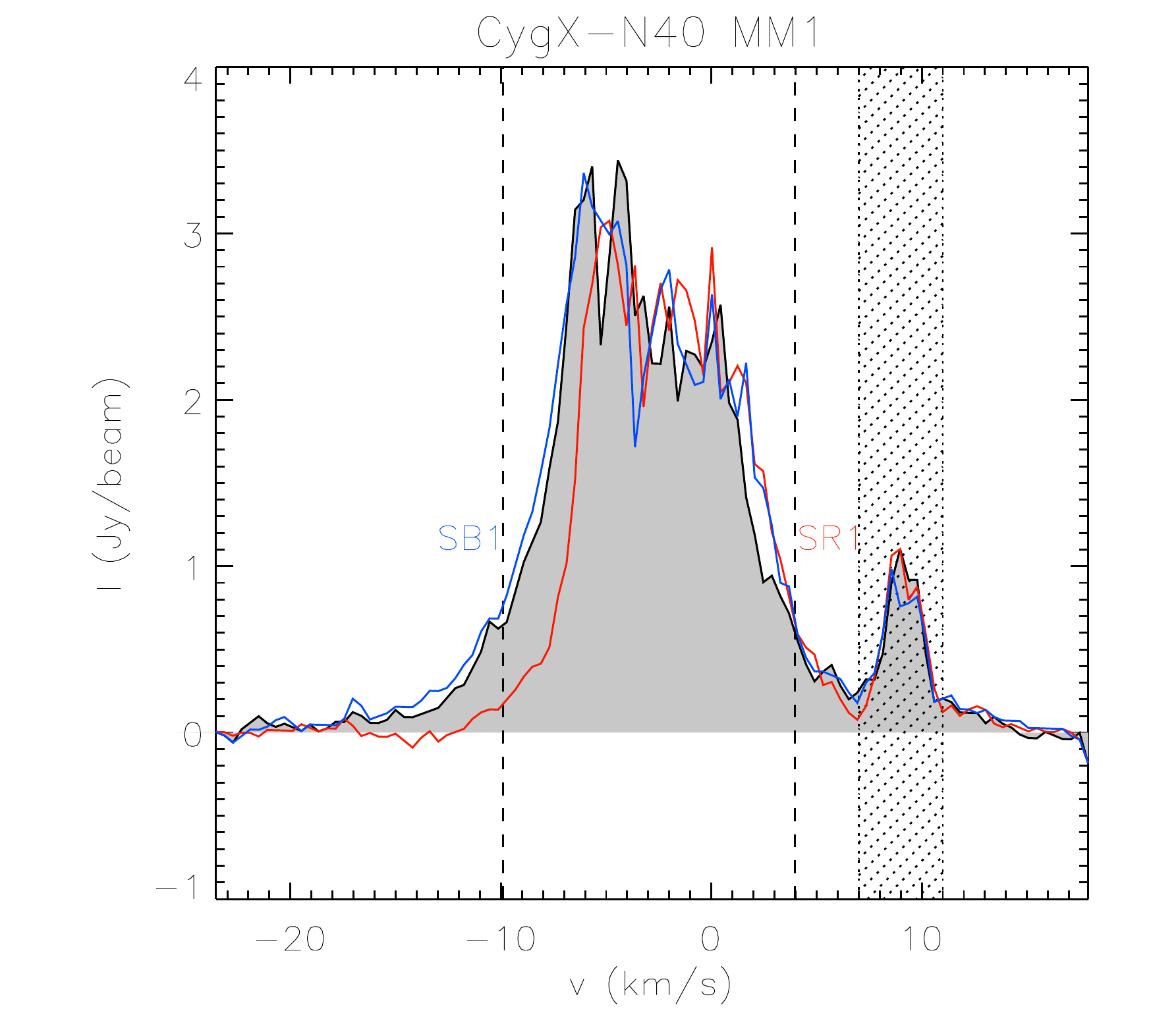}}
	\caption[]{\small{Same as Fig.~\ref{fig:n3_rings}, for CygX-N40 MM1.}}
	\label{fig:n40_rings}}
\end{figure*}

\begin{figure*}[!t]
	\centering
	{\renewcommand{\baselinestretch}{1.1}
	\hbox{\hspace{-0.5cm}\vspace{-6cm} \includegraphics[width=0.625\textwidth]{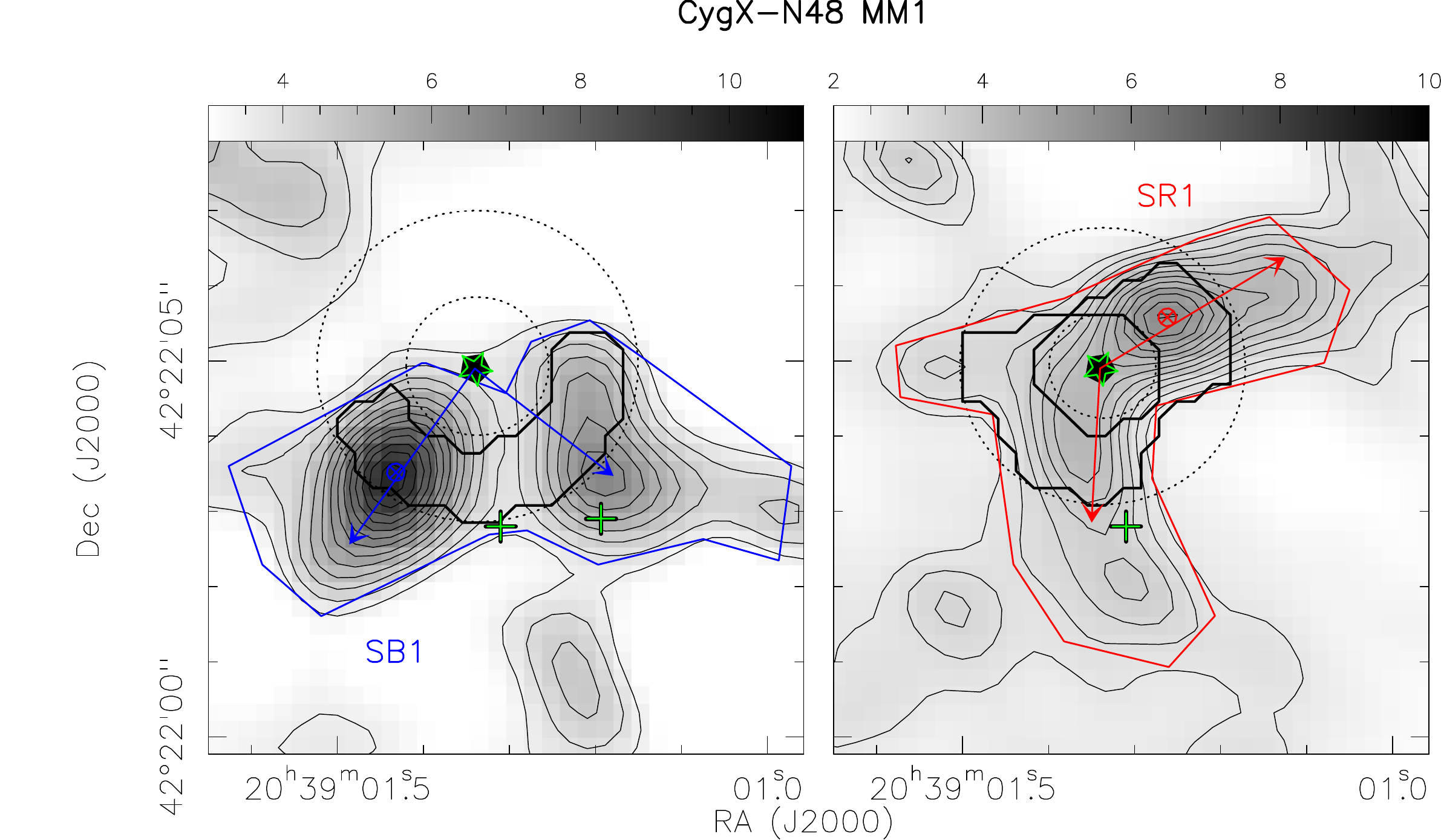}}	
	\hspace{11cm} \hbox{\includegraphics[width=0.37\textwidth]{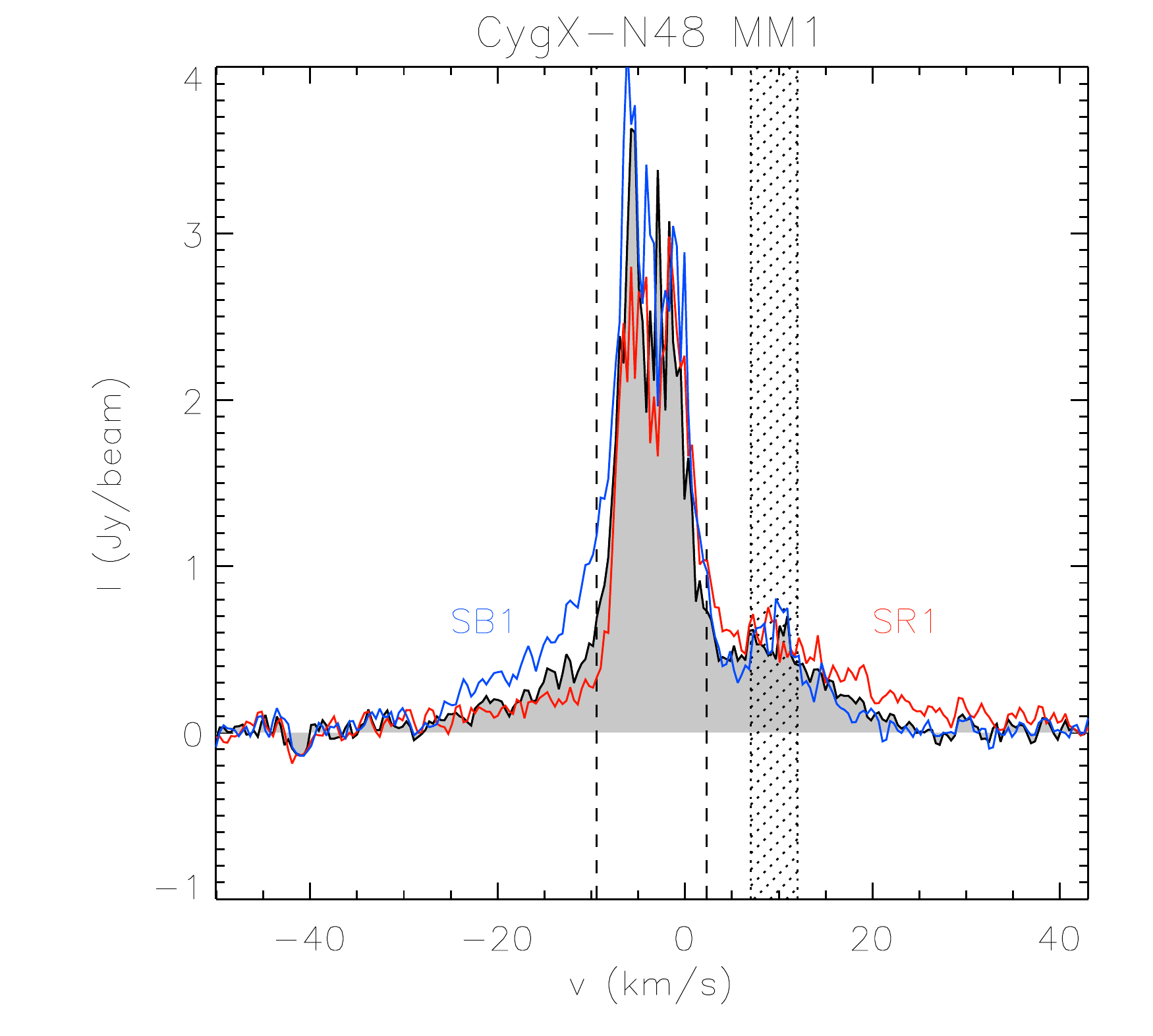}}\\
	\vspace{0.5cm} 
	\hbox{\vspace{-6cm} \includegraphics[width=0.6\textwidth]{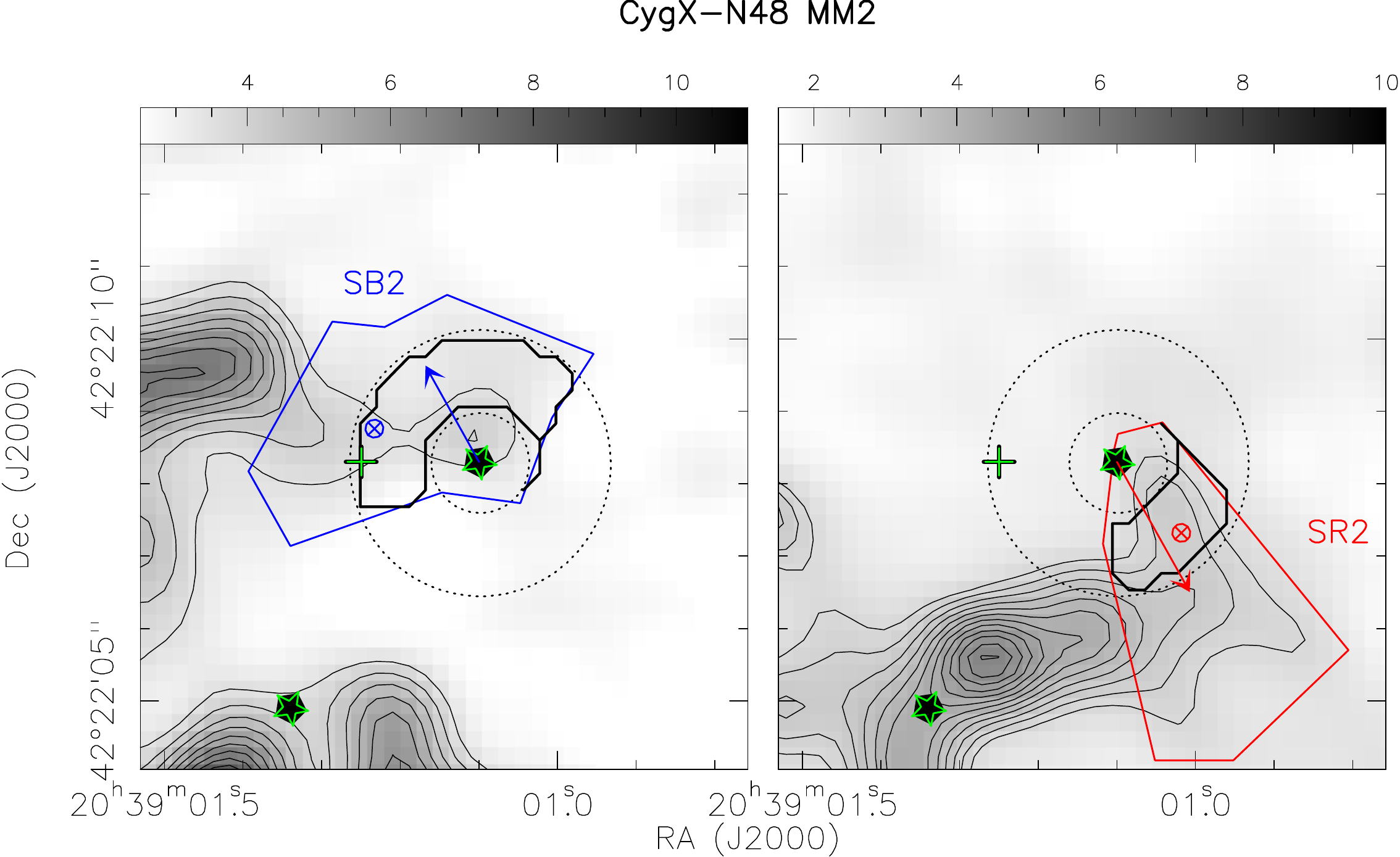}}	
	\hspace{11cm} \hbox{\includegraphics[width=0.37\textwidth]{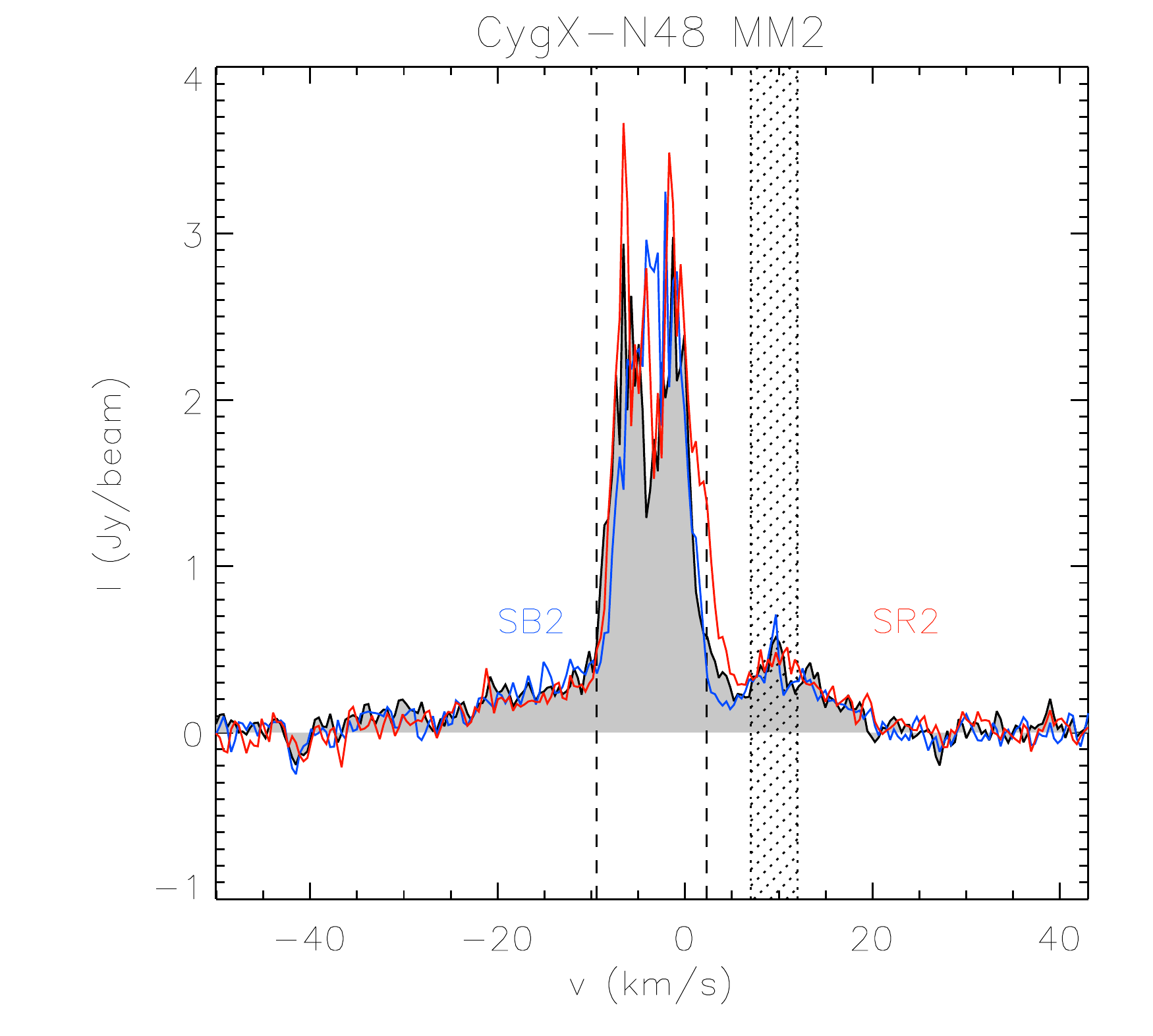}}
	\caption[]{\small{Same as Fig.~\ref{fig:n3_rings}, for CygX-N48 MM1 (top) and MM2 (bottom).}}
	\label{fig:n48_rings}}
\end{figure*}

\begin{figure*}[!t]
	\centering
	{\renewcommand{\baselinestretch}{1.1}
	\hbox{\vspace{-6cm} \includegraphics[width=0.6\textwidth]{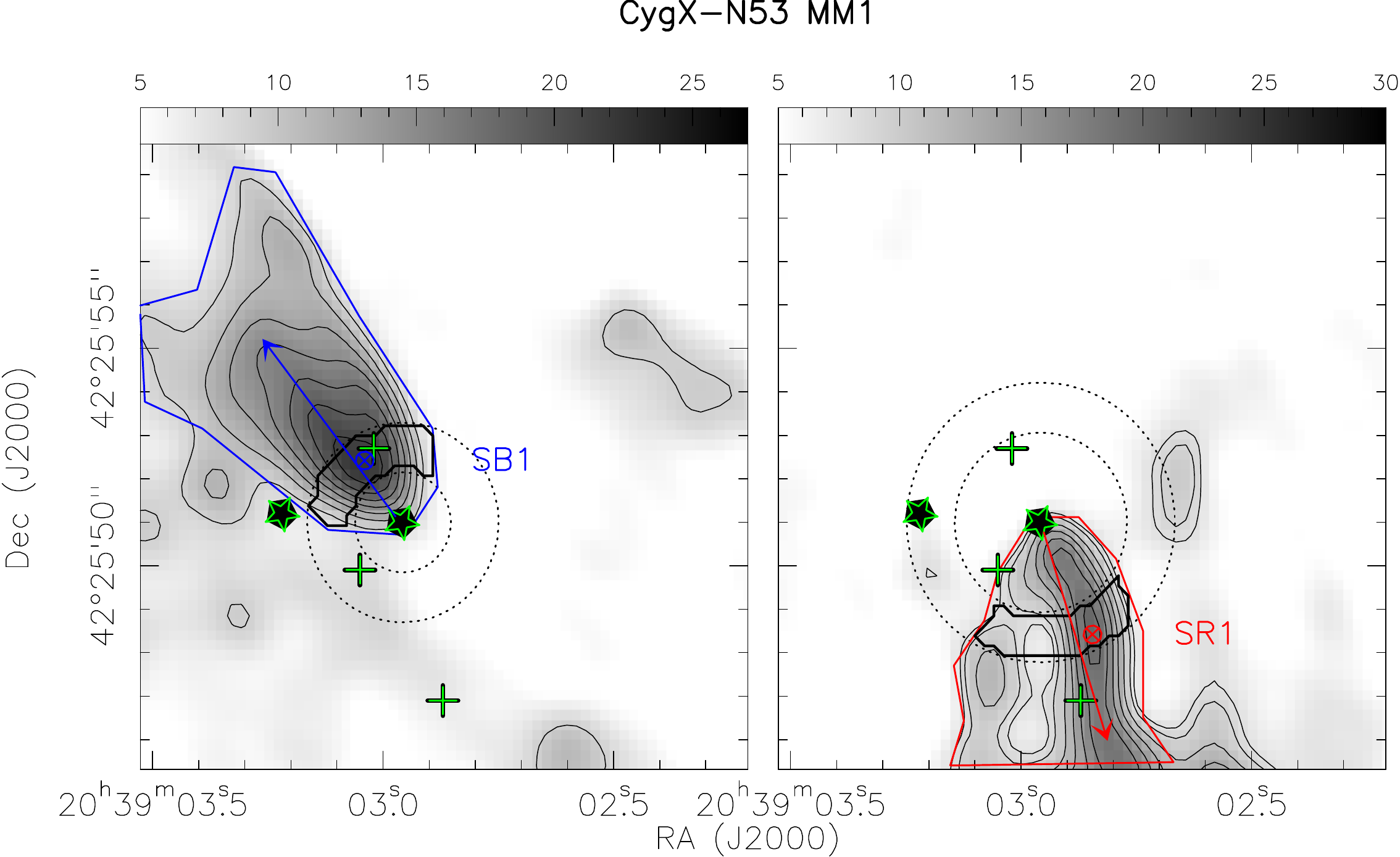}}	
	\hspace{11cm} \hbox{\includegraphics[width=0.37\textwidth]{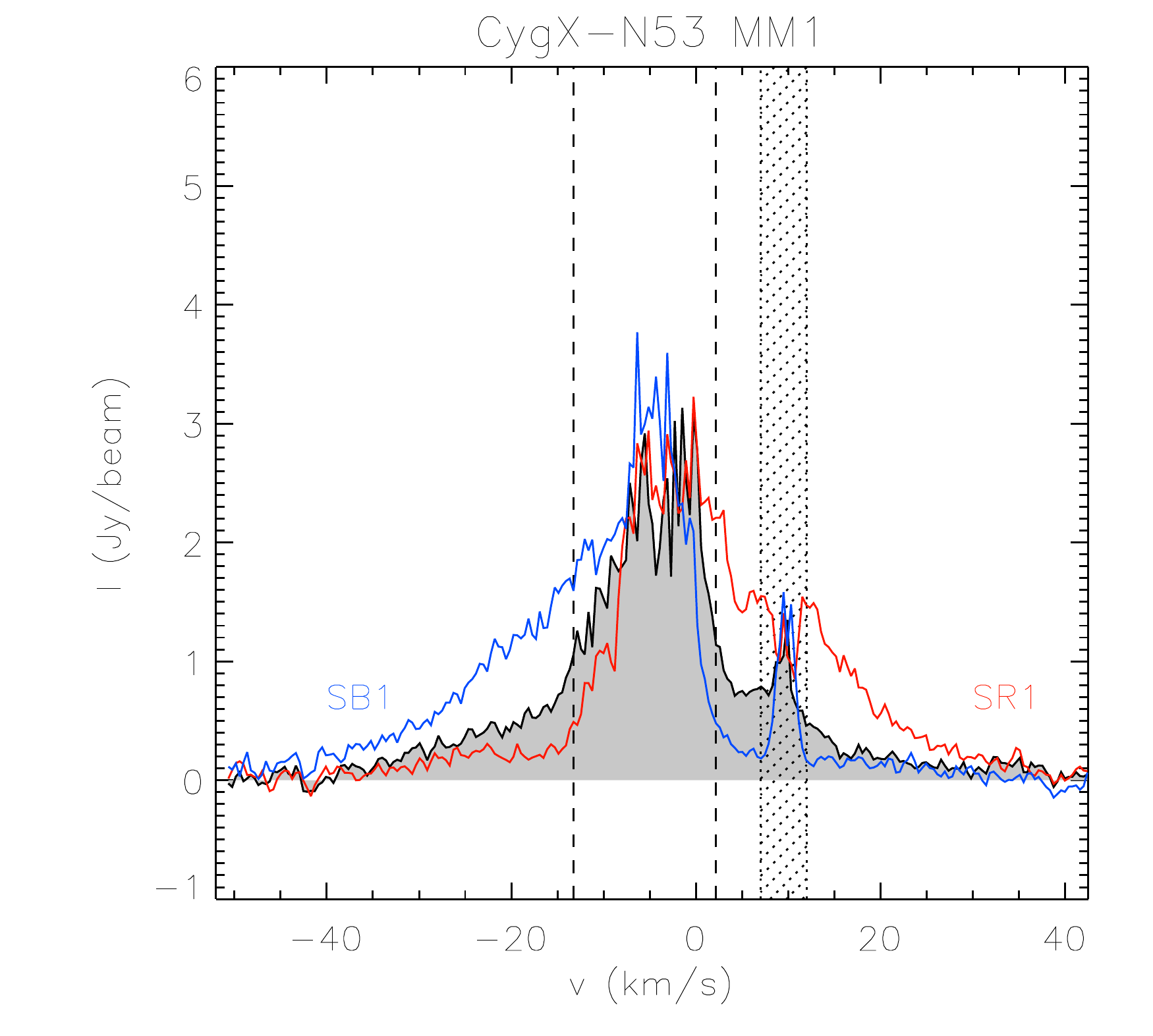}}\\
	\vspace{0.5cm} 
	\hbox{\vspace{-6cm} \includegraphics[width=0.6\textwidth]{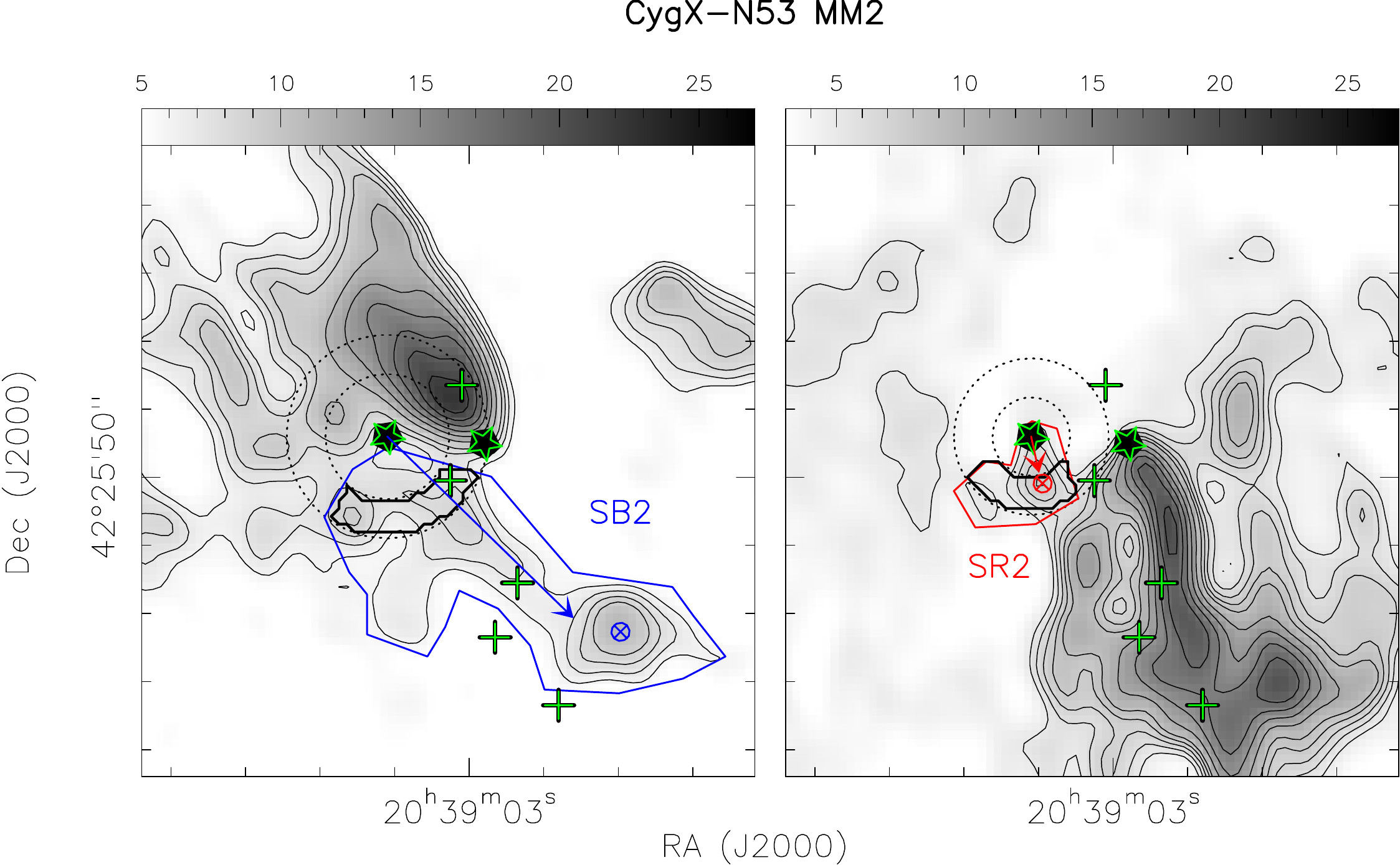}}	
	\hspace{11cm} \hbox{\includegraphics[width=0.37\textwidth]{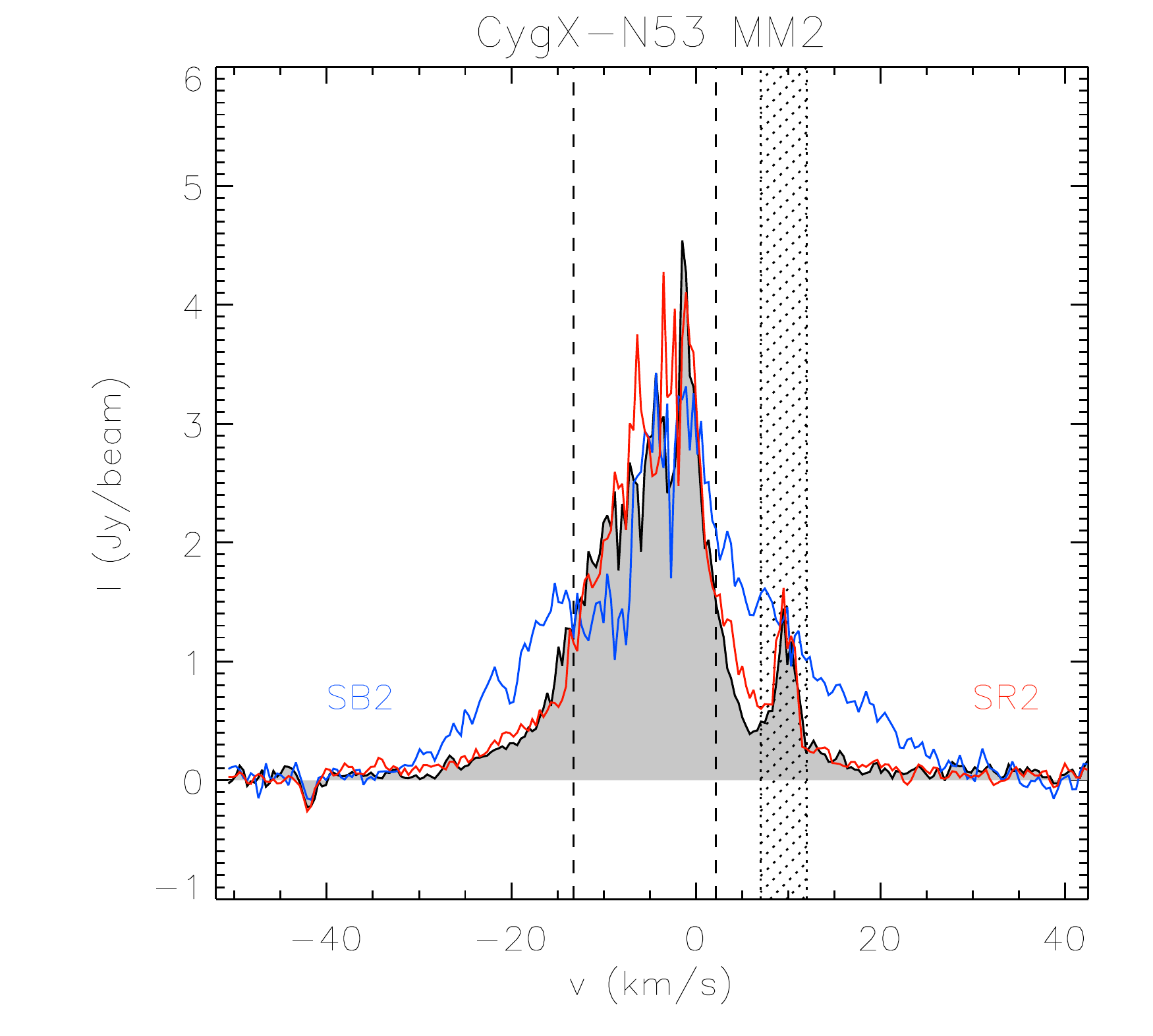}}
	\caption[]{\small{Same as Fig.~\ref{fig:n3_rings}, for CygX-N53 MM1 (top) and MM2 (bottom).}}
	\label{fig:n53_rings}}
\end{figure*}

\begin{figure*}[!t]
	\centering
	{\renewcommand{\baselinestretch}{1.1}
	\hbox{\vspace{-6cm} \includegraphics[width=0.6\textwidth]{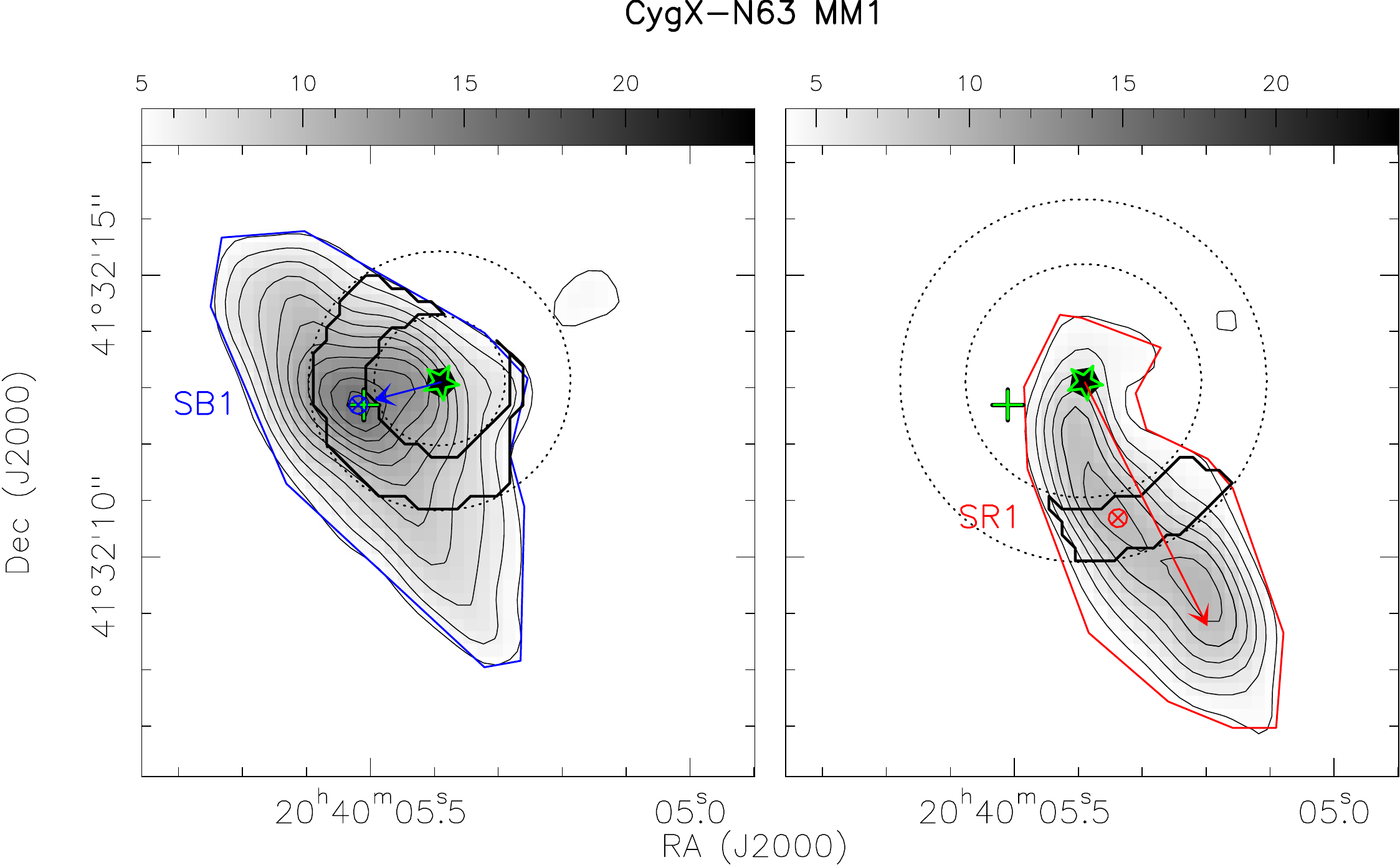}}	
	\hspace{11cm} \hbox{\includegraphics[width=0.37\textwidth]{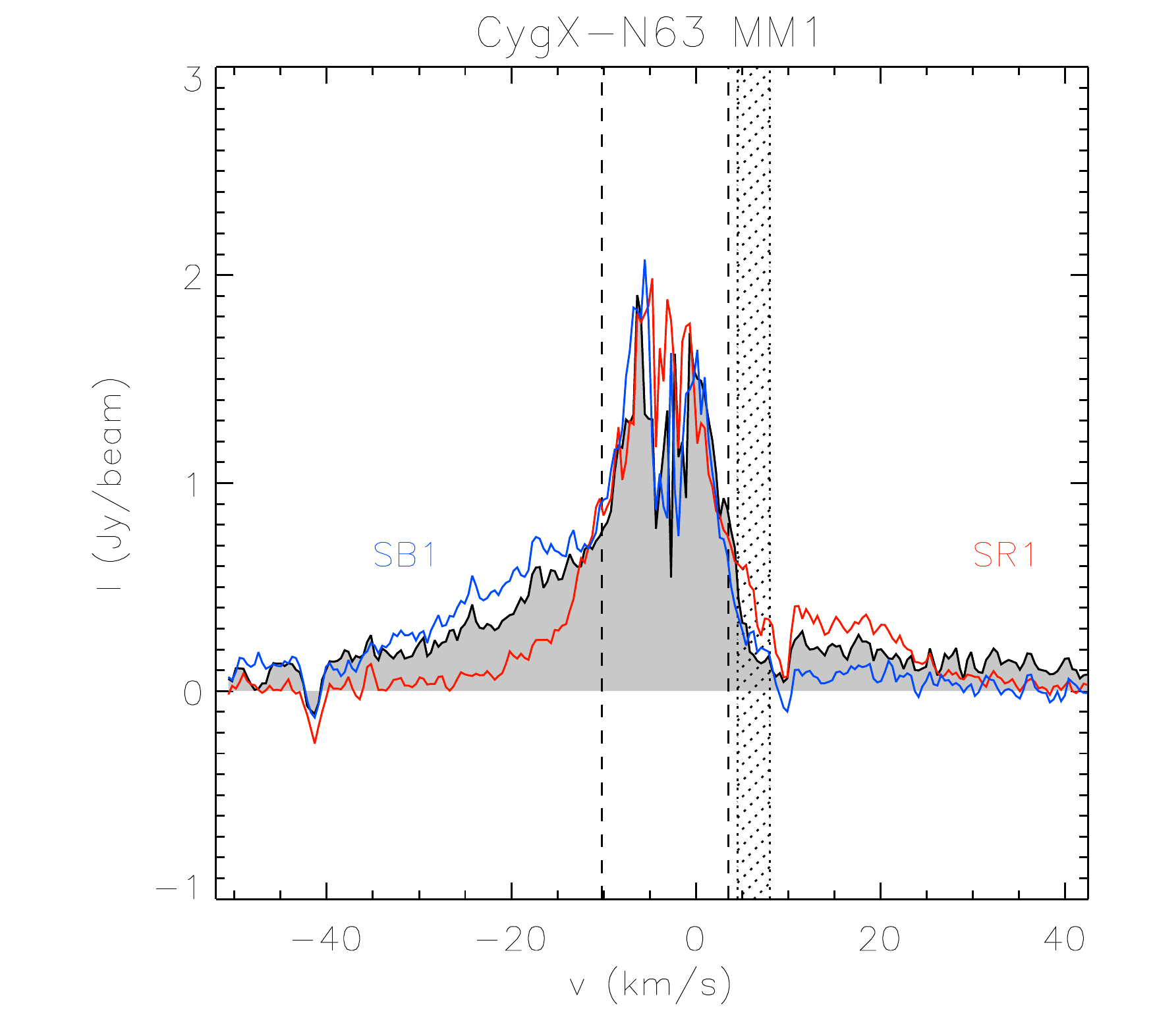}}
	\caption[]{\small{Same as Fig.~\ref{fig:n3_rings}, for CygX-N63 MM1.}}
	\label{fig:n63_rings}}
\end{figure*}

As mentioned in Sect.~\ref{sec:outflow_fco}, to estimate the energetics of the outflows from this sample of sources, we used an approach similar to what \citet[][]{1996A&A...311..858B} used for a sample of low-mass protostellar objects. This method assumes that the momentum flux is conserved along the outflow direction, and it consists of estimating the momentum flux of the outflows on a ring centred on the driving source. The velocity ranges used to integrate the red and blue emission had been estimated using the average spectrum over each region, and they are illustrated in the right-hand panels of Figs.~\ref{fig:n3_rings} to \ref{fig:n63_rings} (with vertical dashed lines). In these panels we show the spectrum at the position of each source, as well as an example of a blue and red-shifted spectra (whose positions are shown in the left-hand panels).

We then estimated the momentum fluxes by integrating across an annulus of fixed width, $\Delta r$, of 1.15'' (i.e. chosen to correspond to the beam FWHM). The inner and outer radii of the annulus, however, are source and lobe-dependent: they are chosen so that the ring comprises the peak of integrated intensity of the respective outflow in the respective lobe (blue or red). The choice of using a variable inner and outer radius aims at maximising the momentum flux estimate, by covering the surroundings of the protostar, which are likely to harbour the most powerful impact of such outflows. To avoid an increase in noise due to non-outflowing gas and contamination from neighbour outflows within the annulus, we defined polygons based on the integrated intensity maps of the blue and the red-shifted emission, to constrain the spacial extent of individual outflows. The areas (rings and polygons) used for each source, and each lobe, are shown in Figs.~\ref{fig:n3_rings} to \ref{fig:n63_rings}. The momentum flux is then estimated for each outflow wing inside the intersection between the rings and the polygon, as
\begin{equation}
F_{\rm{CO}} \propto \int \frac{T \, (v-v_o)^{2} dv }{\Delta r}
\label{eq:mom_flux}
\end{equation}
where $T$ is the main beam intensity (in K) and $v_{o}$ the velocity of the ambient cloud. For these calculations, we assumed a distance of 1.4~kpc, a fractional abundance of CO to H$_{2}$ of $10^{4}$, a temperature of 20~K, a molecular weight of 2.33 and an average correction factor of 3.5 for the opacities of $^{12}$CO wings \citep[][]{1992A&A...261..274C}. At this point we do not, however, correct for any inclination effect of the outflow direction against the line of sight.

The results from these estimates are summarised on Table~\ref{tab:energetics}, where $v_{o}$ is the ambient velocity of the cloud \citep[as from the N$_{2}$H$^{+}$ from][]{2007A&A...476.1243M,2010A&A...524A..18B}, $v$ range is the velocity range used for estimating the momentum flux (chosen using the Gaussian fittings, and excluding velocities where there was emission from another cloud), r$_{\mathrm{in}}$ is the inner radius of the annulus taken for flux measurements, and  $F_{\rm{co}}$ is the momentum flux calculated for the blue and red wing emission. Assuming that the outflows are symmetrically bipolar and that the amount of momentum released through one lobe is the same as released on the other, our best estimate of the total momentum flux ejected is not the sum of the two lobes, but twice the estimate of the wing with the highest momentum flux. Such estimates of the total momentum flux of individual outflows are shown as the total  $F_{\rm{co}}$ (last column of Table~\ref{tab:energetics}), now corrected by a possible effect of the inclination angle \citep[factor 2.9 from a random distribution of angles,][]{1992A&A...261..274C}. Also worth noting is that we may be missing some momentum flux since we do not necessarily pick up the highest velocity material likely carrying a significant part of the momentum flux. Given all the assumptions on estimating the momentum flux, the uncertainties can easily be as high as a factor 2.

\begin{table*}[!t]
\small
\centering
\caption{Outflow energetics}
\begin{tabular}{l | c  | c c c | c c c | c}
\hline 
\hline 
		&			& & Blue & 				& & Red & &Total \\
Source 	& $v_{o}$  	& $v$ range & r$_{\mathrm{in}}$ 	&  $F_{\rm{co}}$ $/ 10^{-5}$ & $v$ range & r$_{\mathrm{in}}$ 		& $F_{\rm{co}}$ $/ 10^{-5}$ &$F_{\rm{co}}$ 	$/ 10^{-5}$ \\
		& (km\,s$^{-1}$)  &  (km\,s$^{-1}$) & (AU) &  (M$_{\odot}$\,km\,s$^{-1}$\,yr$^{-1}$) &  (km\,s$^{-1}$) & (AU) &  (M$_{\odot}$\,km\,s$^{-1}$\,yr$^{-1}$) &  (M$_{\odot}$\,km\,s$^{-1}$\,yr$^{-1}$)\\  
\hline
\scriptsize{CygX-N3\,MM1}	&	14.9	& 	[-20,\,3.5],\,[7.5,\,10]   	& 7762	& $22.6\pm3.3$ 	&	[21,\,51]	& 3105	& $15.8 \pm 1.9$ & 131 \\
\scriptsize{CygX-N3\,MM2}	&	14.9	& 	[-20,\,3.5],\,[7.5,\,10]  	& 1242	& $12.4\pm1.8$	&	[21,\,51] 	& 4347	& $11.8 \pm 2.1$ & 72 \\
\hline
\scriptsize{CygX-N12\,MM1}	&	15.2	& 	[5,\,9] 	& 1552	& $1.7 \pm 0.2$	& 	[24,\,37]	& 3105	& $6.2 \pm 1.1$ & 36 \\
\scriptsize{CygX-N12\,MM2}	&	15.2	& 	[5,\,9]	& 3726 	& $1.7 \pm 0.2$	& 	[24,\,37]	& 3105	& $2.1 \pm 0.3$ & 12 \\
\hline
\scriptsize{CygX-N40\,MM1}	&	-3.5	& 	[-23,\,-10]	& 931	& $0.5 \pm 0.1$	&  [5,\,7], [11,\,18] 	& 931	& $1.3  \pm 0.2$ & 7 \\
\hline
\scriptsize{CygX-N48\,MM1}	&	-3.5	& 	[-30,\,-9.5]	& 1242	& $12.4 \pm 2.3$	&  [2.5,\,7], [12,\,25]	& 931	& $23.3 \pm 3.9$ & 135 \\
\scriptsize{CygX-N48\,MM2}	&	-3.5	& 	[-30,\,-9.5]	& 931 	& $7.8 \pm 1.7$	&  [2.5,\,7], [12,\,25]	& 931	& $7.1 \pm 1.3$ & 35 \\
\hline
\scriptsize{CygX-N53\,MM1}	&	-4.4	& 	[-50,\,-13.5]	& 1552	& $71.0 \pm 9.6$	&  [2.5,\,7], [12,\,40]	& 2794 & $64.6 \pm 8.6$ & 412 \\
\scriptsize{CygX-N53\,MM2}	&	-4.4	& 	[-50,\,-13.5]	& 2484 	& $21.0 \pm 3.7$	&  [2.5,\,7], [12,\,40]	& 1551 & $16.5 \pm 2.2$ & 121 \\
\hline
\scriptsize{CygX-N63\,MM1}	&	-4.5	& 	[-50,\,-9]	& 1552	& $50.1 \pm 8.0$	&  [2,\,6.5], [8,\,40]	& 2794	& $18.2 \pm 3.2$ & 291 \\
\end{tabular}
\label{tab:energetics}
\end{table*}

\end{appendix}

\end{document}